\documentclass{emulateapj}
\usepackage{graphicx}
\newcommand{\secpoint}{\mbox{$''\mskip-7.6mu.\,$}}

\slugcomment{Received DATE; accepted DATE}

\begin{document}

\title{Stellar Populations of UV-Selected Active Galactic Nuclei Host Galaxies at $z \sim 2-3$\altaffilmark{1}}

\shorttitle{UV-SELECTED AGN STELLAR MODELING}
\shortauthors{HAINLINE ET AL.}

\author{\sc Kevin N. Hainline, Alice E. Shapley\altaffilmark{2}}
\affil{Department of Astronomy, University of California,
Los Angeles, 430 Portola Plaza, Los Angeles, CA 90024}

\author{\sc Jenny E. Greene}
\affil{Department of Astrophysical Sciences, Princeton University, Princeton, NJ 08544}

\author{\sc Charles C. Steidel}
\affil{Department of Astronomy, California Institute of Technology, MS 105-24, Pasadena, CA 91125}

\author{\sc Naveen A. Reddy}
\affil{Department of Astronomy, University of California, Riverside, Riverside, CA 92521}

\author{\sc Dawn K. Erb}
\affil{Department of Physics, University of Wisconsin-Milwaukee, Milwaukee, WI 53201}

\author{}

\altaffiltext{1}{Based, in part, on data obtained at the W.M. Keck
Observatory, which is operated as a scientific partnership among the
California Institute of Technology, the University of California, and
NASA, and was made possible by the generous financial support of the W.M.
Keck Foundation.}
\altaffiltext{2}{David and Lucile Packard Fellow}

\begin{abstract}

We use stellar population synthesis modeling to analyze the host galaxy properties of a sample of 33 UV-selected, narrow-lined active galactic nuclei (AGNs) at $z\sim 2-3$. In order to quantify the contribution of AGN emission to host galaxy broadband spectral energy distributions (SEDs), we use the subsample of 11 AGNs with photometric coverage spanning from rest-frame UV through near-IR wavelengths. Modeling the SEDs of these objects with a linear combination of stellar population and AGN templates, we infer the effect of the AGN on derived stellar population parameters. We also estimate the typical bias in derived stellar populations for AGNs lacking rest-frame near-IR wavelength coverage, and develop a method for inferring the true host galaxy properties. We compare AGN host galaxy properties to those of a sample of UV-selected, star-forming non-AGNs in the same redshift range, including a subsample carefully matched in stellar mass. Although the AGNs have higher masses and SFRs than the full non-active sample, their stellar population properties are consistent with those of the mass-selected sample, suggesting that the presence of an AGN is not connected with the cessation of star-formation activity in star-forming galaxies at $z\sim 2-3$. We suggest that a correlation between $M_{BH}$ and galaxy stellar mass is already in place at this epoch. Assuming a roughly constant Eddington ratio for AGNs at all stellar masses, we are unable to detect the AGNs in low-mass galaxies because they are simply too faint. 

\end{abstract}


\section{Introduction}
\label{sec:intro}

Active galactic nuclei (AGNs) may play an important role in the evolution of their host galaxies, as evidenced by the the tight correlation between black hole mass and bulge mass in local galaxies \citep{ferrarese2000, gebhardt2000, gultekin2009}. AGNs and star formation are both fueled by gas accretion processes, and the gas content and temperature in galaxies can be influenced by the energy output from an AGN. This latter process is often used as an explanation for the distribution of luminosities and colors among massive galaxies in the local universe. Feedback from an AGN prevents further star formation, causing galaxy colors to redden as well as contributing to the exponential fall-off at the bright end of the galaxy luminosity function \citep{hopkins2006, croton2006, khalatyan2008,somerville2008}. However, strong evidence linking the presence of an AGN to the cessation of star formation is still missing \citep[but see][]{tremonti2007}. One popular method that has been used to examine this possible connection is the comparison of AGN host galaxy stellar populations to those of similar, non-active galaxies. With a controlled examination of the stellar populations of galaxies hosting AGNs, we can understand both the influence of an AGN on galaxy properties as well as the timescales involved in AGN activity. 

AGN host galaxy properties have been studied across a wide range of redshifts. In the local universe, results from AGNs identified based on their optical properties indicate that these objects are hosted by massive (log($M_{*} / M_{\sun}) > 10$), early-type galaxies \citep{kauffmann2003b}. Similar trends are observed for local radio-loud AGNs; the radio-loud AGN fraction is observed to be $>30\%$ for galaxies with log($M_{*} / M_{\sun}) > 11.7$ \citep{best2005}. Studies of optically- and X-ray-selected AGNs at intermediate redshifts ($z \sim 0.5 - 1.5$) indicate that these AGN host galaxies are massive, with characteristic stellar masses of log($M_{*}/M_{\sun}$) $\sim 11$ \citep{bundy2008,alonso2008,brusa2009}. If AGNs are responsible for quenching star formation, this effect should be reflected in the optical colors of the host galaxies. Optically-selected AGN hosts in the local universe are found to lie on the red sequence \citep{schawinski2007}, while, at higher redshifts, both optically-selected and X-ray selected AGNs reside in galaxies in the transition region between the blue and red sequences \citep{nandra2007,salim2007,coil2009,hickox2009,schawinski2010}. However, \citet{silverman2009a} found that X-ray selected AGNs at $z \sim 1$ often reside in galaxies with higher star-formation rates than similar-mass non-active galaxies, and that the AGN fraction is higher in the blue cloud. These results highlight the importance of comparing AGN hosts with control non-AGN samples matched in stellar mass in order to mitigate the selection effects associated with missing low-mass black holes in small galaxies.
 
To uncover evidence of ongoing processes linking AGNs to their hosts, it is important to examine AGNs at even higher redshifts, during the epoch when the bulge stellar population was forming. At $z\sim2 - 3$, both the star-formation-rate density and the black hole accretion rate peak in the universe, and this era is ideal for examining the relationship between AGN host galaxies and similar non-active hosts \citep{reddy2008,hopkins2007}. At these redshifts, it has proven very challenging to derive robust constraints on the host-galaxy properties of unobscured AGNs \citep{targett2012,santini2012}. Therefore it is preferable to limit the study of host-galaxy properties to obscured AGNs, in which the emission from the central engine does not completely outshine the light from the stellar population \citep{assef2010}. Results at $z \sim 2-3$ for UV-selected AGNs \citep{erb2006b} and near-IR selected AGNs  \citep{kriek2007} indicate that narrow-emission-line selected AGNs at these redshifts are mainly found in high mass galaxies (log($M_{*}/M_{\sun}$) $\sim 11$). These results are similar to those for X-ray selected AGNs from both the Cosmic Assembly Near-infrared Deep Extragalactic Legacy Survey \citep[CANDELS; ][]{grogin2011,koekemoer2011}, and the GOODS and COSMOS fields, which also provide evidence that AGNs are hosted by massive (log($M_{*}/M_{\sun}$) $> 10$) galaxies \citep{mainieri2011, kocevski2012, santini2012}. These authors also estimate host galaxy colors for AGNs that put them both in the blue cloud and the transition region. \citet{xue2010} carefully examined X-ray selected AGNs out to $z \sim 2-3$ and found that, when compared to a mass-matched sample of non-active galaxies, AGNs and non-AGNs occupied similar regions of color space, becoming bluer as redshift increases. These results demonstrate that host galaxy properties for large samples of AGNs need to be carefully compared to the properties of a controlled sample of non-active, star-forming galaxies to understand the effects of an AGN on galaxy evolution. 

In this paper, we focus on a sample of UV-selected AGNs at $z\sim2 - 3$ drawn from the Lyman break galaxy (LBG) ($z\sim3$) and the BX/BM samples ($z\sim2$) \citep{steidel2003, steidel2004}. The objects in our AGN sample were identified as Type II AGNs based on narrow emission lines in the rest-frame ultraviolet. In such objects, emission from the central source is obscured from view as the observer's sightline passes through a central dusty medium. This obscuration provides a better view of the host galaxy surrounding the central AGN. An analysis of the clustering of the UV-selected AGNs indicates that they appear to be hosted by the same dark matter halos as those of the parent population of non-AGN LBGs \citep{steidel2002,adelberger2005a}. The non-AGN LBGs, therefore, provide an ideal control sample with which to compare the AGN LBG host galaxies. Here, we use stellar population synthesis (SPS) modeling to examine the UV through IR SEDs for both UV-selected AGNs and non-AGNs. SPS modeling uses  stellar population templates to match the broad spectral energy distribution (SED) of a galaxy and therefore to estimate the properties of the stellar population such as stellar mass, age, extinction, and star-formation rate. There has been much effort to use SPS modeling to characterize the properties of star-forming galaxies across a wide range in cosmic time \citep[][and references therein]{shapley2011}, and recently the method has been extended to model AGN host galaxies. However, using this process to examine AGN host properties is made difficult by non-stellar emission from the central source, which has to be carefully modeled alongside the stellar population. Host-galaxy SEDs have been decomposed into stellar and AGN contributions for both X-ray selected Type II AGNs at $z\sim 0.5-2.5$ \citep{mainieri2011,santini2012} and IR-selected AGNs at $z\sim 0.7-3.0$ \citep{hickox2007}. We follow a similar methodology here to characterize the UV-selected AGN host galaxy demographics at $z \sim 2 - 3$. Accordingly, we build on earlier work from \citet{erb2006b} and \citet{kriek2007} using a larger sample of AGNs and a more systematic treatment of AGN host galaxy stellar population modeling and selection effects. We compare the distributions of stellar age, dust extinction, SFR, and stellar mass with both a non-AGN sample as well as a non-AGN subsample matched in stellar mass. We also investigate such trends as SFR vs. stellar mass, and $U-V$ color vs. stellar mass, in order to see how the AGNs are distributed relative to the non-AGNs. 

One unique aspect of the UV-selected AGN sample is the set of existing rest-frame UV spectra for both the AGNs and the parent sample of non-AGNs from which they were selected. We can use these spectra to analyze the relationship between the host galaxy stellar population and observed UV spectroscopic properties. In \citet{hainline2011}, we created a composite rest-frame UV spectrum for the UV-selected AGN sample, and studied how the composite varied as a function of UV spectroscopic properties such as emission line equivalent width (EW), UV spectral slope, and galaxy UV luminosity. Here, we use the results from SPS modeling and extend this analysis to examine how the host galaxy properties for this sample of AGNs are reflected in the UV spectra. The UV spectra can also be used to estimate potential selection effects as a function of host galaxy parameters, in order to understand the completeness of our AGN sample, and the relationship between black hole and host galaxy properties.   

We describe the AGN sample selection and the data used in the SPS modeling in \S \ref{sec:sample}. In \S \ref{sec:modeling}, we detail our dual-component (``SPS+AGN'') modeling technique to disentangle the emission from the stellar population and the AGN. Specifically, we use this technique on a subset of AGNs with IRAC data to infer how the presence of AGN emission affects the host galaxy parameters derived from SPS fitting. This analysis leads to the calculation of ``correction factors,'' which are then used to model our full set of AGNs and robustly estimate their host galaxy parameters, as detailed in \S \ref{sec:results}. In \S \ref{sec:trends}, we compare the stellar properties for the AGNs to those of non-active star-forming galaxies, including a subsample of galaxies matched in stellar mass. In detail, we compare the AGNs to non-active galaxies using the SFR vs. stellar mass relation, the rest-frame $U-V$ color vs. stellar mass relation, and UV spectroscopic properties. Finally, in \S \ref{sec:mass_ew}, we consider the origin and implications for the UV spectroscopic trends among AGNs, in terms of sample completeness and the connection between black hole and galaxy mass at $z > 2$. We list our conclusions in \S \ref{sec:conclusions}. Throughout this paper, we adopt the following cosmological parameters: $H_0 = 70$ km s$^{-1}$ Mpc$^{-1}$; $\Omega_{tot}$, $\Omega_{m}$, $\Omega_{\Lambda}$ $= 1, 0.3, 0.7$. 
 
\section{The UV-Selected AGN Sample and Observations}
\label{sec:sample}

\begin{figure}
	\epsscale{1.2} 
	\plotone{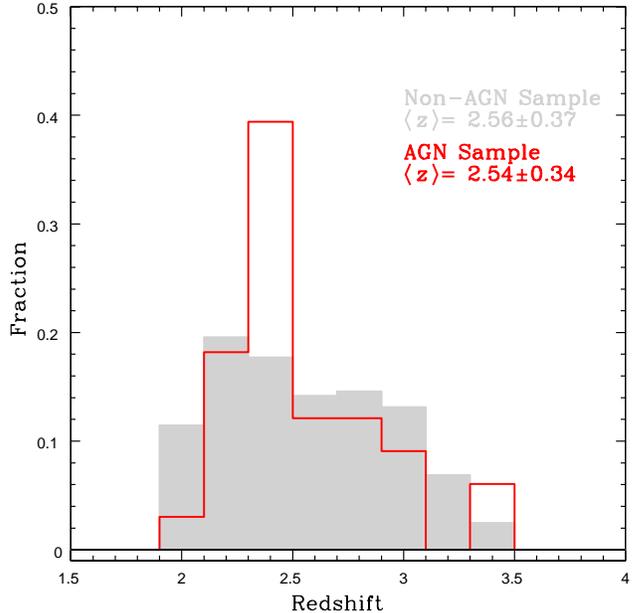} 
	\caption{Redshift distribution for the UV-selected AGNs (red), compared to similarly-selected non-active galaxies (grey) from \citet{reddy2012}. The non-AGN comparison sample spans the same redshift range as the AGN sample.
	\label{fig:redshiftdist}} 
	\epsscale{1.2}
         \end{figure}

The narrow-lined AGNs we model in this paper were selected from a parent sample of 3,059 UV-selected star-forming galaxies at $z \sim 1.5-3.5$. LBGs at $z \sim 3$ are chosen due to the absorption of light shortward of the Lyman break at $912$ \AA\ by the intervening intergalactic medium (IGM). As this break is redshifted to optical wavelengths for $z \sim 3$ galaxies, such objects can be selected based on their position in a $U-G$ vs. $G-{\cal R}$ color-color diagram \citep{steidel2003}. Similar color criteria have been used to identify galaxies at $z \sim 2$, as the $U G {\cal R}$ colors of star-forming galaxies at this redshift correspond to a flat part of the rest-frame UV spectrum redward of the Lyman break \citep{steidel2004, adelberger2004a}. The full sample of $z \sim 2 -3$ UV-selected star-forming galaxies is described in \citet{steidel2003,steidel2004} and \citet{reddy2008}. From this sample, galaxies were identified as having an active nucleus on the basis of features in their rest-frame UV spectrum: strong Ly$\alpha$ emission accompanied by detectable emission in at least one other high-ionization AGN emission line, including \ion{N}{5}$\lambda$1240, \ion{Si}{4}$\lambda \lambda$ 1393,1402, \ion{C}{4}$\lambda$1549, or \ion{He}{2}$\lambda$1640. These objects were selected to be ``narrow-line,'' or ``Type II'' AGNs, as the FWHM for any of the emission features was less than 2000 km s$^{-1}$. The average redshift of the 33 AGNs is $\langle z \rangle = 2.54 \pm 0.34$, with an average ${\cal R}$ magnitude of $\langle {\cal R} \rangle = 24.2 \pm 0.7$, and a range of $22.55 < \cal{R}$$ < 25.72$. We show the redshift distribution for the UV-selected AGNs in Figure \ref{fig:redshiftdist}. We have also assembled a sample of UV-selected non-active galaxies, which have been spectroscopically confirmed at $z \sim 2-3$ \citep{steidel2003,steidel2004,reddy2012} and have near- and/or mid-IR photometric coverage. This non-active comparison sample contains 1727 galaxies when limited to those in the same redshift range as our AGN sample ($1.98 < z < 3.43$) and is plotted with a grey histogram in Figure \ref{fig:redshiftdist}. The non-AGN sample has an average redshift $\langle z \rangle = 2.56 \pm 0.37$, as compared to the average redshift for the AGN sample, $\langle z \rangle = 2.54 \pm 0.34$. In subsequent sections, we will use this non-AGN sample to highlight the impact of the AGNs on stellar population modeling. More details about the AGN sample and its selection are provided in \citet{hainline2011}. 

The UV, optical, and infrared photometry for this AGN sample comes from multiple imaging programs. Objects were initially selected as high-redshift galaxies from their $U$, $G$, and $\cal R$ magnitudes \citep{steidel2003,steidel2004}. The UV and optical imaging for the fields containing our AGNs is detailed in full in \citet{reddy2008}. The galaxies that comprise this sample were selected from imaging characterized by a typical $3\sigma$ photometric depth of $\cal{R}$ $ \sim 27.0$. In order to fully model these objects, we limited our sample to objects with near- and/or mid-infrared photometry. At the redshifts we are studying, important spectral features that indicate age are shifted to the near-IR, such as the 4000 \AA\ break due to metal absorption lines in late type stars, and the Balmer break at 3646 \AA\ from A-type stars. Data at these and longer wavelengths are required in order to constrain stellar population models and give accurate and robust stellar mass and age estimates \citep{shapley2001, reddy2012}. For 15 of our 33 objects (in the Q0100, Q0142, Q0933, Q1217, GOODS, Q1422, Q1623, Q1700, Q2343, and Q2346 fields), near-IR $J$ and $K$ magnitudes were obtained using the WIRC instrument on the Palomar 200'' telescope, with a typical $2''$ aperture $3\sigma$ photometric depth of 22.6 ($K$), and 24.1 ($J$), in Vega magnitudes \citep{erb2006b}.\footnote{24.4 ($K$) and 25.0 ($J$) in AB magnitudes} The WIRC data were reduced using custom IDL scripts as described in \citet{shapley2005a}. When necessary, some of the $K$-band images were smoothed to match the seeing in the $\cal{R}$-band images. Further $K$-band imaging for 11 objects (in the Q0000, Q0201, Q0256, Q2233, DSF2237a, DSF2237b, Q2346 fields) was obtained using the PANIC instrument on the Magellan telescope, as described below. For 16 objects in our sample, we assembled \textit{Spitzer} IRAC \citep{fazio2004} data in Channels 1 (3.6 $\mu$m), 2 (4.5 $\mu$m), 3 (5.8 $\mu$m), and 4 (8.0 $\mu$m). IRAC data for our objects come from five \textit{Spitzer} Programs, as described in \citet{reddy2012} in more detail. These data were reduced with custom IDL scripts that applied artifact correction and flat fielding, after which individual images were mosaiced, and drizzled. Photometry was performed by fitting point-spread functions (PSFs) to the images with positions pre-determined using optical and near-IR data. The full details of the IRAC reduction, PSF fitting, and photometry are found in \citet{reddy2006a}. We also assembled \textit{Spitzer}/MIPS 24 $\mu m$ data for 8 of our objects, as deep surveys were undertaken in the GOODS-N (PI: M. Dickinson) and Westphal (PI: G. Fazio) fields, and the Q1623, Q1700, and Q2343 fields were imaged under GO 1 and 3 \textit{Spitzer} programs. The MIPS data were flat fielded with custom IDL scripts, combined, and photometry was obtained with PSF fitting, as described in \citet{reddy2006a, reddy2010}. 

As only 15 of the narrow-line AGNs in the UV-selected sample had existing coverage at near-IR wavelengths, we imaged 11 additional objects using the PANIC instrument \citep{martini2004} on the 6.5m Magellan I (Baade) telescope at Las Campanas Observatory in September of 2008. These objects were observed over the course of two nights under excellent conditions (median seeing of $0\secpoint5$), with one additional night lost to weather. The data were reduced with IRAF PANIC reduction scripts (``p\_reduce'') (Martini et al. private communication). Each science exposure was linearized, dark subtracted and flat fielded, and sky subtracted using dithered frames adjacent in time. The final images were distortion corrected and stacked. For the purposes of flux calibration, we also obtained and reduced images of \citet{persson1998} near-IR standard stars. The typical $3\sigma$ photometric depths of the final images ranged between 19.5 and 21.5 in $K$ (Vega magnitude), with a median of 21.0. The PANIC $K$-band images were smoothed to match the seeing in the $\cal R$-band images, which were typically characterized by a larger point spread function. Photometry for these data was calculated using SExtractor \citep{Bertin1996}. We used SExtractor ``isophotal'' magnitudes, where the isophotal area adopted to calculate flux in the $\cal R$-band image was applied to the smoothed $K$-band image. With this procedure, we calculated $\cal{R}$$ - K$ colors within a fixed isophotal area, just as the existing optical and near-IR colors had been estimated. 

The optical and near-IR photometric errors for all of our objects were estimated using a Monte Carlo approach, where fake galaxies across a range of magnitudes were added to the data and recovered using SExtractor, with the same input parameters as those used to calculate the actual photometry. This method resulted in distributions of the differences in input and recovered magnitudes in various bins of recovered magnitude. The error that we assigned to each photometric measurement was the standard deviation in the relevant bin. Photometric errors for the IRAC data were obtained with a similar method, where the photometry was calculated using a custom PSF fitting method as described in \citet{reddy2012}. The full $UG$$\cal{R}$$JK+$\textit{Spitzer} photometric datasets for our objects are listed in Table \ref{tab:photometry}. 

\section{SED Modeling}
\label{sec:modeling}

	\begin{figure}
	\epsscale{1.2} 
	\plotone{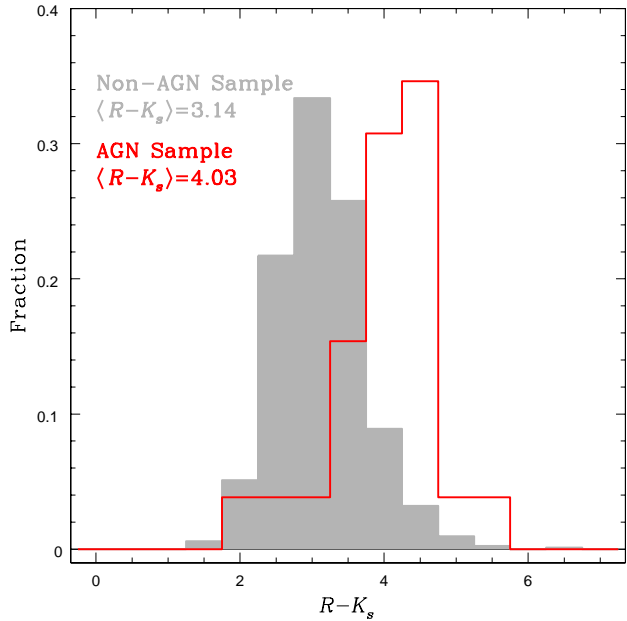} 
	\caption{Fractional $\cal{R}$$-K$ color distribution for our sample of $z\sim 2 - 3$ AGNs, compared with similarly-selected non-active galaxies from \citet{reddy2012}. The $\cal{R}$$-K$ colors for AGNs are, on average, redder than for non-active galaxies, indicating older stellar populations and/or more dust extinction.
	\label{fig:rmkcolors}} 
	\epsscale{1.2}
         \end{figure}

Before discussing the stellar populations of our AGN hosts, we examine their observed colors. The measured $\cal{R}$$-K$ distribution for our full sample of AGNs is shown in Figure \ref{fig:rmkcolors}, along with the $\cal{R}$$-K$ colors for non-active $UG$$\cal{R}$-selected star-forming galaxies \citep{reddy2012}. Although these AGNs and non-active galaxies were selected using the same rest-UV photometric criteria, the redder $\cal{R}$$-K$ colors for the AGNs may be indicative of a difference in the host stellar populations of active and non-active galaxies. Specifically, these colors imply a more evolved stellar population and/or higher dust extinction for the AGN host galaxies, and we can use SPS modeling to interpret the color information for our sample of active galaxies in terms of physical properties.

In SPS modeling, the observed UV through infrared photometry of a galaxy is fit by assuming a population of stars governed by a set of parameters, including star-formation history, metallicity, initial mass function, stellar age and mass, and degree of dust extinction. Stellar population models use a stellar library of the observed spectra of stars of different spectral types, as well as a description of the evolution of stars of different masses and metallicities to produce integrated spectra for a population of co-evolving stars at various ages and for a given star-formation history. These spectra are passed through photometric filters to predict galaxy SED luminosities and colors, which are compared to observations to determine the physical properties of the host galaxies. 

The SPS approach to SED fitting has been widely used to describe the stellar populations of high-redshift galaxies \citep[][and references therein]{sawicki1998,shapley2001,papovich2001,shapley2011}. For our sample of active galaxies, however, we must use caution because of the effect that non-stellar continuum and line emission may have on the fitted parameters. In order to constrain the effect of AGN emission on the stellar population synthesis modeling process, we modeled the observed SED with both a stellar population component as well as an AGN component (``SPS+AGN''). In this approach, the observed colors and magnitudes are decomposed at each wavelength into a sum of the fluxes from the underlying stellar population and emission excited by the central AGN. Our objects are narrow-lined AGNs, and thus the non-stellar emission should be limited to the rest-frame near- to mid-infrared wavelengths \citep{assef2010}. At the same time, we can use our dual-component modeling for objects with photometric information at infrared wavelengths both to understand the true stellar populations of AGN host galaxies, and also to understand how the presence of an AGN affects the best-fit stellar parameters. 

In this section, we describe our dual-component modeling in detail. We start by discussing our sample of AGNs with IRAC data, which we can use to describe the AGN contribution to the SED (\S \ref{sec:irac}). We then use our dual-component modeling procedure to calculate corrections to host galaxy parameters that account for broadband AGN emission (\S \ref{sec:procedure}). We also correct our photometry for the presence of strong, discrete AGN emission lines (\S \ref{sec:correcting}), and conclude with a section detailing our systematic uncertainties (\S \ref{sec:uncertainties}).
 
\subsection{IRAC Data}
 \label{sec:irac}
 
 	\begin{figure*}
	\epsscale{1.21} 
	\plotone{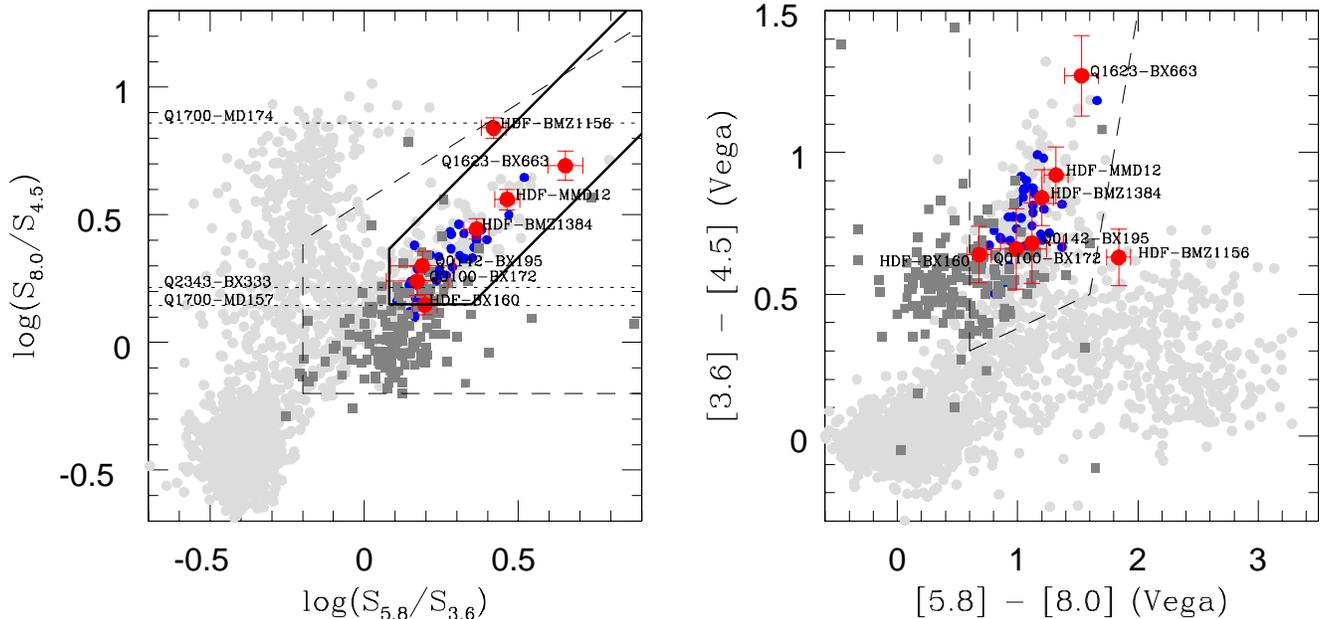} 
	\caption{IRAC color-color plots from \citet{lacy2004} (left) and \citet{stern2005} (right). The $z\sim2-3$ UV-selected AGNs with IRAC detections are shown in red. Dotted lines on the left are for objects with data only in Channels 2 and 4. The dashed lines show the color criteria used by Lacy et al. (left) and Stern et al. (right) to select AGNs by virtue of their power-law emission. The heavy black line indicates the revised selection criteria from \citet{donley2012}. The light grey points are \textit{Spitzer} First Look Survey objects with clean detections in all four IRAC Channels \citep{lacy2005}. This sample likely spans a wide range in redshift, though not all sources have spectroscopic coverage. Blue circles are \textit{Spitzer} power-law galaxies from Park et al. (2010) at $z > 1.4$ which demonstrate the range of power-law slopes spanned by the UV-selected IRAC AGNs. Dark grey squares are non-AGN star-forming galaxies (as classified by their rest-frame UV spectra) in the same redshift range as the UV-selected AGNs (Reddy et al. 2012). The UV-selected non-AGNs and AGNs occupy largely distinct regions in IRAC color-color space, highlighting the contribution of non-stellar emission to the AGN rest-frame near-IR SEDs. The bulk of the UV-selected non-AGNs fall outside of the \citet{stern2005} and \citet{donley2012} selection areas, but would be selected by the  \citeauthor{lacy2004} criteria. The UV-selected AGNs would be selected as AGNs under the \citeauthor{lacy2004} criteria, while all but HDF-BMZ1156 would be selected as AGNs under the \citet{stern2005} and \citet{donley2012} criteria. Predictions of AGN colors as a function of redshift from \citet{donley2012} find that strong AGNs inhabit the space occupied by HDF-BMZ1156 at $z \sim 2 - 3$. 
	\label{fig:iracplot}} 
	\epsscale{1.21}
         \end{figure*}
 
As described in \S \ref{sec:sample}, there is $UG\cal{R}$$JK$ photometry for the majority of our sample, but only 16 AGNs have IRAC measurements. Of these 16 objects, 11 are characterized by ``power-law'' emission in the rest-frame near-IR, where the fluxes in the IRAC bands increase monotonically towards longer wavelengths. This emission is hypothesized to arise in an AGN from thermal or non-thermal emission near the dusty active nucleus \citep{rees1969, neugebauer1979, elvis1994}. An additional two objects with IRAC coverage (Q1623-BX454 and Westphal-MM47, only including Channels 1 and 2) have flat mid-IR slopes, indicating a weak AGN as compared to the stellar population. As these objects do not have $K$-band data, we fit their SEDs solely with a stellar population. The final three objects (Q1217-BX46, Q1623-BX747, and Q2233-MD21) only have IRAC data in Channel 1, which was not sufficient for the purpose of discerning the presence of a power-law in the rest-frame near-IR. We fit the SEDs for these objects, including the IRAC Channel 1 data, with a stellar population only. 

To place our sample of UV-selected IRAC AGNs in context, we compare it to other galaxy samples with coverage in all four IRAC channels. In Figure \ref{fig:iracplot}, we plot the 11 objects with IRAC power-law slopes on the color-color diagrams from \citet{lacy2004} and \citet{stern2005}, which were initially used to select galaxies with red IR colors as possible AGNs. The AGNs with detections in all four IRAC bands are plotted as red points, and, in the left figure, we use dashed lines to indicate the positions of three AGNs with only Channel 4 and Channel 2 photometry. In this left panel, we show the AGN selection box from \citet{lacy2004} with long dashed lines. Each of our AGNs lies in the detection area. \citet{donley2012} revised the selection criteria to account for contamination due to star formation, and we indicate the revised AGN selection area with a dark line. All but one of our objects fall within this revised area. On the right, we show the AGN selection criteria from \citet{stern2005} with long dashed lines, and all but one object falls within this area. HDF-BMZ1156 falls outside of the \citeauthor{stern2005} and \citeauthor{donley2012} selection areas (although it is consistent with being included within the \citeauthor{donley2012} selection area when its photometric uncertainty is taken into account)\footnote{We note that HDF-BMZ1156, formerly known as HDF-oMD49, has previously been discussed in \citet{steidel2002} and \citet{reddy2006b}. This object was only weakly detected in the Chandra 2 Msec data, and has been identified as an example of a Compton-thick AGN by \citet{alexander2008}, based on its combined mid-IR and X-ray properties.}. The grey points on this diagram are those objects from the \textit{Spitzer} First Look Survey with clean detections in all four IRAC Channels across all redshifts, and illustrate the full range in color-color space spanned by extragalactic sources \citep{fadda2004}. We have also plotted those objects from our non-active UV-selected $z \sim 2-3$ comparison sample with IRAC coverage in all four channels. These points are shown as dark grey squares. Quite strikingly, UV-selected non-AGNs and AGNs occupy largely distinct regions in IRAC color-color space. Only a small fraction of the UV-selected non-AGNs fall within the Donley et al. (2012) selection area, and these objects may comprise a sample of additional AGNs without strong high-ionization emission lines in their rest-frame UV spectra.

To compare to high-redshift power-law galaxies, we have plotted a sample of $z > 1.4$ power-law sources from \citet{park2010} on both diagrams as blue points. Our active sample spans a large range in IRAC power laws. The specific values for the power-law slope for our objects ($\alpha$, where $f_{\nu} \propto \nu^{\alpha}$) range between $-0.62$ and $-3.50$, with an average (median) of $\alpha = -1.75\, (-1.79)$. These values show the same range as those presented in \citet{park2010}, although the objects in our sample with the most negative slope values, Q1623-BX663 and Q1700-MD174, would be among the reddest of the \textit{Spitzer} power-law galaxies (the power law for Q1700-MD174 is fit from only the IRAC Channel 2 and Channel 4 points). Our galaxies are thus representative of the full range of high-redshift \textit{Spitzer} power-law galaxies, and should illustrate AGN properties for a range of SED types. It is notable that the majority of the UV-selected AGNs in our sample with IRAC data have power-laws, supporting their identification as AGNs, consistent with the observation in \citet{reddy2006b} that all but one UV-identified AGN in the GOODS-N field had significant excess flux at 8 $\mu$m.

\subsection{Stellar Population and AGN Modeling Procedure}
\label{sec:procedure}

In order to accurately describe the SEDs for our sample, we require a suite of stellar population models and an AGN template. For our sample of  $z\sim2-3$ AGNs we chose \citet[][hereafter BC03]{bc2003} stellar population models. Recently, much attention has been focused on the degree to which thermally pulsating asymptotic giant branch (TP-AGB) stars contribute to the near-IR luminosity of a population of $0.5 - 2$ Gyr age stars \citep{maraston2005,maraston2006,eminian2008}. Different treatments of TP-AGB stars systematically affect derived stellar mass and ages \citep{maraston2005}. In order to gauge how the presence of TP-AGB stars affects our inferred stellar parameters, we also applied the stellar population models of \citet[][hereafter Mar05]{maraston2005} and Charlot \& Bruzual (2012) (private communication, hereafter CB12) to our SEDs. For the BC03 and CB12 models, we adopted the Padova 1994 stellar evolution tracks. We selected solar metallicity models and a \citet{chabrier2003} initial mass function (IMF) extending from 0.1 to 100 M$_{\sun}$. We adopted solar metallicity models as non-active galaxies in a similar mass range appear to have metallicities only slightly lower than solar \citep{shapley2004}. It should be noted that using a Salpeter IMF increases the best-fit stellar masses and star-formation rates by a factor of 1.8 with little effect on the predicted spectral shape. To account for dust extinction, we used a \citet{calzetti2000} starburst attenuation law, which on average provides an accurate description of the reddening and attenuation of the UV stellar continuum in both nearby and distant starburst galaxies \citep{reddy2004, reddy2006a}. 

For the BC03 and CB12 models, we use constant star formation (CSF) models. For the Mar05 models, we used an exponentially declining SF model of the form $\mathrm{SFR}(t) \propto \mathrm{exp}(-t/\tau)$, with an e-folding time of $\tau = 20$ Gyr as the closest approximation of a CSF model.  We chose to not model with more complex star-formation histories, such as exponentially declining star-forming models, short-duration bursts superposed on top of an exponentially declining star-formation history, or exponentially increasing star formation models. Recent work presented in \citet{reddy2012} indicates that exponentially declining star-formation models result in SED-derived SFR values that are lower on average than those derived from a combination of UV and IR data\footnote{\citeauthor{reddy2012} derive the SFR$_{IR+UV}$ by combining SFRs measured from photometry probing the rest-frame UV and IR. Observed-frame optical photometry is used to calculate the rest-frame UV (1700 \AA) luminosity, and MIPS 24 $\mu$m photometry is used to calculate the rest-frame IR (8 $\mu$m) luminosity. These luminosities are converted to SFRs assuming typical galaxy ages and star-formation histories, and then combined to form SFR$_{IR+UV}$.}. Exponentially increasing star formation models have been shown to apply to galaxies at $z > 2$, and naturally arise in hydrodynamic and semi-analytic models of star formation, and are suggested by the relationship between SFR and stellar mass observed at high redshift. \citep{maraston2010, papovich2011, reddy2012}. These histories give SFRs that are similar to the SFRs derived from CSF models, along with older ages \citep{reddy2012}.

	\begin{figure}
	\epsscale{1.2} 
	\plotone{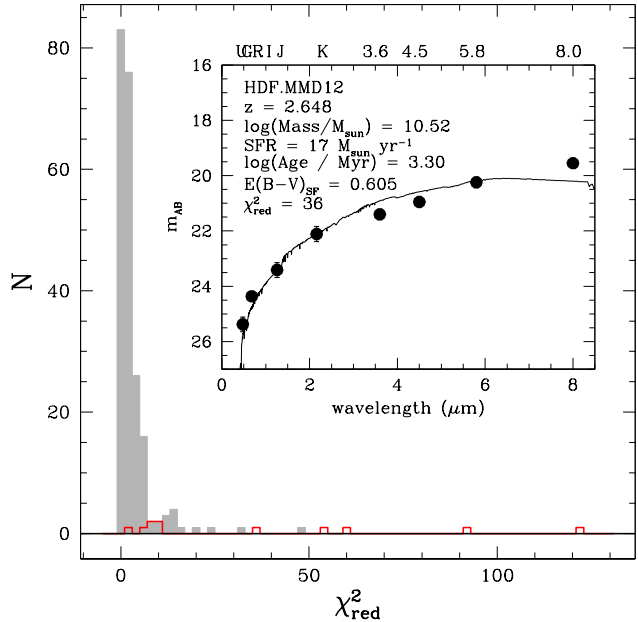} 
	\caption{Histogram of reduced $\chi^2$ values from SPS-only model fits to UV-selected IRAC AGNs (red), and UV-selected non-AGNs (grey). The IRAC AGNs are not well-fit by SPS-only models due to emission in the IRAC Channels, while the bulk of the non-AGNs are well fit ($\langle \chi^{2}_{red} \rangle = 2.3 \pm 1.8$). In the inset, we show the best-fit SPS-only SED model for an example AGN, HDF-MMD12, using a BC03 CSF model. The power law seen in the IRAC bands is indicative of the presence of an AGN, which results in a poor fit ($\chi^{2}_{red}= 36$) for a stellar emission-only model. The errors on the IRAC points are smaller than the size of the data points. The solid curve is the fit derived taking into account all of the data. 
	\label{fig:badfit}} 
	\epsscale{1.2}
         \end{figure}
         
Stellar population models, even those with large contributions from old, red stars, cannot successfully model the infrared emission seen in an IRAC power-law object. In Figure \ref{fig:badfit}, we show in red the distribution in reduced $\chi^{2}$ ($\chi^{2}_{red}$) found by modeling our IRAC AGNs with only a stellar population. We also plot the $\chi^{2}_{red}$ distribution for the non-AGN comparison sample in grey, using the subsample of 231 non-AGNs with SED coverage extending out to 8$\mu$m. The AGNs have higher $\chi^{2}_{red}$ values than the bulk of the non-AGN comparison population, indicating poor fits. For the non-AGNs, $\langle \chi^{2}_{red} \rangle = 2.3 \pm 1.8$, which is lower than all but 1 of our IRAC AGNs (Q0142-BX256, which does not have IRAC Channel 4 data, but has large errors on the Channel 3 data point). In the inset we show a CSF fit for one of our objects, HDF-MMD12, an AGN at $z = 2.648$. The best fit is very poor, with a $\chi^{2}_{red} = 36$, due to the presence of the strong AGN power-law emission in the IRAC bands. Often, the SEDs of Type II AGN hosts are modeled excluding points redwards of the rest-frame optical, as the AGN emission for narrow-lined objects should be minimal at rest-frame UV and optical wavelengths \citep{barger2005a, bundy2008}. With our sample of IRAC AGNs, we can model the power-law AGN emission and the stellar emission simultaneously to see exactly how the presence of an AGN affects the best-fit stellar population parameters. We also note that there exists a tail of galaxies without signs of AGN activity in their rest-frame UV spectra with large $\chi^{2}_{red}$ values from SPS-only model fits. We will discuss these objects in \S \ref{sec:iracresults}, and in a forthcoming paper.

We then require a prescription for the non-stellar AGN emission in our dual-component modeling. Often, the non-stellar continuum contribution to the near-IR emission is removed by assuming that this portion of the spectrum can be represented by a power law of the form $f_{\nu} \propto \nu^{-\alpha}$, where $f_{\nu}$ is the flux density at a frequency $\nu$ \citep{alonso2003, hainlinel2011}. Since this method of modeling the AGN emission has no explicit basis in the physics of the AGN, we instead opt for the AGN model from \citet{assef2010} to describe our AGN infrared emission\footnote{We also explored the use of multiple different empirical Type II AGN templates from \citet{polletta2007}, but found that the reddened \citeauthor{assef2010} model was better able to reproduce the observed photometry of our UV-selected AGN sample (see also \citet{mainieri2011} for a similar methodology).}. The \citeauthor{assef2010} AGN model was created using an iterative method, starting with the \citet{richards2006} average quasar SED template, modified to better replicate the behavior in the UV predicted by an accretion disk model (in an attempt to minimize, in part, contamination by host galaxy emission). This template was applied to the SEDs of AGNs from the NOAO Deep Wide-Field Survey in the Bo\"{o}tes field, and iteratively updated to fit the full sample of data. The final AGN template is thus physically motivated, and, as the template has been both normalized to a distance of 10 pc and created to have an integrated luminosity of $10^{10} L_{\sun}$, its application to our data gives us an estimate of the bolometric luminosity of the AGN. The \citet{assef2010} AGN template describes a Type I AGN, with strong emission in the rest-frame ultraviolet thought to be emitted from the accretion disk, as well as emission in the rest-frame near- to mid-IR from warm and hot dust emission. At the same time, the AGNs that we model with the \citeauthor{assef2010} template were selected to be Type II by virtue of narrow emission lines seen in the rest-frame UV, where accretion disk emission should be obscured by dust. To model our Type II AGNs, following the analysis of \citet{assef2010}, we apply dust extinction to the AGN template using a reddening curve derived from the Small Magellanic Cloud for $\lambda < 3300$\AA$\,$\citep{gordon1998}, and following the Galactic curve at longer wavelengths \citep{cardelli1989}. The resulting template should then consist of strong IR emission from hot dust near the active nucleus, which we use to model the IRAC power-law. 

Since the objects we are modeling were selected to be Type II AGNs, we made a prior assumption of $E(B-V)_{AGN}$ $> 1.0$. We model these objects by taking a Type I AGN template and applying extinction in order to produce a Type II AGN template to fit the SEDs. For the Type II AGNs discussed here, the rest-frame UV and optical light should be dominated by stellar emission, given the narrow emission lines in the rest-frame UV suggesting significant obscuration of the central engine \citep{hainline2011}. Low values of AGN extinction, on the other hand, would result in SEDs with prominent AGN continuum emission even at rest-frame UV wavelengths, as well as broad-line region emission. Although the initial modeling results for two systems, HDF-BMZ1384 and Q2343-BX333, indicated best fit values of $E(B-V)_{AGN}$ less than 1.0, we rejected such fits on the basis that these are Type II objects. Furthermore, the $\chi^{2}$ values for these low $E(B-V)_{AGN}$ fits were not significantly lower than for fits with $E(B-V)_{AGN}$ at values similar to the bulk of the sample ($E(B-V)_{AGN} \sim 2 - 7$). The same lower limit on $E(B-V)_{AGN}$ is also adopted in \citet{mainieri2011} for a similar SPS+AGN fitting procedure used to fit X-ray selected Type II quasars at $z > 0.8$. Throughout this paper, we report only those fits where $E(B-V)_{AGN}$ $> 1.0$. We also limit $E(B-V)_{AGN} < 8.0$, as larger values for the AGN extinction did not improve the fits for the objects we examined.

We used the \citeauthor{assef2010} AGN template, along with the stellar population models, to fit the observed SEDs of our IRAC subsample. The overall procedure we used is similar to that of \citet{shapley2001, shapley2004, shapley2005a}, but here the predicted colors are the sum of stellar emission and non-stellar emission from our AGN model. The best-fit model therefore yields constraints on both stellar population as well as AGN parameters. The stellar parameters that we modeled included dust extinction [parameterized as $E(B-V)_{SF}$], stellar population age ($t_{SF}$), star-formation rate (SFR), stellar mass ($M_{*}$), and in the case of the Mar05 models, star-formation history ($\tau$). The ages that we used ranged between 50 Myr and the age of the universe at the redshift of the modeled galaxy. This lower limit of 50 Myr was chosen to exclude models with ages younger than the dynamical timescales estimated for typical $z \sim 2 - 3$ LBGs \citep{erb2006a,law2007}.  We considered stellar extinctions between $E(B-V)_{SF}$ = 0.0 and 0.7\footnote{It is important to note that SPS-only models with arbitrarily high values of $E(B-V)_{SF}$ do not accurately fit the UV-selected AGN SEDs. Due to the specific shape of the SED of a star-forming stellar population, the application of high values of $E(B-V)_{SF}$ cannot not simultaneously replicate the rest-frame UV and optical photometry as well as the rest-frame near-IR power-law that is typical of our UV-selected AGNs. In the best-case scenario, such fits vastly underpredict the data at shorter wavelengths while only roughly fitting those in the rest-frame near-IR.}. The AGN parameters included a central source extinction, $E(B-V)_{AGN}$, and a normalization constant, $N_{\mathrm{AGN}}$.

In the first step of the SED fitting procedure, we created total SPS+AGN models for each potential combination of input parameters. The stellar model flux at each wavelength (with a specific star-formation history, age, and extinction value) was summed with the AGN model flux (with a specific AGN extinction value, and multiplied by the normalization factor $N_{\mathrm{AGN}}$). In order to add the AGN template to our SPS models in a meaningful way, we used values of $N_{\mathrm{AGN}}$ that reflected how the SPS models are normalized. The \citeauthor{assef2010} AGN template is normalized to an AGN bolometric luminosity $L_{bol} = 10^{10} L_{\sun}$, where this luminosity is calculated by integrating the entire template longward of Ly$\alpha$. In terms of the SPS models, the BC03 and CB12 CSF templates are normalized at 1 M$_{\sun}$ yr$^{-1}$ of star formation. Thus, for the BC03 and CB12 CSF fits, N$_{\mathrm{AGN}} = L_{bol} / (\mathrm{SFR}\times 10^{10})$. The Mar05 models are all normalized to 1 $M_{\sun}$ of stars, and so N$_{\mathrm{AGN}} = L_{bol} / (M_{*} \times 10^{10})$ for these fits. While we report the best-fitting $N_{\mathrm{AGN}}$ values, it is useful also to compare $L_{bol}$ between fits, which indicates the strength of the AGN for each object. 

The combined AGN and stellar model spectrum was further attenuated by IGM absorption from neutral hydrogen \citep{madau1995}, which only affects the predicted $U$ and/or $G$ magnitudes, depending on the redshift. These output SPS+AGN templates for a given set of parameters were compared in magnitude space to the optical through mid-IR SED for a specific object (though, not including \textit{Spitzer} MIPS data, see \S \ref{sec:iracresults}), and a value of $\chi^2$ for the fit was calculated. In these fits, we only modeled photometric detections. In a few cases, there were upper limits in IRAC Channels that were not modeled (Channel 4 for Q0142-BX256, Channel 3 and Channel 4 for Q1623-BX454, Channel 2 for Q1623-BX74, and Channel 3 for Q2233-MD21). In all cases except that of Q0142-BX256, the IRAC upper limits are entirely consistent with the best-fit SED model. We sought to scale our total models to the observed SEDs such that the calculated $\chi^2$ was minimized, and this normalization naturally led to an estimation of the star-formation rate and stellar mass. The parameters [$E(B-V)_{SF}$, $t_{SF}$, SFR, $M_{*}$, $E(B-V)_{AGN}$, and $N_{\mathrm{AGN}}$] resulting in the minimum $\chi^2$ with respect to the measured photometry for an object are referred to as the ``best-fit'' parameters.

\subsection{Correcting the Photometry for the Presence of Strong Emission Lines}
\label{sec:correcting}

Flux contribution from strong nebular emission lines excited by the presence of an AGN can be a significant source of contamination in broadband SEDs \citep{schaerer2011}. As described in \citet{hainline2011}, the rest-frame UV spectra for the $z\sim2-3$ Type II AGNs described here include emission lines such as Ly$\alpha$, with an average rest-frame EW of $80$\,\AA. This strong emission line, as well as others in the rest-frame UV, such as \ion{C}{4}$\lambda$1549 and \ion{He}{2}$\lambda$1640, can enhance the flux in the $U$ or $G$ bands (depending on redshift), and alter the observed $U-G$ or $G-$$\cal{R}$ colors. Rest-frame optical features such as H$\alpha$ or [\ion{O}{3}]$\lambda$5007 can similarly add flux in the near-IR, and subsequently affect the measured $\cal{R}$$-K$ colors. We adopt a method of correcting for these emission lines outlined in \citet{papovich2001}, where a rest-frame EW, $EW_{0}$, for an object at a redshift, $z$, introduces a change in the magnitude:

\begin{equation}
 \Delta m = -2.5\, \mathrm{log} \left[1 + \frac{EW_0 (1+z)}{\Delta \lambda} \right] 
\end{equation}

\noindent In this equation, $\Delta \lambda$ is the width of the filter bandpass. We calculated $ \Delta m $ for each measured emission line, and, where its absolute value was larger than the photometric error, we corrected for this magnitude difference to produce the final photometry. 

Strong emission lines do not affect the $U$ or $G$ bands for 19 out of 33 AGNs in our sample at $2.17 \le z \le 2.48$. For the remainder, we used UV spectra \citep[described in][]{hainline2011} to measure EW values for the strong emission lines such as Ly$\alpha$, \ion{C}{4}, and \ion{He}{2}. Ly$\alpha$ was the chief contaminant to the $G$ band for this redshift range, with the average (median) value of the magnitude difference  $ \Delta m = 0.36 \, (0.34)$. The other UV lines were not strong enough to significantly contaminate the photometry: in objects where \ion{C}{4} was detected, the EW was on average (median) 48\% (17\%) of the EW of Ly$\alpha$, and similarly, in objects where \ion{He}{2} was detected, the EW was on average (median) 25\% (13\%) of the EW of Ly$\alpha$. However, for one object (HDF-BMZ1156), very strong \ion{C}{4} and \ion{He}{2} contaminated the $G$ band (at the redshift of this object, Ly$\alpha$ fell bluewards of the $G$ band), with a combined $ \Delta m = 0.37$. 

The $K$ band contains H$\alpha$ for 20 out of 33 AGNs, and [\ion{O}{3}]$\lambda$5007 for an additional 4 objects. Near-IR spectra have only been obtained for a subset of our AGNs, as part of an ongoing project to measure the rest-frame optical spectroscopic properties for this sample \citep[][Hainline et al., in prep]{erb2006b}. For this analysis, there are 9 objects with near-IR spectral coverage (8 objects with H$\alpha$ measurements, and one with a measurement of [\ion{O}{3}]), and we have determined the H$\alpha$ and [\ion{O}{3}] fluxes for the available lines. These spectra do not show a significant continuum, so we used the $K$-band magnitude in order to estimate the continuum flux density (corrected for line flux), and therefore the EW for H$\alpha$ and [\ion{O}{3}]. The average (median) magnitude difference from the H$\alpha$ emission lines is $\Delta m  = 0.18 \, (0.12) $. It should be cautioned that the majority of our objects do not have rest-frame optical spectra, and H$\alpha$ or [\ion{O}{3}] could contribute to the $K$ band photometry for these AGNs. However, we expect that the effect should not be large compared to the $K$-band uncertainties, based on objects with such measurements.

\subsection{Systematic Uncertainties}
\label{sec:uncertainties}

	\begin{figure*}
	\epsscale{0.8} 
	\plotone{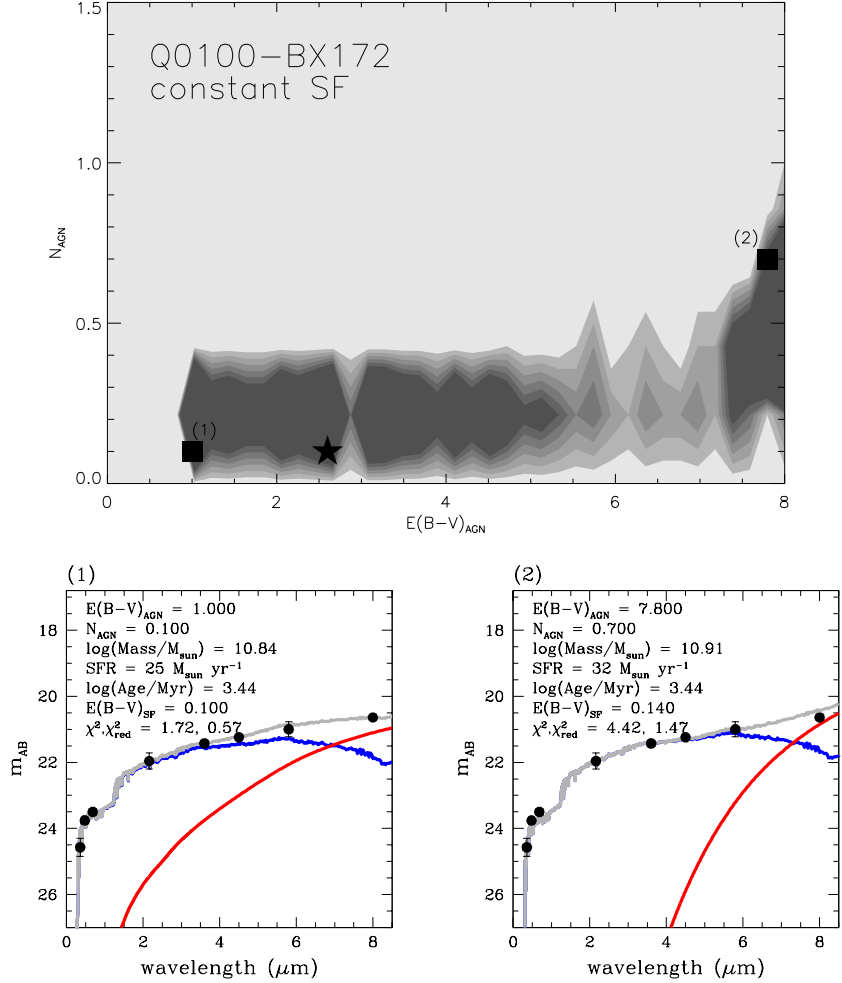} 
	\caption{Confidence intervals for SED fits to Q0100-BX172, showing the covariance between $N_{AGN}$ and $E(B-V)_{AGN}$ in our fitting results. This plot shows the density of points from the Monte Carlo error results, where darker shading indicates a higher density of points. The black star is the best-fit value from the SED fitting as shown in Table \ref{tab:iracparam}. The majority of fits show low values for $N_{AGN}$ across a range of $E(B-V)_{AGN}$, with both increasing together proportionally at larger values. Below the confidence interval plot, we show two sample fits for the black squares, indicating how the stellar parameters (with units of log(Age) in Myr, SFR in M$_{\sun}$ yr$^{-1}$, and log(Mass) in M$_{\sun}$) change with $N_{AGN}$ and $E(B-V)_{AGN}$. The blue line is the best-fit stellar model (BC03), while the red line is the best-fit AGN model. The sum of the models in each panel is given as the grey line.
	\label{fig:confint}} 
	\epsscale{1.}
         \end{figure*}
         
While the uncertainties relating to single-component stellar population fitting are well described in \citet{shapley2005a}, the addition of an AGN component to the fits introduces new model degeneracies and systematics. For example, without an AGN component, the inferred extinction, $E(B-V)$, depends on the chosen star-formation history and age. For CSF models, a given set of galaxy $G-\cal{R}$ and $\cal{R}$$-K$ colors can be produced by models with older ages and less dust extinction, or younger ages and more dust extinction. 

With the addition of an AGN template to the SPS modeling, we find covariances between AGN model extinction ($E(B-V)_{AGN}$) and normalization ($N_{AGN}$), as well as between the AGN parameters and the stellar parameters. Within the IRAC sample described in \S \ref{sec:irac}, a given set of photometric values for our individual AGNs can typically be explained by model fits with lower values of $E(B-V)_{AGN}$ along with lower values of $N_{\mathrm{AGN}}$, or higher values of $E(B-V)_{AGN}$ along with higher values of $N_{\mathrm{AGN}}$. This degeneracy has an effect on the best-fit stellar parameters. For a given object, lower $E(B-V)_{AGN}$ and $N_{\mathrm{AGN}}$ model fits result in lower values of $E(B-V)_{SF}$ and correspondingly lower SFRs. As described above, lower $E(B-V)_{SF}$ fits are also accompanied by older ages. We can explain these covariances qualitatively. For a given intrinsic AGN power-law, lower values of $E(B-V)_{AGN}$ indicate more AGN contamination at shorter wavelengths, and these fits have low values of $N_{\mathrm{AGN}}$ in order to fit the observed SED without overpredicting the $UG$$\cal{R}$ flux. The increased flux at shorter wavelengths from a low $E(B-V)_{AGN}$ AGN template is accompanied by lower $E(B-V)_{SF}$ values, as the increased AGN flux requires less dust to account for the red UV to optical continuum ($\cal{R}$$-K$ color). Our modeling procedure works by adding the AGN template (modified by some dust obscuration, and multiplied by $N_{\mathrm{AGN}}$) to a stellar population model and then normalizing the result to the observed SED to determine the SFR. As a low-$E(B-V)_{AGN}$, low-$N_{\mathrm{AGN}}$ dual-component model has more flux from the AGN at shorter wavelengths, such a model predicted that young stars contribute less flux in the rest-frame UV, which results in lower SFR values. We demonstrate this covariance using Q0100-BX172 as an example in Figure \ref{fig:confint}.

In order to account for covariance among the best-fit stellar and AGN parameters, errors in best-fit parameters were estimated using a Monte Carlo analysis. We created a large set ($N = 500$) of fake SEDs by varying the observed colors and magnitudes in a manner consistent with the photometric errors. Each fake SED was modeled using the same method as for the actual data, where the best-fit stellar and AGN parameters were those that resulted in a minimization of $\chi^{2}$. We estimated the 1$\sigma$ errors for the best-fit model parameters by calculating the standard deviation of the distribution on each parameter from our Monte Carlo simulation results. 

\section{Stellar Population Modeling Results}
\label{sec:results}

\subsection{AGN and Stellar Population Modeling Results}
\label{sec:iracresults}
 
 	\begin{figure*}
	\epsscale{1.17} 
	\plotone{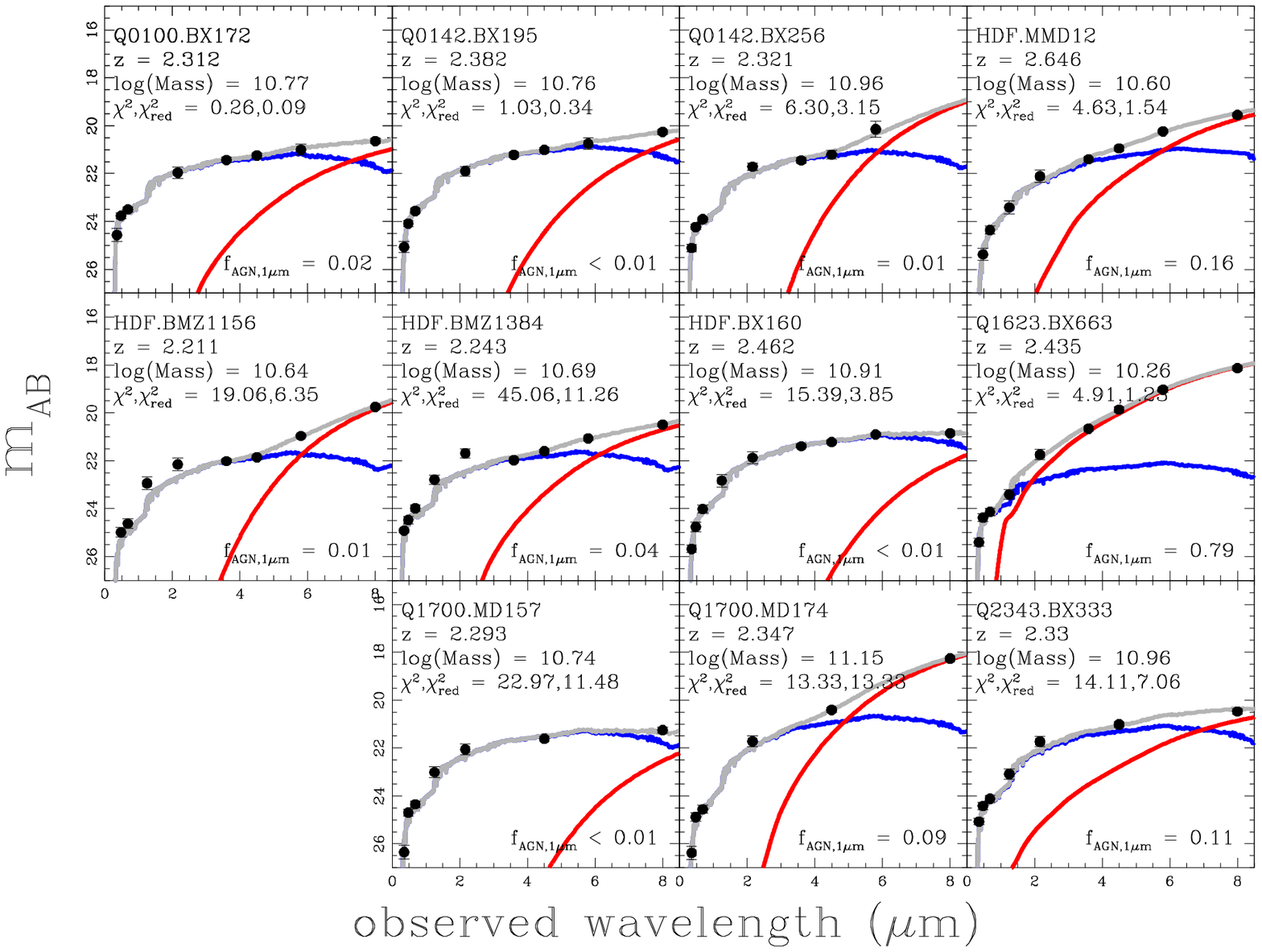} 
	\caption{SED models for the 11 objects with IRAC power-law fits. The blue line is the best-fit stellar model (BC03), while the red line is the best-fit AGN model. The sum of the models is shown as the grey line. The best-fit parameters for these models are given in Table \ref{tab:iracparam}, and the log(Mass) shown is in units of solar masses. In the bottom right corner, we show the fraction of AGN flux to total flux measured from the best-fitting models at rest-frame 1 $\mu$m. Values listed as $<0.01$ have a negligible AGN contribution at $1\mu m$. 
	\label{fig:iracconst}} 
	\epsscale{1.17}
         \end{figure*}
 
The dual-component SPS+AGN fits successfully reproduce the range in SEDs found in the IRAC AGN subsample. We show the SED fits for the 11 IRAC AGNs in Figure \ref{fig:iracconst}. In each panel, the grey curve represents the best-fitting dual-component model, which is decomposed into the sum of fluxes from a stellar population (blue curve) and an \citeauthor{assef2010} AGN template (red curve). Also shown are the $\chi^2$ and $\chi^2_{red}$ values output from the modeling program. The poorest fits, with the highest values of $\chi^2_{red}$, are for those objects where the $K$-band flux was greater than predicted from the fit, as in HDF-BMZ1156, HDF-BMZ1384, and to a lesser extent Q1700-MD174 and Q2343-BX333. The underprediction of $K$-band flux could be a result of the choice of star-formation history, or the contamination from H$\alpha$ emission to the flux in the $K$-band. It should be noted that we have a rest-frame optical spectrum for HDF-BMZ1156, and have made a correction for the presence of H$\alpha$ in this object ($\Delta m = 0.07$), and yet the best-fitting model still underpredicts the $K$ band. We have also calculated the fraction of the total emission in the best-fit SPS+AGN model at rest-frame 1 $\mu$m that is contributed by the AGN template.  These values of $f_{AGN, 1 \mu m}$ are reported in the bottom right corner of each panel in Figure 6, and express the relative strength between the emission from the AGN and the stellar population. The values of $f_{AGN, 1 \mu m}$ range from $<0.01$ for HDF-BX160 and Q1700-MD157 to 0.79 for Q1623-BX663. The best-fitting parameters for the IRAC AGNs are reported in Table \ref{tab:iracparam}. The IRAC AGNs have average (median) values and standard deviations for the best-fit parameters of $\langle \mathrm{log(Mass)_{CSF}} \rangle = 10.82\,(10.76) \pm 0.22$, $\langle \mathrm{SFR_{CSF}} \rangle = 56\,(37) \pm 51$ M$_{\sun}$ yr$^{-1}$, $\langle \mathrm{E(B-V)_{SF,CSF}} \rangle = 0.23\,(0.24) \pm 0.08$, and $\langle \mathrm{log(Age)_{CSF}} \rangle = 3.23\,(3.21) \pm 0.25$. 

Based on the range in best-fitting parameters, we can split the IRAC AGN host galaxies into two populations based on the slope of the IRAC power law (Figure \ref{fig:iracplot}). The first population contains those objects in our IRAC subsample with shallower observed power-law slopes. Most of the galaxies that comprise this population (Q2343-BX333, Q0100-BX172, Q0142-BX195, HDF-BX160, and Q1700-MD157) have dual-component fits where the AGN only becomes the dominant source of emission at $6 - 8 \mu$m observed-frame (Fig. \ref{fig:iracconst}). These objects are best fit by models with low values of $N_{\mathrm{AGN}}$ along with a range of $E(B-V)_{AGN}$ values. The one exception is HDF-MMD12, which has a low value for $N_{\mathrm{AGN}}$ accompanied by strong AGN emission at observed-frame wavelengths shorter than 8 $\mu$m. HDF-MMD12 is best fit by a relatively young stellar population ($t_{SF} \sim 200$ Myr). This model SED has a weak Balmer break with respect to those of the other objects, and less flux from stars at longer wavelengths, in comparison to the \citet{assef2010} AGN template.

The second population of IRAC AGNs consists of objects with much steeper (redder) power-law slopes. These objects (HDF-BMZ1384, HDF-BMZ1156, Q1700-MD174, Q0142-BX256, and Q1623-BX663) are best fit by models with larger values of $N_{\mathrm{AGN}}$, in which the AGN emission already becomes dominant at wavelengths as short as $5 - 6 \mu$m in the observed frame. Correspondingly, the rest-frame UV AGN emission lines in the objects with steeper power-law slopes are stronger on average than those measured in the shallow power-law objects. The AGN bolometric luminosities of the steep power-law objects are also larger, on average, than those of the shallow power-law objects. Q1623-BX663 is the most extreme example. This galaxy has a steep power-law ($\alpha = -2.24$), with a large value of $N_{\mathrm{AGN}}$ accompanied by a low value for the AGN extinction. As AGN emission in this object is dominant even to the $K$ band for CSF models, Q1623-BX663 represents a ``worst-case scenario'' for modeling the AGNs using $UG$$\cal{R}$$JK$ photometry alone and not including an AGN component. Q1623-BX663 has one of the steepest slopes observed in our IRAC subsample, and it is unlikely that many objects in the full AGN sample will have similar AGN-dominated fits. Together, these AGN populations show a wide range of AGN contamination, and should provide a representative subset for correcting the full sample for AGN emission.

As a sanity check, we apply our dual-component modeling procedure to the subsample of non-AGNs (discussed in \S \ref{sec:sample} and \S \ref{sec:procedure}) that have coverage extending to 8 $\mu m$. Half of these objects have best-fit models with $N_{AGN} = 0.00$, and the full distribution is highly skewed towards either zero or extremely low values of $N_{AGN}$, in contrast to our UV-selected AGNs with IRAC coverage, all 11 of which are fit by models with a non-zero value for $N_{AGN}$. We assessed the probability of drawing a sample of 11 non-zero values from the non-AGN $N_{AGN}$ distribution, using $10^6$ random trials. This test revealed that the likelihood of randomly drawing such a sample from the non-AGN distribution is only 0.05\%, again demonstrating the distinct nature of the SEDs of the UV-selected AGNs. We do note that there is a small population of objects in the non-AGN sample that is best fit by models with $N_{AGN}$ significantly greater than zero. While a discussion of these objects is outside of the scope of the current work, their strong IRAC power-law emission indicates that they may be hosting an AGN, but do not show significant emission lines in their UV spectra. These objects will be followed up in a paper analyzing LBG/BX/BM AGNs selected on the basis of their infrared emission. As these objects comprise only a small fraction of the total non-AGN sample, they will not significantly affect our conclusions regarding the distribution of non-AGN stellar populations.

As discussed in \S \ref{sec:procedure}, we also used the CB12 and Mar05 models in our dual-component SPS+AGN fitting. Here we report the differences in derived parameters for the CSF models. For the CB12 modeling, the stellar masses and ages from the fitting are typically 70\% of those derived using the BC03 models, while the CB12 SFR  and $E(B-V)_{SF}$ values are statistically equivalent to the BC03 values. The Mar05 models produced similar differences. For the Mar05 modeling, the stellar masses from the fitting are typically 60\% of those derived using the BC03 models, and the ages are 50\% of those derived using BC03. The $E(B-V)_{SFR}$ values derived using both models were statistically the same, and the Mar05 SFR values were larger than the BC03 values by 25\%. Despite these small systematic differences among stellar population models, our main conclusions regarding the AGN contamination do not depend on the specific stellar population model adopted.

We have MIPS photometry for eight of our objects with IRAC data. We do not use the MIPS data to constrain our SPS modeling, and find that for the majority of the objects with MIPS data, the best-fit AGN template overpredicts the MIPS flux in our objects by a factor of two on average. Examining the range of stellar and AGN parameters that best fit our optical through IRAC data, we find that attempting to fit for the MIPS data leads to best-fit models that systematically underpredict the IRAC fluxes, due to the shape of the reddened \citeauthor{assef2010} template. Analysis of of the mid-IR SED of HDF-BMZ1156 by \citet{alexander2008} sheds light on the origin of this discrepancy. \citeauthor{alexander2008} find that a Type II AGN template (i.e., the SED of NGC 1068) provides a good fit to the photometry of HDF-BMZ1156 at wavelengths from 16 through 70 $\mu$m. However, an additional hot ($\sim 1000$K) dust component, often found in obscured AGNs \citep[e.g.,][]{alonso2001}, is required to fit the IRAC datapoints at shorter wavelengths. If the other AGNs in our sample are similar to HDF-BMZ1156, our use of a single component to describe both IRAC and longer-wavelength data will naturally overpredict the MIPS 24$\mu$m flux, since we have forced a Type II template to fit the IRAC data. At the same time, this difference is not significant for most of the sample. The range of SPS+AGN templates that fit the observed $UG$$\cal{R}$$JK$+IRAC data within the photometric errors is within 1$\sigma$ of the MIPS uncertainty for the majority (five out of the eight) of our objects. Since we do not use the MIPS data in our SED fitting, the discrepancies between predicted and observed fluxes suggest that the $L_{bol}$ values reported in Table \ref{tab:iracparam} may slightly overpredict the true values.

\subsection{Correcting for AGN Emission}
\label{sec:agncorrections}

To understand the true AGN host galaxy stellar populations, it is very important to correct for AGN emission in the SED of those objects without IRAC data, where we can't directly quantify the AGN contribution. Many authors fit AGN host galaxy SEDs using SPS models alone \citep[see][for discussions]{bundy2008,bluck2011}, and find that while the presence of an AGN does affect stellar mass estimates, it is not a significant difference. Here we adopt a different approach, and use our results from the SPS+AGN modeling to quantify how much the presence of an AGN affects the stellar modeling parameters. For the 11 objects in the IRAC AGN sample, we refit the SED but only through the $K$-band, and using an SPS-only model. The difference in best-fit parameters based on $K$-band and SPS-only vs. $K$-band+IRAC and SPS+AGN modeling indicates how the majority of the objects in our sample without IRAC coverage must be corrected. It is valid to apply the correction factors derived from the IRAC AGN sample to our entire sample, given that the objects with IRAC coverage have similar average $\cal{R}$ magnitudes and $\cal{R}$$-K$ colors to those without IRAC coverage. The procedure, star-formation models, and parameters are the same as those discussed in $\S \ref{sec:procedure}$, but here we do not include the AGN template in the $K$-band-only fits.

The resulting stellar masses and SFRs calculated with an AGN component are lower than those calculated without the AGN component. The best-fit ages and values for $E(B-V)_{SF}$ are consistent between fits with and without the AGN component. For a few objects, such as Q1623-BX663, the best-fit parameters are quite different, as this galaxy suffers from the most AGN contamination at the shorter wavelengths (Fig. \ref{fig:iracconst}). The results for HDF-BMZ1156 and HDF-BMZ1384 are also discrepant, due to the bright $K$ magnitude for these objects, which is not well-fit in the dual-component modeling extending to IRAC wavelengths.

We analyze the AGN contamination by calculating the ratio between the parameters ($M_{*}$, SFR, $E(B-V)_{SF}$, and age) resulting from fits with the SPS+AGN modeling to the average parameters resulting from the SPS-only modeling, which we refer to as ``correction factors.'' We calculated the ratios for the best-fit parameters, and we also analyze the AGN contamination using results from the Monte Carlo error analysis (\S \ref{sec:uncertainties}). In this method, we calculated the average value for the full distribution of Monte Carlo output parameters, and then divided these averages for our SPS+AGN modeling by the averages from the SPS-only modeling. This second method gives a better understanding of the full set of models that fit an object's SED within the error on the photometry and therefore we only report correction factors calculated in this manner. We note however that the correction factors do not depend significantly on the method used. The correction factors calculated for the 11 IRAC AGNs are shown in Table \ref{tab:corrections}.  While our results suggest that the $UG\cal{R}$$JK$ photometry for Type II AGNs is largely free from AGN emission, the best-fit masses and SFR values derived without including an AGN component are larger by a factor of $1.4$ (although, on an object-by-object basis, these factors for mass and SFR are not always the same). The values for $E(B-V)_{SF}$ are only larger by a factor of 1.1, and the ages are consistent. We use these factors to correct the best-fit parameters based on SED modeling results for those objects where we do not have IRAC data under the assumption that the SPS+AGN model is more appropriate. 
 
We also calculated AGN correction factors using the CB12 and Mar05 modeling. The correction values calculated for both the CB12 and Mar05 modeling are similar to those calculated using the BC03 modeling. While the mass and age correction values are smaller than than those derived from the BC03 models, they differ by less than one standard deviation of the distribution of BC03 correction factors. The corrected best-fit parameters are not significantly dependent on the choice of model (e.g., BC03, CB12, or Mar05).

\subsection{Full Sample Modeling Results}
\label{sec:fullresults}

	\begin{figure*}
	\epsscale{1.17} 
	\plotone{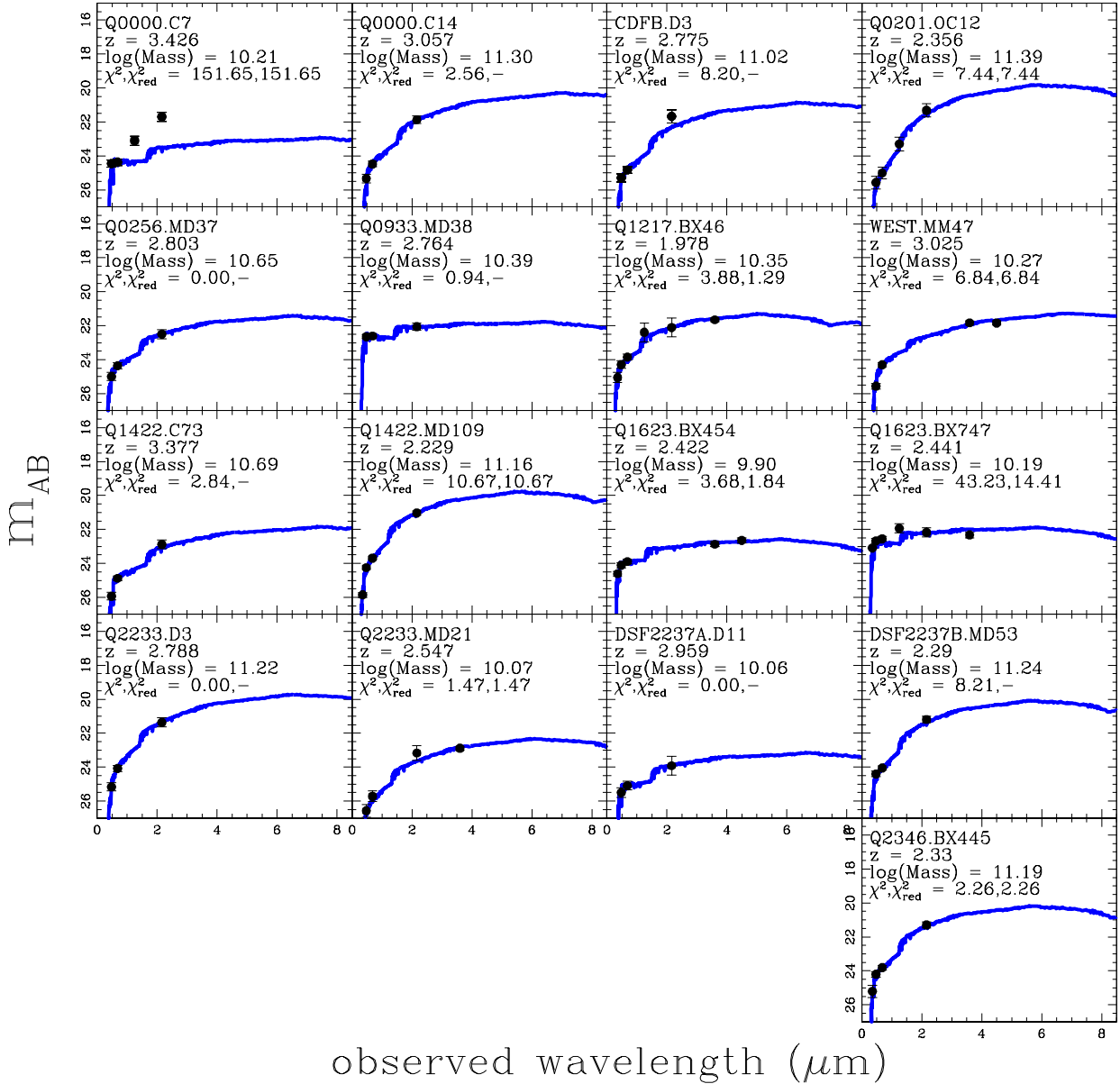} 
	\caption{SED models for the 17 objects that do not have IRAC data, or have IRAC data that is not sufficient for the purpose of discerning the presence of a power-law in the rest-frame near-IR. The models shown here are BC03 models, and do not include any contribution from an AGN. The best-fit mass for each object is given, corrected for the presence of an AGN as described in \S \ref{sec:agncorrections}.
	\label{fig:noniracconst}} 
	\epsscale{1.17}
         \end{figure*}

To fit the objects in our AGN sample lacking multiple IRAC photometric measurements, we used SPS-only models (using the same procedure described in \S \ref{sec:procedure}, but without the addition of an AGN model). For each of these objects, we corrected the best-fit mass, SFR, age, and $E(B-V)_{SF}$ for the presence of an AGN using the average correction factors reported in the previous section, and list the best fit values (both uncorrected and corrected) in Table \ref{tab:noiracparam}. The errors on the corrected parameters reflect both the errors derived using the Monte Carlo bootstrapping described in \S \ref{sec:uncertainties} as well as the standard deviation of the correction factors, given in Table \ref{tab:corrections}. Objects with only three photometric data points have reduced $\chi^{2}$ values given by a dash. SED fits for each of these objects are shown in Figure \ref{fig:noniracconst}. The masses indicated in these figures are the corrected best-fit masses. The $\chi^{2}$ values indicate that many of the objects are well fit by the SPS models. In the case of Q0000-C7 and Q1623-BX747, the SEDs were not well fit by the CSF models. For Q0000-C7, a blue $G-$$\cal{R}$ color leads to models that are unable simultaneously to fit both the UV and near-IR portions of the SED. For Q1623-BX747, the $J$, $K$, and Channel 1 photometric data\footnote{The data set for this object has coverage in one IRAC Channel, which is not enough to determine an AGN power-law; see \S \ref{sec:irac}.} show a decline to larger wavelengths, which is not reflected in the best-fit stellar models. 

	\begin{figure*}
	\epsscale{1.2} 
	\plotone{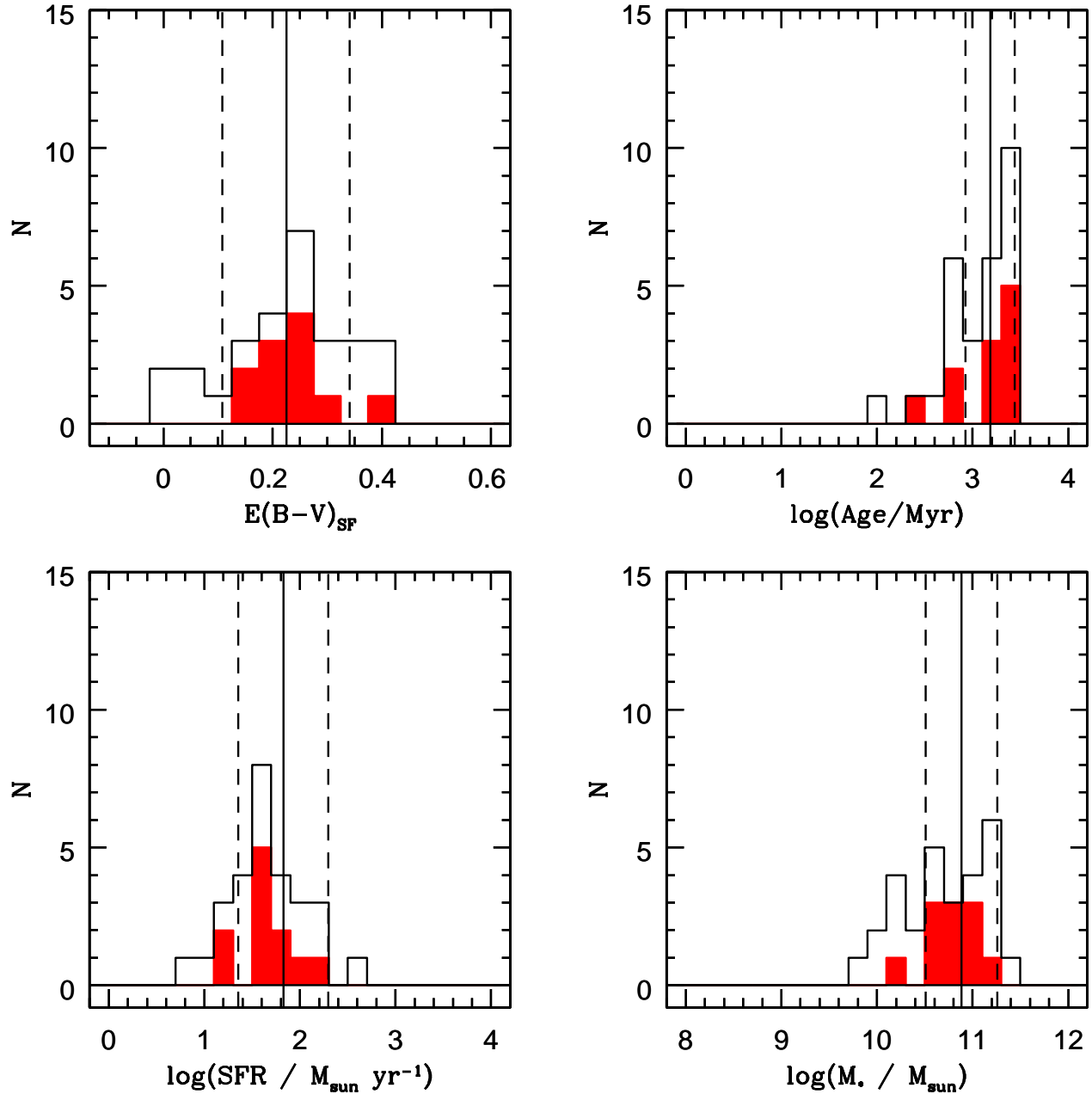} 
	\caption{Histograms of best-fit stellar parameters for our sample of UV-Selected AGNs. The histogram represents the total number of UV-selected AGNs, and we highlight the IRAC AGNs in red. Vertical solid lines indicate the averages for each parameter, and the dashed lines indicate the standard deviations of the distributions. The average (median) values for the best-fit parameters from all of the AGNs are, $\langle \mathrm{log(M_{*}/M_{\sun})} \rangle = 10.85\,(10.71) \pm 0.36$, $\langle \mathrm{SFR} \rangle = 63\,(37) \pm 67$ M$_{\sun}$ yr$^{-1}$, $\langle \mathrm{log(Age/Myr)} \rangle = 3.19\,(3.18) \pm 0.26$, and $\langle \mathrm{E(B-V)_{SF}} \rangle = 0.22\,(0.23) \pm 0.12$.
	\label{fig:bestfit_hist}} 
	\epsscale{1.2}
         \end{figure*}

         \begin{figure}
	\epsscale{1.2} 
	\plotone{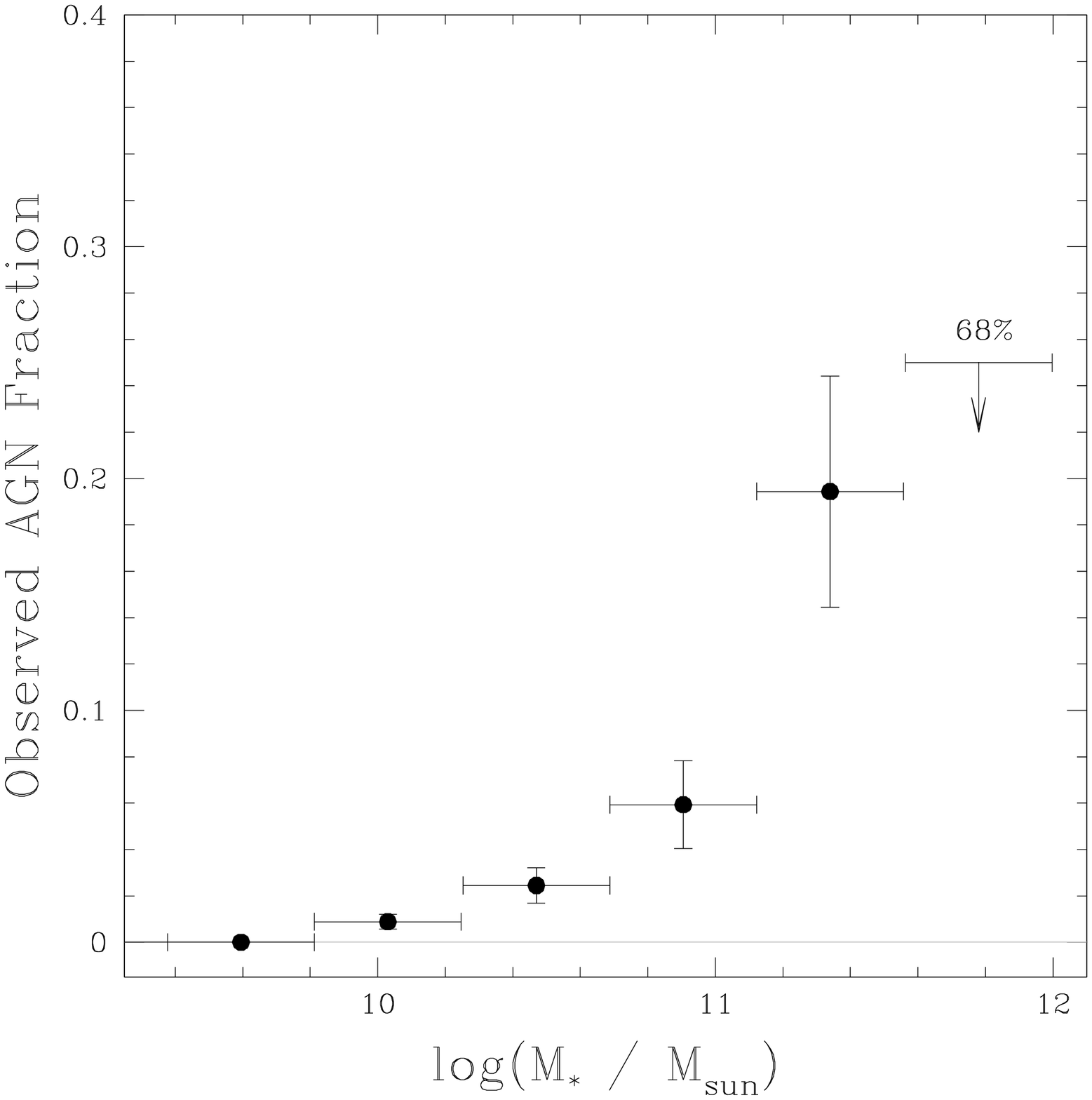} 
	\caption{Observed AGN fraction as a function of stellar mass. The fractions were calculated by comparing the number of AGNs in a bin to the number of non-active galaxies in the full comparison sample. The AGN fraction rises steadily from log($M_{*} / M_{\sun}) = 10$ to log ($M_{*} / M_{\sun}) = 11.5$, where it reaches a maximum of 19\%. Uncertainties in observed AGN fraction were calculated by varying the masses of the AGNs by their mass errors and recalculating the AGN fractions for each bin. Horizontal error bars indicate the width of the bins. As there are 0 AGNs out of 4 objects in the highest-mass bin, we estimate the 68\% confidence limit on the AGN fraction (0.25) using binomial statistics. At the 95\% confidence level, this fraction is 0.53.
	\label{fig:agnfraction}} 
	\epsscale{1.2}
         \end{figure}

We summarize the full statistics for our sample of UV-selected AGNs in Figure \ref{fig:bestfit_hist}. All 28 of the AGNs with fitted SEDs are represented in these histograms. The AGNs with IRAC coverage are plotted with their best-fit parameters from the dual-component modeling, and the AGNs without IRAC coverage are plotted with the corrected best-fit parameters, calculated as described in \S \ref{sec:fullresults}. The average values for the best-fit parameters and standard deviations are indicated on the figure with vertical lines. There is a large spread in $E(B-V)_{SF}$ values, with an average (median) of $\langle \mathrm{E(B-V)_{SF}} \rangle = 0.22\,(0.23) \pm 0.11$, and a range of $0.00 - 0.40$. The best-fit star-formation rates, with a mean of $\langle \mathrm{SFR} \rangle = 63\,(37) \pm 67$ M$_{\sun}$ yr$^{-1}$ similarly span a large range of values between 5.2 - 309 M$_{\sun}$ yr$^{-1}$. The most striking results, however, are the ages and masses for this sample. As suggested by the $\cal{R}$$-K$ colors shown in Figure \ref{fig:rmkcolors}, the best-fit masses are high: log$(\langle \mathrm{M_{*}/M_{\sun}} \rangle) = 10.85\,(10.71) \pm 0.36$. The AGNs that comprise the sample have old ages, with an average age of log$(\langle \mathrm{age/Myr} \rangle) = 3.19\,(3.18) \pm 0.26$. These average stellar masses and ages are large compared to the typical values quoted for UV-selected non-active galaxies \citep{erb2006b,shapley2001}, and will be discussed further in \S \ref{sec:trends} and \S \ref{sec:mass_ew}.

We can use the $L_{bol}$ values derived from the dual-component fitting to estimate accretion rates for the IRAC AGNs. It should be noted that, as described in Section \ref{sec:iracresults}, the values of $L_{bol}$ for our sample are potentially too high as a result of the overprediction of the observed MIPS flux, so the derived accretion rates and Eddington ratios should be viewed as upper limits. The accretion rate is described by the equation $\dot{M} = L_{bol} / \epsilon_{rad}  c^{2}$, where we assume a standard accretion efficiency $\epsilon_{rad} = 0.1$. Using the values for $L_{bol}$ from our fitting, we calculate an average (median) accretion rate for our sample of $0.3 (0.1)$ M$_{\sun}$ yr$^{-1}$ with a range of $0.04 - 0.9$ M$_{\sun}$ yr$^{-1}$. These rates are similar to the range of accretion rates found for optically selected Type I AGNs at $1 < z  < 2.2$ from \citet{merloni2010} (with a median accretion rate of $\dot{M} \sim 0.4 M_{\sun} \, \mathrm{yr}^{-1}$), and a factor of several lower than those estimated for X-ray selected Type II QSOs from \citet{mainieri2011} (with a median accretion rate of $\dot{M} \sim 1$ M$_{\sun}$ yr$^{-1}$).

The results from the SED fitting were also used to derive the properties of the central active galactic nucleus in each of our galaxies. Black hole masses are typically estimated in Type I AGNs using broad emission line widths and AGN continuum luminosities by setting $M_{BH} \propto v^2 r$, where $r$, the size of the AGN broad line region, is estimated from the continuum luminosity using the relationship between luminosity and BLR size based on reverberation mapping studies \citep[e.g.,][]{bentz2009,barth2011}. For our sample of Type II AGNs, black hole masses were estimated using the $M_{BH} - M_{\mathrm{bulge}}$ scaling relation from \citet{haring2004}:

\begin{equation}
 \mathrm{log}(M_{BH} / M_{\sun}) = (8.20 \pm 0.10) + (1.12 \pm 0.06) \, \mathrm{log}(M_{bulge} / 10^{11} M_{\sun})  
\end{equation}

For this calculation, we assumed that the stellar masses derived from our SED fitting roughly corresponded to the bulge masses for these objects. We adjusted the black hole masses for redshift evolution following the parameterization derived for Type I AGNs out to $z = 2.2$ in \citet{merloni2010} \citep[but see][]{lauer2007,alexander2008b}:

\begin{equation}
\Delta \mathrm{log}(M_{BH}/M_{*})(z) = (0.68 \pm 0.12)\, \mathrm{log}(1 + z) 
\end{equation}

The values for log($M_{BH}$) for our sample of AGNs have a range of $7.7 - 8.7$, with an average (median) of log($M_{BH}/M_{\sun}$)$ = 8.36 (8.29)$. The typical uncertainty in $M_{BH}$ is $\sim36\%$. For the eleven AGNs with IRAC data and $L_{bol}$ measurements, we calculated Eddington ratios ($\lambda_{\mathrm{Edd}} = L_{bol} / L_{\mathrm{Edd}}$, where $L_{\mathrm{Edd}}/L_{\sun} = 3.2\times10^{4}\, M_{BH}/M_{\sun}$)\footnote{We calculated an Eddington ratio for Q1623-BX663 of 0.49, but this value is not meaningful due to the large calculated error on the stellar mass for this object.}. The range is lower than that presented for X-ray selected Type II QSOs in \citet{mainieri2011} (in which $\lambda_{\mathrm{Edd}}$ values were calculated in the same manner). The median Eddington ratio for our sample is 0.03 (with a typical uncertainty of $\sim45\%$), while the median Eddington ratio for the \citeauthor{mainieri2011} sample is 0.1. Therefore the black holes hosted by the AGNs in our sample are accreting at significantly sub-Eddington rates. 

\section{AGN Host Galaxy Trends}
\label{sec:trends}

In order to understand the impact of an AGN on its host galaxy, we must compare the UV-selected AGN host galaxy properties to those of a representative sample of non-active galaxies. As the AGNs studied here appear to be hosted by galaxies drawn from the same parent population as the non-AGN LBG sample \citep{steidel2002,adelberger2005a}, a comparison of the AGN and non-AGN host galaxy stellar populations in principle provides a means of testing the impact of AGN feedback. In this section, we examine a non-active sample of UV-selected galaxies, within which we isolate a sample matched in stellar mass for a controlled comparison between AGNs and non-AGNs (\S \ref{sec:comparison}). Using these samples, we examine the relationship between SFR and stellar mass (\S \ref{sec:sfrmass}), as well as rest-frame $U-V$ color and stellar mass (\S \ref{sec:colormass}). Finally, making use of our unique spectroscopic data set, we search for connections between best-fit stellar population parameters and rest-frame UV spectroscopic features (\S \ref{sec:uvcomposite}). 

\subsection{Non-Active Comparison Sample}
\label{sec:comparison}

A careful examination of the demographics of the UV-selected AGNs requires the selection of a comparison sample of non-active galaxies. In \S \ref{sec:sample}, we highlighted a sample of non-AGNs that lie in the same redshift range. Recent results have demonstrated the importance of controlling for stellar mass in understanding the relationship between AGNs and their host galaxies \citep{silverman2009a,xue2010,aird2012}, and so we also constructed a comparison sample that spans a similar mass range to that of the AGN host galaxies. For each AGN in our sample, we chose six non-active sources at random, without any duplicate objects, and characterized by masses within the uncertainty of the selected AGN host mass. The size of the mass-matched sample was limited by the nature of the stellar mass distribution of the parent non-AGN sample, in which there are only six unique objects for each AGN at the massive end. Thus, the final mass-matched sample has 168 objects, as compared to the sample of 28 AGNs with stellar population fits, and has similar redshift properties to the AGN and non-AGN samples. The redshift distribution of the mass-matched sample is characterized by $\langle z \rangle = 2.52 \pm 0.37$, which closely follows that of the UV-selected AGN sample. We will refer to the larger sample of non-AGNs as the ``full non-active comparison sample,'' and the subsample matched in stellar mass as the ``mass-matched comparison sample.'' The average mass for the mass-matched comparison sample is log($\langle \mathrm{M}_{*} \rangle) = 10.85 \pm 0.36$, which, by construction, is the same as for the AGNs. We modeled the photometry of non-AGNs using BC03 CSF SPS-only models as described in \S \ref{sec:fullresults}. 

We used the full non-active comparison sample to calculate the observed fraction of AGNs as a function of stellar mass. In Figure \ref{fig:agnfraction}, we show the results, with errors derived from the mass errors for the AGNs that went into each bin. These results indicate that the fraction of AGNs detected rises steadily as a function of stellar mass, peaking at 19\% for $\mathrm{log}(M_{*}/M_{\sun}) = 11.25$. Above this mass, the number of non-AGNs in the full non-active sample drops to only 4 objects, preventing the determination of tight constraints on the AGN fraction. We observe AGNs predominantly in higher mass galaxies, although as described in \S \ref{sec:mass_ew}, this trend likely reflects our increasing incompleteness towards lower stellar masses. Accordingly, these results may indicate that at least 19\% of UV-selected star-forming galaxies to $\cal{R}$$ = 25.5$ host actively accreting black holes. We return to this discussion in \S \ref{sec:mass_ew}.

\subsection{SFR - Stellar Mass Relation}
\label{sec:sfrmass}

	\begin{figure}
	\epsscale{1.2} 
	\plotone{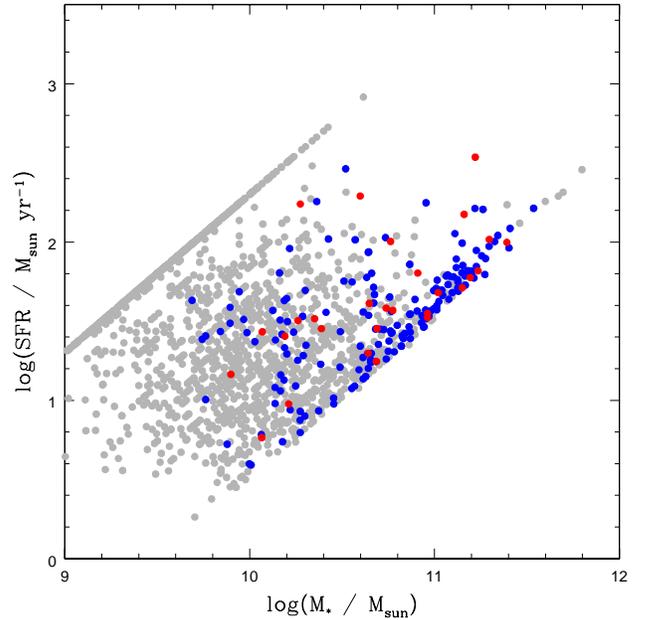} 
	\caption{SFR vs. stellar mass for the AGNs (red points) along with the full comparison sample of non-active star-forming $UG$$\cal{R}$-selected galaxies (grey points) and the mass-matched comparison sample (blue points). AGNs are found at the high mass end of the full relation. However, both AGNs and mass-matched non-AGNs span a similar range of SFR values at a given stellar mass. The ridge of grey points at high SFRs is a result of imposing a lower age limit of $t_{SF} > 50$ Myr (see \S \ref{sec:sfrmass}).  
	\label{fig:sfr_mass}} 
	\epsscale{1.2}
         \end{figure}

Star-forming galaxies follow a tight positive correlation between SFR and stellar mass, which has been observed locally \citep{brinchmann2004}, as well as out to intermediate \citep[][$z \sim 1$]{noeske2007,elbaz2007} and high redshifts \citep[][$z \sim 2-3$]{reddy2006b,daddi2007,reddy2012}. This trend is used to argue for the smooth star-formation histories of the majority of star-forming galaxies in the universe \citep{noeske2007,daddi2007}. This relation is characterized by roughly the same slope across a wide redshift range (although results from \citet{noeske2007} indicate a non-linear relation for star-forming galaxies at intermediate redshifts), while the normalization has decreased by an order of magnitude from $z \sim 2$, reflecting the overall decrease in the global star-formation rate density with time \citep{daddi2007}. We plot the SFR and stellar mass for our AGNs with red points in Figure \ref{fig:sfr_mass}. We find a positive correlation for the AGN host galaxies with a best-fit slope of $0.20 \pm 0.04$, shallower than derived for non-AGNs. 

We compared the AGN population to our non-active comparison samples in order to examine the origin of the shallow slope. In Figure \ref{fig:sfr_mass} we plot $\mathrm{log(SFR)}$ against $\mathrm{log}(M_{*})$ for the non-active comparison sample in grey and for the mass-matched comparison sample in blue. The non-active galaxies display a positive trend with a large scatter. The ridge of points seen at the high SFR side of the relation is an artifact of imposing a minimum age of $t_{SF} = 50$ Myr on our fits. Given the linear relation between $M_{*}$ and SFR for CSF models, all points with best-fit ages of $t_{SF} = 50$ Myr will fall on a line of constant $M_{*}$/SFR. We note that the sharp cutoff for galaxies with low SFRs at a fixed stellar mass is due to the restriction on the maximum allowed age, which must be younger than the age of the universe at the redshift of each object. The lack of points at the low-mass, low-SFR end is due to a bias in the $UG$$\cal{R}$ selection technique towards selecting those galaxies with the highest SFR at a given mass, causing the slope of the relation to become shallower at the low-mass end \citep{reddy2012}. At the high-mass end, the shallow slope reported for the AGNs is similar to what is observed among the non-active sources. Above $\mathrm{log}(M_{*}/M_{\sun}) = 10.5$, the $UG$$\cal{R}$ selection is biased against selecting those galaxies with the largest SFR values at a given stellar mass, as such objects will have large amounts of dust extinction and red $UG$$\cal{R}$ colors, placing them outside the $UG$$\cal{R}$ selection window. This bias tends to flatten the $\mathrm{log}(M_{*}) - \mathrm{log(SFR)}$ relation at the high-mass end. Of particular interest, the SFR values spanned by the UV-selected AGN sample ($(\langle \mathrm{SFR} \rangle) = 63\,(37) \pm 67$ M$_{\sun}$ yr$^{-1}$) are similar to those in the mass-matched comparison sample, where the average (median) SFR is $\langle \mathrm{SFR} \rangle = 45\,(35) \pm 38$ M$_{\sun}$ yr$^{-1}$. These similarities are observed over the full AGN and mass-matched comparison samples, as well as in bins of stellar mass. We quantify them using the Kolmogorov-Smirnov two-sample test, which yields a 44\% probability that the AGN and mass-matched non-AGN SFR distributions are drawn from the same parent population.

The above trends can also be considered in terms of the specific star-formation rate ($\mathrm{sSFR} = \mathrm{SFR} / M_{*}$, which, for CSF models is equal to $1 / t_{SF}$). We calculated an average (median) sSFR $= 1.36\, (0.66)$ Gyr$^{-1}$ for the AGN hosts, with a range of sSFR: $0.4 - 9.2$ Gyr$^{-1}$, indicating active star-formation\footnote{It should be noted that the minimum and maximum allowed age limits for our SED fitting translate into limits on the range of sSFR values that can be recovered for both AGNs and non-AGNs. These limits are apparent in the plot of SFR vs. $M_{*}$ shown in Figure \ref{fig:sfr_mass}.}. We find that the sSFR values for our mass-matched non-active sample span a similar range, with an average (median) sSFR $= 1.10\, (0.47)$ Gyr$^{-1}$, and a range of sSFR: $0.33 - 8.78$ Gyr$^{-1}$. The sSFR values for our Type II UV-selected AGNs are similar to those obtained for a sample of X-ray selected Type II QSOs with active star-formation at $z \sim 2$ from \citet{mainieri2011} (sSFR = $0.8 - 3$ Gyr$^{-1}$ for the high-mass objects at $z \sim 2$). These authors found consistent evolution for the sSFRs of the QSO host galaxies and those of non-active BzK-selected star-forming galaxies from \citet{pannella2009}. Similarly, once we control for stellar mass, the SFR and sSFRs of AGNs are indistinguishable from those of non-AGNs. 

\subsection{Color - Stellar Mass Relation}
\label{sec:colormass}

	\begin{figure}
	\epsscale{1.2} 
	\plotone{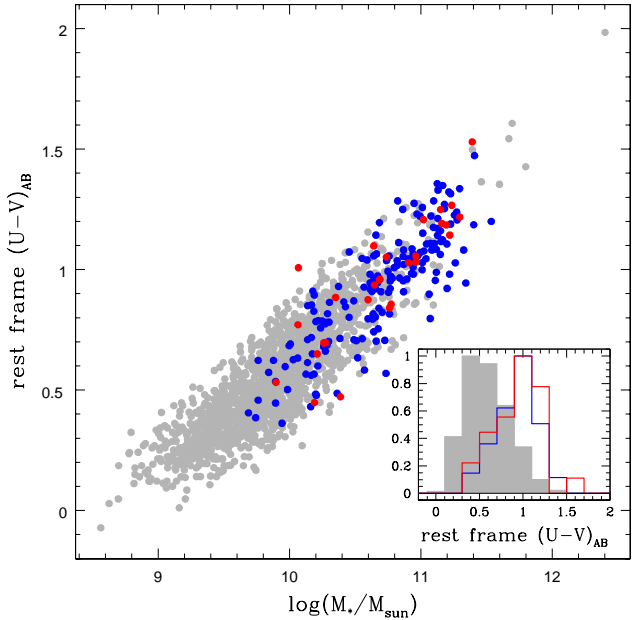} 
	\caption{Rest-frame $U-V$ color vs. stellar mass. The rest-frame $U-V$ colors are calculated from the best-fit SEDs as described in \S \ref{sec:colormass}. AGNs are indicated in red, the non-active comparison sample in grey, and the mass-matched sample in blue. The objects plotted were initially selected by their observed $UGR$ colors to be strongly star forming, which leads to the tight blue sequence \citep{labbe2007}. The UV-selected AGN hosts are located in a similar color-mass space to the mass-matched sample, but are absent below log($M_{*}/M_{\sun}$)$\sim 10$. In the inset, we plot the histograms of $U-V_{rest}$ for the three samples, using the same colors. These histograms have been normalized to have the same peak values for clarity. 
	\label{fig:umv_mass}} 
	\epsscale{1.2}
         \end{figure}

A basic probe of the demographics of AGN activity is provided by the optical color magnitude diagram, which can also be cast in terms of galaxy color as a function of stellar mass. In the local universe, galaxies display a bimodality in their observed colors, separating into a population of lower-mass blue star-forming galaxies and a sequence of higher-mass red quiescent galaxies \citep[e.g.,][]{baldry2004}. This same bimodality has been observed in galaxies at intermediate \citep[$z \sim 1$,][]{bell2004} and higher redshifts \citep[$z \sim 2$,][]{cassata2008, kriek2008, williams2009}. The existence of these two populations out to $z \sim 2$ suggests that the mechanisms driving the evolution of galaxies from blue to red began early in the history of the universe. AGNs have been proposed as a potential driver of the evolution in the color-mass relation by providing energy to the ISM that quenches star-formation, causing a galaxy to transition from the blue population of galaxies to the red \citep{croton2006,hopkins2008}. In support of this scenario, X-ray and optically selected AGNs at low and intermediate redshifts are found to reside in galaxies in the transition region in optical color space between the blue and red sequences \citep{nandra2007,salim2007,coil2009,hickox2009,schawinski2010}. However, recent AGN host galaxy studies indicate that both mass selection and dust obscuration effects may bias conclusions about the role of AGNs in galaxy evolution. \citet{silverman2008} demonstrated that a sample of X-ray and mass selected AGNs were preferentially found in galaxies with ongoing star formation and blue optical colors. Similarly, \citet{xue2010}, using a mass-matched sample of X-ray selected AGNs, showed that the fraction of AGNs remains constant as a function of color. \citet{aird2012} found that, when stellar-mass selection effects are taken into account, AGN hosts are preferentially found in galaxies with blue and green optical colors, but can exist in galaxies of any color. \citet{cardamone2010} corrected AGN host colors for the presence of dust, revealing that AGN hosts cleanly separate into a population of young, dusty, star-forming galaxies, and older, red, passive galaxies.

\begin{figure}
	\epsscale{1.2} 
	\plotone{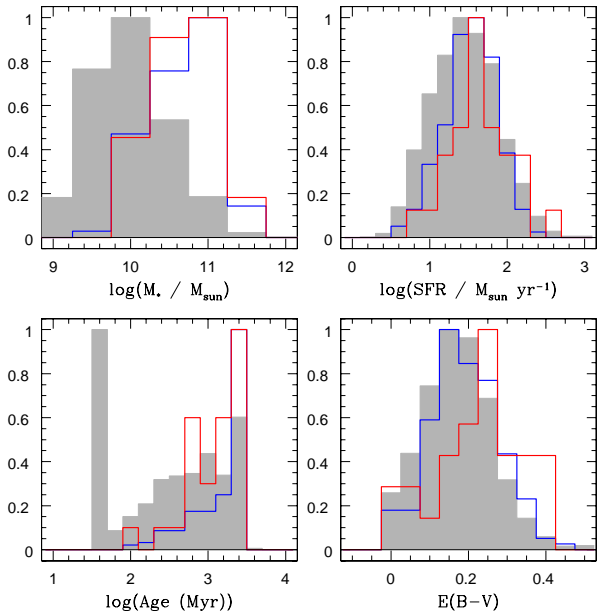} 
	\caption{Histograms of best-fit stellar population parameters. Distributions for AGNs are shown in red, those for the full non-AGN sample in grey, and for the mass-matched sample in blue. These histograms have been normalized to have the same peak values for clarity. The AGNs have similar properties to those of the mass-matched sample on average. The AGNs have a slightly higher average extinction ($\langle E(B-V )_{SF,AGN} \rangle = 0.22 \pm 0.11$) than the mass-matched sample ($\langle E(B-V )_{SF,MM} \rangle = 0.20 \pm 0.09$). 
	\label{fig:mass_matched_histogram}} 
	\epsscale{1.2}
         \end{figure}

We used the results from AGN SED modeling (\S \ref{sec:results}) to examine the positions of the UV-selected AGNs in color-mass space. We calculated rest-frame colors of AGN host galaxies using the best-fit SPS models, along with the Johnson UBV transmission curves from \citet{maiz2006}. The estimation of the rest-frame colors requires an assumption that the models accurately represent the stellar populations in the rest-frame UV and optical, which is supported by the good SPS+AGN dual-component modeling fits to the IRAC AGNs. In order to compare the AGN host colors to those of the non-active UV-selected star-forming galaxies, we also calculated the rest-frame colors for the non-active comparison samples described in the previous section. The $(U-V)_{rest,AB}$ vs. $M_{*}$  relation is shown in Figure \ref{fig:umv_mass}, where the AGNs are plotted in red and the non-active galaxies in grey. The non-AGN hosts follow a trend where the most massive galaxies are also the reddest, similar to the blue sequence observed for galaxies out to $z \sim 3$ \citep{labbe2007}. The sample of UV-selected AGNs are preferentially hosted by high-mass galaxies and these objects are found to span the same range of $(U-V)_{rest,AB}$ colors as those of the non-AGNs in the mass-matched sample. The average (median) color for the AGN sample is $(U-V)_{rest,AB} = 0.9\, (0.9) \pm 0.2$, while for the mass-matched sample, it is $(U-V)_{rest,AB} = 1.0\, (1.0) \pm 0.3$ (inset of Figure \ref{fig:umv_mass}). \citet{xue2010}, \citet{mainieri2011}, and \citet{aird2012} found similar results for the demographics of their AGN host galaxy samples. All of these results demonstrate the importance of mass-selection in the comparison of AGN hosts to non-active galaxies. The redder colors of the AGN hosts relative to the full non-AGN comparison sample do not indicate a lack of ongoing star formation, but, rather, as we show below, more dust and perhaps older stellar populations characteristic of the massive end of the UV-selected star-forming population. The UV-selected AGNs were initially chosen based on their $UG$$\cal{R}$ colors to be star-forming, and, as discussed in the previous section, we found that all of the AGNs have sSFR $> 0.4$ Gyr$^{-1}$, indicating active star formation. Figure \ref{fig:mass_matched_histogram} demonstrates that the SFR, age, color, and dust extinction properties are similar between the AGN sample and the mass-matched comparison sample.  The presence of an AGN does not seem to affect the global properties of their host galaxies in comparison to the mass-matched sample. The lack of UV-selected AGNs in host galaxies below $10^{10} M_{\sun}$ will be discussed in \S \ref{sec:mass_ew}.

We can use the mass-matched sample to explore the $\cal{R}$$-K$ distribution shown in Figure \ref{fig:rmkcolors} in more detail. The fact that AGNs occupy the red portion of this distribution can be explained by their massive host galaxies, and the relation between rest-UV/optical color and stellar mass. The average $\cal{R}$$-K$ color for the UV-selected AGN sample is $4.0 \pm 0.7$, while for the mass-matched sample, the average is $3.6 \pm 0.6$. We used the results from our dual-component modeling to estimate that the presence of an AGN caused the $\cal{R}$$-K$ color to be 0.3 mag redder than it would be in the absence of an AGN. If we subtract this difference from those AGNs where we did not use dual-component modeling, the average UV-selected AGN host galaxy $\cal{R}$$-K$ color becomes $3.6 \pm 0.7$. Therefore, at fixed stellar mass, the AGN host galaxies are typical in terms of their UV/optical (i.e., $\cal{R}$$-K$) colors. 

\subsection{UV Composite Spectra}
\label{sec:uvcomposite}

	\begin{figure*}
	\epsscale{1.} 
	\plotone{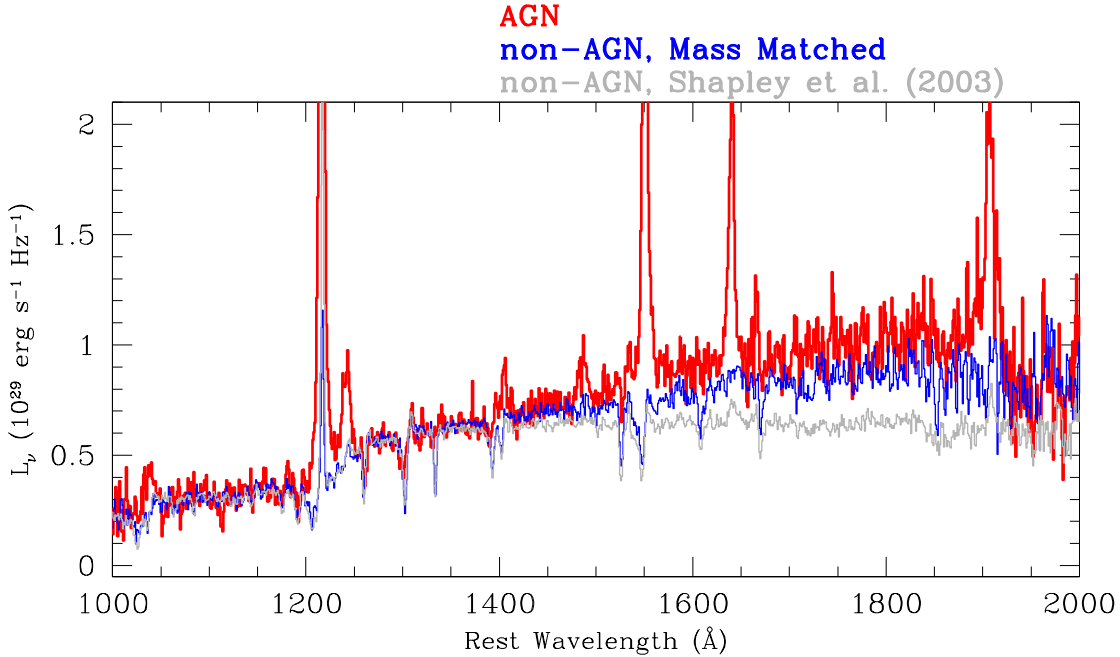} 
	\caption{The UV composite spectra for the UV-selected active and non-active samples. We plot the AGN composite in red, the mass-matched composite in blue, and the non-AGN LBG composite from \citet{shapley2003} in grey. The power-law slope, $\beta$, as measured directly from the AGN spectrum, is $\beta = -0.3 \pm 0.2$, while for the mass-matched composite, the slope is $\beta = -0.7 \pm 0.1$. The similar slopes may be due to the similar levels of dust extinction in these two samples. The mass-matched composite shows a much redder slope than the full non-active composite from \citet{shapley2003}, which has a measured power-law slope of $\beta = -1.49$. These spectra confirm our conclusion from the photometry that the AGN UV continuum is primarily due to stellar light, rather than AGN emission. 	
	\label{fig:massmatchcomposite}} 
	\epsscale{1.}
         \end{figure*}

In \citet{hainline2011}, we investigated the UV spectral properties for our sample of UV-selected AGNs. The UV portion of the galaxy spectrum contains strong emission and absorption features that we used to probe both AGN and outflow activity. At $z \sim 2 - 3$, the individual UV spectra have low S/N, and, to overcome this limitation, we averaged the individual spectra to create higher S/N composite spectra. One of the most striking results from \citet{hainline2011} consisted of the extremely red rest-frame UV continuum of the AGN composite spectrum, relative to that of the $z\sim 3$ non-AGN composite spectrum from \citet{shapley2003}. We can now use the results from SED modeling to investigate the physical origin of this difference. Furthermore, we can analyze how the rest-frame UV spectroscopic properties of AGN vary as a function of stellar population properties. 

For this analysis, we created composite spectra from sub-samples of AGNs separated by the host galaxy properties. Due to the small sample size of objects with SED fitting parameters (28 objects) and our desire to maximize the S/N of the resulting subsample composite spectra, we simply divided the sample in half for these analyses. We constructed composite spectra following the methodology of \citet{hainline2011}. Each individual spectrum was first shifted to the rest frame using an estimate of the systemic redshift from \ion{He}{2}$\lambda$1640. We converted each spectrum from the units of flux density to those of luminosity density, scaled each one to a common median in the wavelength range of $1250 - 1380$ \AA, and combined the spectra, with the four highest and lowest outliers removed at each wavelength. 

In order to investigate the red UV continuum of the AGN composite spectrum, we consider the properties of the mass-matched comparison sample. Using existing rest-frame UV spectroscopy for the full non-active sample, we created a UV composite spectrum from the objects in the mass-matched comparison sample. We used the same method to create this non-AGN composite as for the AGN composite. In Figure \ref{fig:massmatchcomposite}, we compare the full AGN composite spectrum to the mass-matched non-AGN composite as well as the non-AGN LBG composite at $z\sim3$ from \citet{shapley2003}. The UV continuum shape is commonly described by $\beta$, the slope of a power law of the form $L_\lambda \propto \lambda^{\beta}$, which is fit to the continuum. For the AGN composite spectrum, $\beta = -0.3 \pm 0.2$ \citep{hainline2011}, which is a much redder slope compared to the LBG composite from \citet{shapley2003}, where $\beta = -1.5$. We measure a value of $\beta = -0.7 \pm 0.1$ for the mass-matched composite, redder than the full non-AGN composite\footnote{Errors on the $\beta$ values were calculated following a bootstrap technique where 500 fake composite spectra were constructed from the sample of actual spectra and we measured $\beta$ for each fake composite. The error represents the standard deviation of the distribution of measured $\beta$ values}. The comparably red slopes of the AGN and mass-matched UV composite spectra \citep[in contrast to the significantly bluer slope of the total LBG composite from][]{shapley2003} confirms that the underlying UV continuum seen for the AGN sample is predominantly due to stellar light, and not AGN emission. Furthermore, since scattered light from a buried AGN would have an intrinsically \textit{bluer} rest-frame UV spectrum than that of a stellar population \citep{zakamska2006}, the slight difference in UV slopes can not be attributed to AGN emission. We can naturally explain the redder slopes observed in the high-mass active and inactive samples by looking at the distributions of reddening for each subsample. In Figure \ref{fig:mass_matched_histogram}, we show histograms of the best-fit properties of the AGN sample along with the full non-AGN and mass-matched comparison samples. As discussed in \S \ref{sec:fullresults}, $\langle E(B-V)_{SF,AGN} \rangle = 0.22 \pm 0.11$ for the AGNs, while for the mass-matched comparison sample, $\langle E(B-V)_{SF,MM} \rangle = 0.20 \pm 0.09$. For the sample of non-AGNs at $z \sim 3$ from \citet{shapley2003}, the estimated $\langle E(B-V)_{SF,Shapley} \rangle = 0.15 \pm 0.09$\footnote{These $E(B-V)_{SF}$ values were calculated directly from the UV photometry as described in \citet{shapley2003}.}. The errors on these values represent the standard deviation of the distribution of values. The average extinction values for the AGN and mass-matched samples are significantly different from the value for the non-AGNs in \citet{shapley2003}, reflecting what is observed for the UV power-law slopes. 

	\begin{figure*}
	\epsscale{1.} 
	\plotone{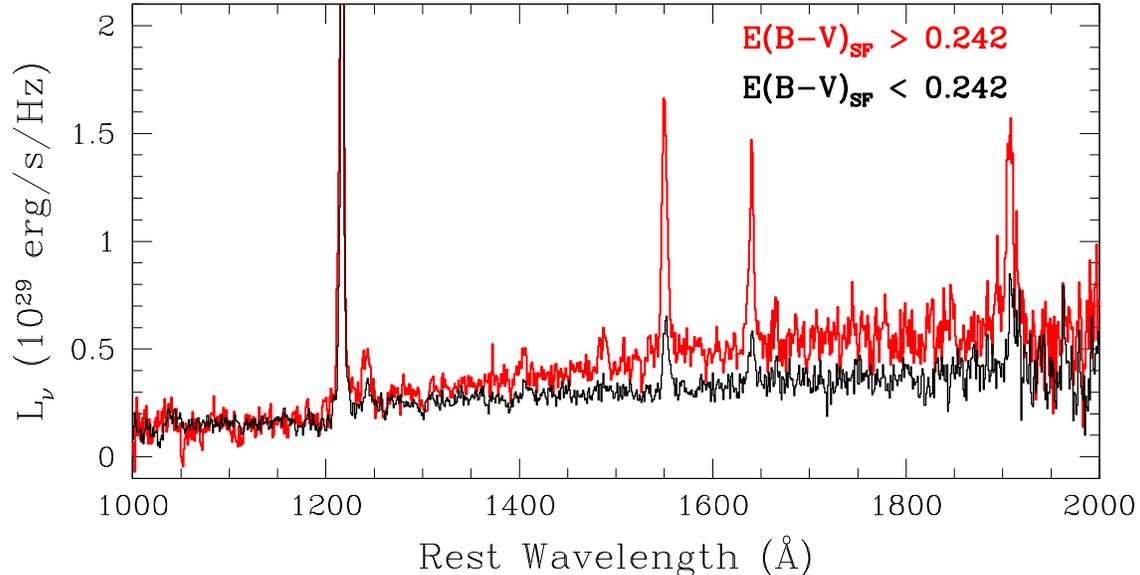} 
	\caption{Comparison of the continuum slope between the AGN composite spectrum for objects separated by $E(B-V)_{SF}$. The composite with larger $E(B-V)_{SF}$ is significantly redder than the smaller $E(B-V)_{SF}$ composite. The UV power-law slope $\beta$, as measured directly from the $E(B-V)_{SF} > 0.242$ spectrum, is $\beta = 0.0 \pm 0.3$, while for the $E(B-V)_{SF} < 0.242$ spectrum, $\beta = -1.0 \pm 0.3$. The large average $E(B-V)_{SF}$ for our AGN sample indicated in Figure \ref{fig:bestfit_hist} leads to the red UV continuum seen in AGN UV composite spectrum.
	\label{fig:ebvcomp}} 
	\epsscale{1.}
         \end{figure*}

We further investigated the role that extinction plays in the slope of the UV spectrum by separating our objects into two bins based on $E(B-V)_{SF}$. The AGNs were separated at the median extinction from our fits, $E(B-V)_{SF} = 0.242$, and both composites are plotted in Figure \ref{fig:ebvcomp}. In red, we plot the composite spectrum for those objects with $E(B-V)_{SF} > 0.242$, and in black, the composite spectrum for those objects with $E(B-V)_{SF} < 0.242$. The composite spectrum for objects with higher $E(B-V)_{SF}$ is much redder, with $\beta = 0.0 \pm 0.3$. For the AGNs with lower $E(B-V)_{SF}$, the composite spectrum has a $\beta = -1.0 \pm 0.3$. This trend is also seen in the mass-matched comparison sample separated by $E(B-V)_{SF}$, where the high-extinction mass-matched composite is redder, with $\beta = -0.4 \pm 0.2$, while the low-extinction mass-matched composite has $\beta = -1.1 \pm 0.1$. The slope of the low-extinction mass-matched composite spectrum ($\langle E(B-V)_{SF,MM} \rangle = 0.12 \pm 0.05$) is similar to the slope of the low-extinction AGN composite ($\langle E(B-V)_{SF,AGN} \rangle = 0.13 \pm 0.08$), which provides more evidence that dust extinction significantly modulates the UV power-law slope in our sample of AGNs. 

\begin{figure*}
	\epsscale{1.} 
	\plotone{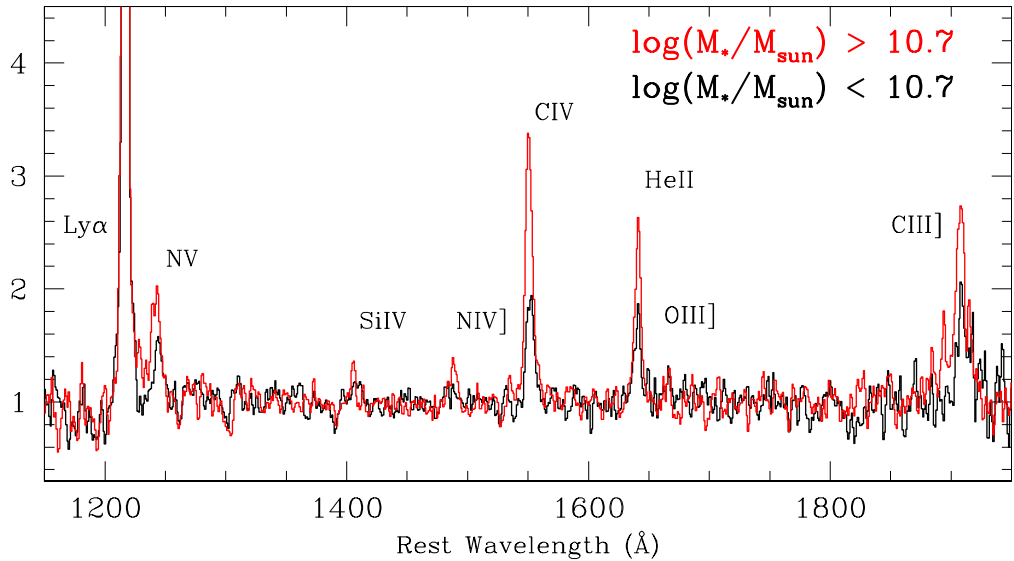} 
	\caption{Comparison of the strong emission lines between the AGN composite spectrum for objects separated by $M_{*}$. These spectra are continuum normalized to better illustrate the line EW differences. The high-mass composite spectrum is shown in red, while the low-mass composite spectrum is shown in black. The AGN emission lines of \ion{N}{5}, \ion{C}{4}, \ion{He}{2} are stronger in the high-mass composite, with specific EW value given in Table \ref{tab:mass_separated}.
	\label{fig:masscomp}} 
	\epsscale{1.}
         \end{figure*}

We also consider how rest-frame UV AGN spectroscopic properties vary with stellar mass. The continuum normalized composite spectra for the AGNs separated by stellar mass are shown in Figure \ref{fig:masscomp}\footnote{Given the correlation between stellar mass and $E(B-V)_{SF}$, a separation by stellar mass divides the sample in almost the same way as a separation by $E(B-V)_{SF}$.}. We measured the luminosities and EW for the strongest UV emission lines (\ion{H}{1} Ly$\alpha$, \ion{N}{5}$\lambda$1240, \ion{C}{4}$\lambda$1549, and \ion{He}{2}$\lambda$1640) in both the high-mass and low-mass composites. Uncertainties on these values were calculated following a bootstrap technique where 500 fake composite spectra were constructed from the sample of spectra used in creating the real composite spectra \citep[see][]{shapley2003,hainline2011}. The results are shown in Table \ref{tab:mass_separated}. We find that the EW values for the \ion{C}{4} and \ion{He}{2} emission lines are stronger in the high-mass composite than the low mass composite, while the \ion{N}{5} and Ly$\alpha$ lines have statistically equivalent EW values. The EW of an emission line was calculated by integrating the luminosity in the line and dividing by the average of the luminosity density of the continuum on either side of the line. As these are Type II AGNs, the observed UV continuum arises from stellar emission, which is supported by the similar UV slopes seen in the AGN UV composite and the mass-matched UV composite. We find that the emission lines used to identify the presence of an AGN (\ion{C}{4} and \ion{He}{2}) are stronger in the high-mass composite spectrum. This trend of decreasing high-ionization EW with decreasing stellar mass may explain the dearth of low-mass UV-selected AGNs in our sample. We will explore the origins and implications for this trend in the following section. 

\section{The Demographics of UV-selected AGNs at $z \sim 2-3$}
\label{sec:mass_ew}

	\begin{figure*}
	\epsscale{0.9} 
	\plotone{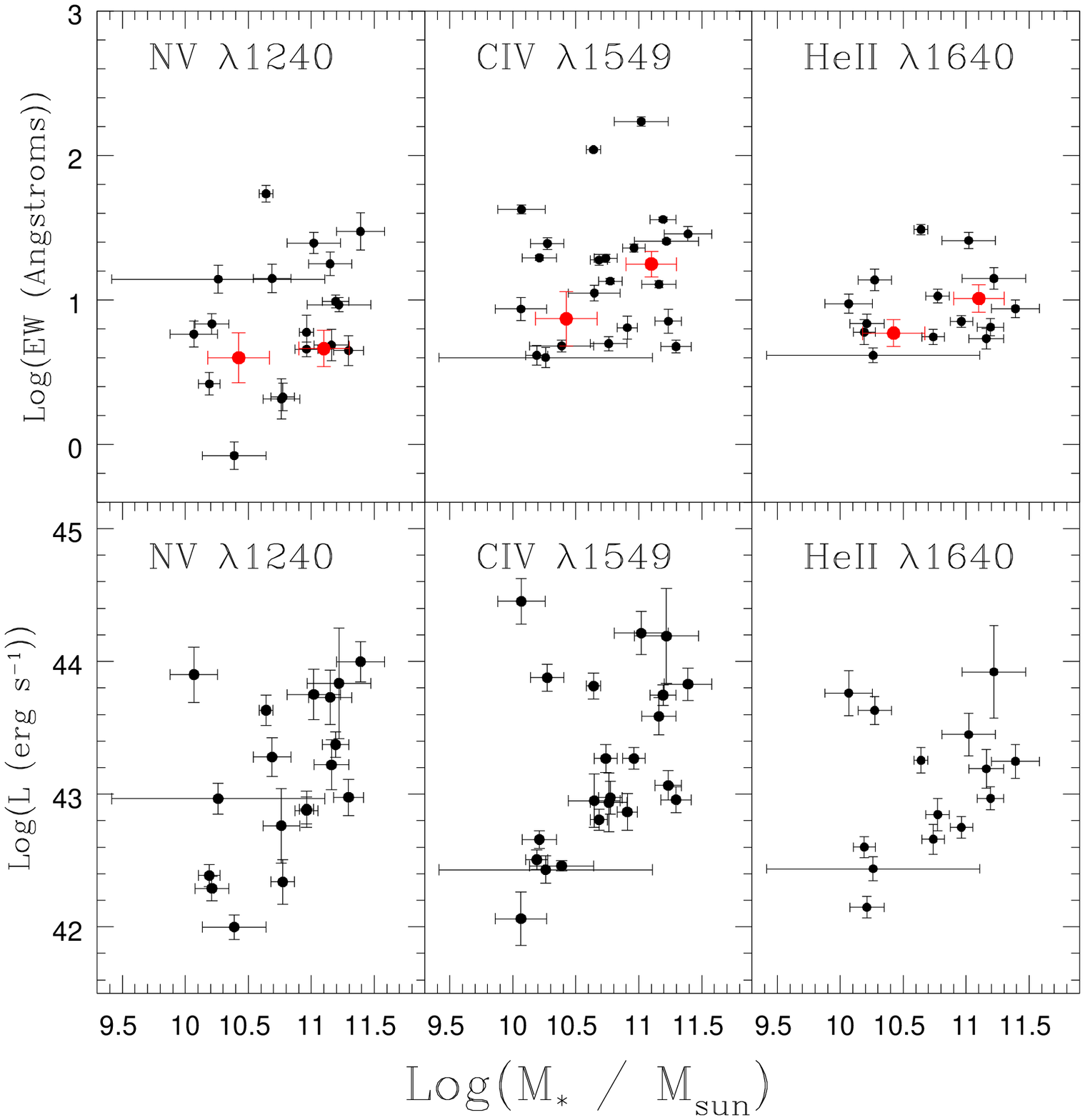} 
	\caption{EW (top three panels) and line luminosities (bottom three panels) for the strong UV emission lines in the AGN sample plotted against stellar mass. The black points are measured from the individual AGN spectra, and the red points in the EW plots represent values measured from the composite spectra as separated into two mass bins. For \ion{C}{4} and \ion{He}{2}, the points from the composite spectra indicate that the EW values are higher in the high mass composites. The luminosity values are corrected for dust extinction using the $E(B-V)_{SF}$ values from the SED fitting, and show a positive trend with mass, such that the higher mass galaxies have intrinsically brighter emission lines. 
	\label{fig:ewlum_vs_mass}} 
	\epsscale{0.9}
         \end{figure*}
         
We find an apparent segregation of AGN hosts at the high-mass end of the $z\sim 2-3$ galaxy population, as shown in Figures \ref{fig:sfr_mass}, \ref{fig:umv_mass}, and \ref{fig:mass_matched_histogram}. At lower-redshift ($z\sim 0.2-1.0$), the preferential incidence of X-ray-selected AGNs in higher-mass host galaxies is explained by \citet{aird2012} as a selection effect, arising from a constant Eddington-ratio distribution and an X-ray flux limit. For a fixed Eddington ratio, AGNs in higher-mass galaxies have higher AGN luminosities (assuming a linear correlation between $M_{BH}$ and $M_{host}$), and are therefore easier to detect. We use the extinction-corrected \ion{C}{4} luminosity as a proxy for AGN luminosity. Without direct X-ray or UV-optical continuum measurements, we must resort to less direct measures of luminosity for our obscured sample. Here we assume that the extinction-corrected \ion{C}{4} luminosity from the narrow-line region traces the accretion rate onto the black hole, as shown for narrow AGN optical emission lines (e.g., [\ion{O}{3}]$\lambda$5007) by \citet{yee1980} and \citet{shuder1981}. With this assumption, we can apply the arguments from \citet{aird2012} to our higher-redshift AGN sample, in order to investigate the cause of the apparent segregation in host galaxy stellar mass, and the detectability of AGN activity in lower-mass host galaxies. The lack of detected AGN activity in the low-mass UV-selected sample may be explained simply as a similar selection effect. That is, if black holes are smaller on average in low-mass galaxies, then their emission may fall below our detection threshold.

In the first place, we must establish empirically how emission-line EW and luminosity vary with stellar mass. The EW of an emission feature depends on both the intrinsic luminosity of the feature as well as the continuum luminosity. In Figure \ref{fig:ewlum_vs_mass}, we plot emission line EW against stellar mass for objects in our AGN sample. The \ion{N}{5}, \ion{C}{4}, and \ion{He}{2} emission lines were measured from individual spectra. In the top panels, we also plot the EW values measured from the composite spectra separated by mass. For the luminosity values, the individual spectra were flux-normalized such that the $1500$ \AA\ continuum luminosity matched the estimated rest-frame $1500$ \AA\ luminosity calculated from $G$ and/or $\cal{R}$-band photometry for these objects. We then corrected each spectrum for extinction under the assumption that the stellar extinction estimated from the SED modeling applies to the narrow-line region. The \ion{N}{5} and \ion{C}{4} emission line luminosities shown in Figure \ref{fig:ewlum_vs_mass} demonstrate a weak trend with mass (Spearman's rank coefficients of 0.45 and 0.43 for the \ion{N}{5} and \ion{C}{4} distributions respectively, corresponding to $\sim 2 \sigma$ significance) such that the higher mass galaxies have higher luminosity emission lines\footnote{This trend is strengthened considerably by removing Q2233-MD21, which has very luminous emission lines and a low best-fit stellar mass. The Spearman rank coefficients rise to 0.68 and 0.62, for the \ion{N}{5} and \ion{C}{4} distributions respectively,  corresponding to $\sim 3 \sigma$ significance.}. 

Of the three emission lines used to select an AGN, \ion{C}{4} is typically the strongest, and we focus on that line in order to test the detectability of lower-mass AGNs. As discussed above, results from \citet{aird2012} suggest that X-ray selected AGNs out to $z \sim 1$ have a universal distribution of Eddington ratios independent of host galaxy stellar mass. If AGNs at $z\sim 2-3$ are also characterized by a universal Eddington ratio distribution, and, furthermore, if $M_{BH}$ and $M_*$ are correlated as in the low-redshift universe, galaxies with higher stellar masses should exhibit AGN-related emission lines with larger luminosities (i.e., a constant ratio of $L_{CIV}/M_*$). The assumption of a constant $L_{CIV}/M_*$ ratio is supported by the observed correlation between \ion{C}{4} luminosity and stellar mass in our sample. Using this assumption along with the assumption of the correlation between $M_{BH}$ and $M_*$, we can test our ability to detect UV-selected AGNs in host galaxies with a given log$(M_*/M_{\sun})$.

Since AGNs are easiest to find at the highest stellar masses, we use the spectra of the highest-mass AGNs in our sample to estimate the ratio of $L_{CIV}/M_*$. We averaged the \ion{C}{4} line luminosities for all AGNs in our sample with log($M_{*}/M_{\sun}) > 11$. Under the above assumption that emission line luminosity scales linearly with stellar mass, we used the average line luminosity at log($M_{*}/M_{\sun}) \sim 11$ to compute the average line luminosity for two bins of stellar mass below the minimum mass in our AGN sample: $\mathrm{log}(M_{*}/M_{\sun}) = 9.0 - 9.5$ and $\mathrm{log}(M_{*}/M_{\sun}) = 9.5 - 10.0$. Based on the results of the SED modeling for the objects in the non-AGN comparison sample in these two stellar mass bins, we applied an average level of extinction ($\langle \mathrm{E(B-V)_{SF}} \rangle = 0.10$ for the lower mass bin, and $\langle \mathrm{E(B-V)_{SF}} \rangle = 0.12$ for the higher mass bin) to predict extinction corrected \ion{C}{4} luminosities. Finally, we used the objects in the lower-mass bins from the non-AGN comparison sample to estimate an average rest-frame $1500$ \AA\ continuum luminosity density from the $G$ and $\cal{R}$ photometry for the objects. Based on the predicted \ion{C}{4} luminosities and typical UV continuum luminosity densities, we estimated the typical \ion{C}{4} EWs for objects in these lower-mass bins. The resulting EW values are $\mathrm{EW}_{\mathrm{CIV}} = 3.4$ \AA\ for the $\mathrm{log}(M_{*}/M_{\sun}) = 9.0 - 9.5$ bin, and $\mathrm{EW}_{\mathrm{CIV}} = 6.6$ \AA\ for the $\mathrm{log}(M_{*}/M_{\sun}) = 9.5 - 10.0$ bin. 

With average \ion{C}{4} EWs in hand, we can now ask whether we would have identified these lines in galaxies with log($M_{*}/M_{\sun}) < 10$. To test whether these lines would be detectable in lower-mass galaxies, we used the individual UV spectra for 15 representative non-AGNs in each of these lower-mass bins. We calculated the uncertainty on the EW in the spectral regions around the \ion{C}{4} line using the rms noise in the individual spectra, and estimated that in order for the line to be detected, $\mathrm{EW}_{\mathrm{CIV}} > 8.5$ \AA\ for the $\mathrm{log}(M_{*}/M_{\sun}) = 9.0 - 9.5$ bin, and $\mathrm{EW}_{\mathrm{CIV}} > 9.0$ \AA\ for the $\mathrm{log}(M_{*}/M_{\sun}) = 9.5 - 10.0$ bin. These estimates represent a 5$\sigma$ detection limit that was used for identifying \ion{C}{4} in the AGN spectra. For typical spectra of objects at log$(M_{*}/M_{\sun}) < 10.0$, the predicted \ion{C}{4} EW values would therefore be too low to be detected. In conclusion, if galaxies with $\mathrm{log}(M_{*}/M_{\sun}) < 10.0$ hosted black holes accreting at the same average Eddington ratio as in higher mass galaxies, we would be unable to detect them given the emission line sensitivity in our spectroscopic sample. 

The segregation of UV-selected AGNs in high-mass ($\log M_*/M_{\sun} \geq 10.0$) host galaxies is naturally explained based on two key assumptions: (1) that $\langle L_{CIV}/M_{BH} \rangle$ is a constant as a function of $M_{BH}$ (i.e., a constant Eddington ratio distribution), as demonstrated for X-ray-selected AGNs at $z< 1$ \citep{aird2012} and supported by the observed correlation between $L_{CIV}$ and $M_*$ in our sample, and (2) that $M_{BH}$ is proportional to $M_*$. Even if lower-mass AGN host galaxies were characterized by similar Eddington ratios to those at higher-mass, we would be unable to detect them based on their high-ionization rest-frame UV emission lines. Viewed from a different perspective, the very segregation of UV-selected AGNs in high-mass hosts suggests that $M_{BH}$ and $M_*$ are already correlated at $z> 2$, during the epoch when both bulges and black holes are actively growing. Alternatively, our results can be explained if there are no black holes at $z\sim 2-3$ in galaxies at $\log(M_*/M_{\sun}) < 10$, or if the average Eddington ratio decreases in less massive systems. This latter scenario is inconsistent with results at lower redshift from \citet{aird2012}. 

The discussion thus far has focused entirely on Type II AGNs. As presented in \citet{steidel2002}, \citet{adelberger2005e}, and \citet{adelberger2005d}, within the UV-selected galaxy population, broad-line or ``Type I" AGNs have a similar frequency to that of the Type II AGNs discussed here. It is important to consider if any of our conclusions regarding the demographics of AGN activity at $z\sim2-3$ are impacted by including a description of these Type I AGNs. Although it is not possible to make robust models of the host galaxy stellar populations of Type I objects, due to significant contamination from the active nucleus, we can investigate the typical Eddington ratios for Type I objects spanning the same range in $L_{bol}$ as that of our Type II sample. We estimated $L_{bol}$ for the Type I objects from their rest-frame 1350~\AA\ luminosities (i.e., $\lambda L_{\lambda}$(1350\AA), calculated using the observed $G$- and $\cal{R}$-band photometry for these objects), which we then multiplied by the bolometric correction factor of 4.3 from \citet{vo2009}. We used the relationship between $L_{bol}$ and $M_{BH}$ implied by the correlations presented in \citet{adelberger2005d} to estimate $M_{BH}$ values. We find typical black hole masses of $\log(M_{BH}/M_{\sun}) \sim 8$ -- comparable to the black hole masses presented section 4.3. Accordingly, the Type I objects in the a matched range of $L_{bol}$ must have similar Eddington ratios to those of the Type II objects presented here, and appear to represent their unobscured analogs. It is worth noting that the $L_{bol}$ distribution of Type I AGNs identified within the the UV-selected survey extends to higher luminosities than those spanned by the Type II objects, with roughly one third of Type I sample characterized by $L_{bol} >10^{46}$ erg s$^{-1}$. Eddington-limited accretion in these systems corresponds to black hole masses of $\log(M_{BH}/M_{\sun}) \sim 8.4$, which represents a lower limit in black hole mass.  Accordingly, it seems unlikely that these black holes are hosted by significantly lower-mass galaxies than the Type II AGNs discussed in this work. It is remarkable that both types of UV-selected AGNs at fixed bolometric luminosity are characterized by Eddington ratios $<0.1$ for typical black hole masses of $\log(M_{BH}/M_{\sun}) \sim 8$. At some earlier time, these black holes must have been accreting at significantly higher Eddington ratios. An important goal will be the identification of the more active progenitors of the UV-selected AGNs at $z\sim 2-3$.

\section{Conclusions}
\label{sec:conclusions}

Using SPS modeling, we have examined the host galaxy populations for a sample of $z \sim 2-3$ Type-II AGNs identified by the presence of rest-frame UV emission features. The results of our AGN+SPS dual-component modeling demonstrate the importance of using long wavelength coverage to properly model the SEDs of type II AGNs. Without data longwards of $1 - 2 \mu$m (rest frame), the masses and SFRs for our sample of AGNs would be overpredicted by an average factor of $1.4$. We have used the results of our modeling to examine the host galaxy trends for our sample. Our primary results are the following:

\begin{enumerate}
  \item The host galaxies for the UV-selected AGNs have very high average (median) masses (log$(\langle \mathrm{M_{*}/M_{\sun}} \rangle) = 10.85\,(10.71) \pm 0.36$) and high SFRs ($\langle \mathrm{SFR} \rangle = 63\,(37) \pm 67$ M$_{\sun}$ yr$^{-1}$).
  \item We have constructed a mass-matched non-AGN sample with which to compare to the AGN hosts. The AGNs lie at the high-mass edge of the $U-V_{rest,AB}$ vs. $M_{*}$ relation compared to the full non-AGN sample, but span the same range of color as the mass-matched non-AGN sample. Additionally, the AGNs have similar SFRs to the mass-matched sample, indicating that the presence of an AGN does not affect the colors or star-formation properties of these highly star-forming hosts at these redshifts.
  \item A primary result from \citet{hainline2011} is that the UV composite spectrum for the AGNs has a redder continuum than the full non-AGN composite from \citet{shapley2003}. The UV composite constructed for the objects in the mass-matched comparison sample has a red continuum that is very similar to that of the AGN composite, and we argue that this result is due to more dust extinction in the higher-mass galaxies hosting AGNs as well as the mass-matched sample. This similarity between the UV continua of the mass-matched non-AGNs and AGNs confirms that the UV continuum of AGN hosts is dominated by starlight. 
  \item We separated our AGNs by stellar mass into two subsamples and created composite spectra for each of these subsamples. The \ion{C}{4} and \ion{He}{2} EWs are larger in the high-mass composite spectrum than in the low-mass composite spectrum. This trend provides evidence that the EWs of the UV emission lines used to identify the AGNs depends on stellar mass. We also observe that the dust-corrected luminosities for the \ion{N}{5}, \ion{C}{4}, and \ion{He}{2} emission lines scale with host galaxy stellar mass. These trends would be predicted if the mass of the host galaxy scaled with the black hole mass and emission line luminosity.  
  \item We estimate the \ion{C}{4} line EW for potential AGNs hosted by galaxies of mass $\mathrm{log}(M_{*}/M_{\sun}) = 9.0 - 9.5$ and $\mathrm{log}(M_{*}/M_{\sun}) = 9.5 - 10.0$, under the assumption that line luminosity scales linearly with galaxy mass. The predicted \ion{C}{4} EWs ($3.4$ \AA\ for the lowest mass bin, and $6.6$ \AA\ for the higher mass bin) would not be detectable at the 5$\sigma$ level required for AGN identification in our sample, indicating that we cannot identify UV-selected AGNs in galaxies with stellar masses $\mathrm{log}(M_{*}/M_{\sun}) < 10.0$. Additionally, we can use the segregation of UV-selected AGN hosts to argue that $M_{BH}$ and $M_{*}$ are already correlated at $z > 2$.
\end{enumerate}

While we have demonstrated that AGNs hosted by lower-mass galaxies would not be detected in our sample based on the presence of UV emission lines, AGNs can be identified using X-ray luminosity and IR colors. In future work, we will investigate the demographics of AGNs selected using these additional methods and obtain a more complete picture of AGN activity at $z > 2$. 

\acknowledgments 

The authors wish to thank Roberto Assef, James Aird, and Alison Coil for helpful discussions.
A.E.S. acknowledges support from the David and Lucile Packard Foundation. C.C.S. acknowledges 
additional support from the John D. and Catherine T. MacArthur Foundation and the Peter and 
Patricia Gruber Foundation. We wish to extend special thanks to those of Hawaiian ancestry on 
whose sacred mountain we are privileged to be guests. Without their generous hospitality, most 
of the observations presented herein would not have been possible.

\bibliographystyle{apj}
\bibliography{apj-jour,lbgrefs}

\begin{thebibliography}{114}
\expandafter\ifx\csname natexlab\endcsname\relax\def\natexlab#1{#1}\fi

\bibitem[{{Adelberger} \& {Steidel}(2005{\natexlab{a}})}]{adelberger2005e}
{Adelberger}, K.~L. \& {Steidel}, C.~C. 2005{\natexlab{a}}, \apj, 630, 50

\bibitem[{{Adelberger} \& {Steidel}(2005{\natexlab{b}})}]{adelberger2005d}
---. 2005{\natexlab{b}}, \apjl, 627, L1

\bibitem[{{Adelberger} {et~al.}(2005){Adelberger}, {Steidel}, {Pettini},
  {Shapley}, {Reddy}, \& {Erb}}]{adelberger2005a}
{Adelberger}, K.~L., {Steidel}, C.~C., {Pettini}, M., {Shapley}, A.~E.,
  {Reddy}, N.~A., \& {Erb}, D.~K. 2005, \apj, 619, 697

\bibitem[{{Adelberger} {et~al.}(2004){Adelberger}, {Steidel}, {Shapley},
  {Hunt}, {Erb}, {Reddy}, \& {Pettini}}]{adelberger2004a}
{Adelberger}, K.~L., {Steidel}, C.~C., {Shapley}, A.~E., {Hunt}, M.~P., {Erb},
  D.~K., {Reddy}, N.~A., \& {Pettini}, M. 2004, \apj, 607, 226

\bibitem[{{Aird} {et~al.}(2012){Aird}, {Coil}, {Moustakas}, {Blanton},
  {Burles}, {Cool}, {Eisenstein}, {Smith}, {Wong}, \& {Zhu}}]{aird2012}
{Aird}, J., {Coil}, A.~L., {Moustakas}, J., {Blanton}, M.~R., {Burles}, S.~M.,
  {Cool}, R.~J., {Eisenstein}, D.~J., {Smith}, M.~S.~M., {Wong}, K.~C., \&
  {Zhu}, G. 2012, \apj, 746, 90

\bibitem[{{Alexander} {et~al.}(2008{\natexlab{a}}){Alexander}, {Brandt},
  {Smail}, {Swinbank}, {Bauer}, {Blain}, {Chapman}, {Coppin}, {Ivison}, \&
  {Men{\'e}ndez-Delmestre}}]{alexander2008b}
{Alexander}, D.~M., {Brandt}, W.~N., {Smail}, I., {Swinbank}, A.~M., {Bauer},
  F.~E., {Blain}, A.~W., {Chapman}, S.~C., {Coppin}, K.~E.~K., {Ivison}, R.~J.,
  \& {Men{\'e}ndez-Delmestre}, K. 2008{\natexlab{a}}, \aj, 135, 1968

\bibitem[{{Alexander} {et~al.}(2008{\natexlab{b}}){Alexander}, {Chary}, {Pope},
  {Bauer}, {Brandt}, {Daddi}, {Dickinson}, {Elbaz}, \& {Reddy}}]{alexander2008}
{Alexander}, D.~M., {Chary}, R.-R., {Pope}, A., {Bauer}, F.~E., {Brandt},
  W.~N., {Daddi}, E., {Dickinson}, M., {Elbaz}, D., \& {Reddy}, N.~A.
  2008{\natexlab{b}}, \apj, 687, 835

\bibitem[{{Alonso-Herrero} {et~al.}(2008){Alonso-Herrero},
  {P{\'e}rez-Gonz{\'a}lez}, {Rieke}, {Alexander}, {Rigby}, {Papovich},
  {Donley}, \& {Rigopoulou}}]{alonso2008}
{Alonso-Herrero}, A., {P{\'e}rez-Gonz{\'a}lez}, P.~G., {Rieke}, G.~H.,
  {Alexander}, D.~M., {Rigby}, J.~R., {Papovich}, C., {Donley}, J.~L., \&
  {Rigopoulou}, D. 2008, \apj, 677, 127

\bibitem[{{Alonso-Herrero} {et~al.}(2003){Alonso-Herrero}, {Quillen}, {Rieke},
  {Ivanov}, \& {Efstathiou}}]{alonso2003}
{Alonso-Herrero}, A., {Quillen}, A.~C., {Rieke}, G.~H., {Ivanov}, V.~D., \&
  {Efstathiou}, A. 2003, \aj, 126, 81

\bibitem[{{Alonso-Herrero} {et~al.}(2001){Alonso-Herrero}, {Quillen},
  {Simpson}, {Efstathiou}, \& {Ward}}]{alonso2001}
{Alonso-Herrero}, A., {Quillen}, A.~C., {Simpson}, C., {Efstathiou}, A., \&
  {Ward}, M.~J. 2001, \aj, 121, 1369

\bibitem[{{Assef} {et~al.}(2010){Assef}, {Kochanek}, {Brodwin}, {Cool},
  {Forman}, {Gonzalez}, {Hickox}, {Jones}, {Le Floc'h}, {Moustakas}, {Murray},
  \& {Stern}}]{assef2010}
{Assef}, R.~J., {Kochanek}, C.~S., {Brodwin}, M., {Cool}, R., {Forman}, W.,
  {Gonzalez}, A.~H., {Hickox}, R.~C., {Jones}, C., {Le Floc'h}, E.,
  {Moustakas}, J., {Murray}, S.~S., \& {Stern}, D. 2010, \apj, 713, 970

\bibitem[{{Baldry} {et~al.}(2004){Baldry}, {Glazebrook}, {Brinkmann},
  {Ivezi{\'c}}, {Lupton}, {Nichol}, \& {Szalay}}]{baldry2004}
{Baldry}, I.~K., {Glazebrook}, K., {Brinkmann}, J., {Ivezi{\'c}}, {\v Z}.,
  {Lupton}, R.~H., {Nichol}, R.~C., \& {Szalay}, A.~S. 2004, \apj, 600, 681

\bibitem[{{Barger} {et~al.}(2005){Barger}, {Cowie}, {Mushotzky}, {Yang},
  {Wang}, {Steffen}, \& {Capak}}]{barger2005a}
{Barger}, A.~J., {Cowie}, L.~L., {Mushotzky}, R.~F., {Yang}, Y., {Wang}, W.-H.,
  {Steffen}, A.~T., \& {Capak}, P. 2005, \aj, 129, 578

\bibitem[{{Barth} {et~al.}(2011){Barth}, {Pancoast}, {Thorman}, {Bennert},
  {Sand}, {Li}, {Canalizo}, {Filippenko}, {Gates}, {Greene}, {Malkan}, {Stern},
  {Treu}, {Woo}, {Assef}, {Bae}, {Brewer}, {Buehler}, {Cenko}, {Clubb},
  {Cooper}, {Diamond-Stanic}, {Hiner}, {H{\"o}nig}, {Joner}, {Kandrashoff},
  {Laney}, {Lazarova}, {Nierenberg}, {Park}, {Silverman}, {Son}, {Sonnenfeld},
  {Tollerud}, {Walsh}, {Walters}, {da Silva}, {Fumagalli}, {Gregg}, {Harris},
  {Hsiao}, {Lee}, {Lopez}, {Rex}, {Suzuki}, {Trump}, {Tytler}, {Worseck}, \&
  {Yesuf}}]{barth2011}
{Barth}, A.~J., {Pancoast}, A., {Thorman}, S.~J., {Bennert}, V.~N., {Sand},
  D.~J., {Li}, W., {Canalizo}, G., {Filippenko}, A.~V., {Gates}, E.~L.,
  {Greene}, J.~E., {Malkan}, M.~A., {Stern}, D., {Treu}, T., {Woo}, J.-H.,
  {Assef}, R.~J., {Bae}, H.-J., {Brewer}, B.~J., {Buehler}, T., {Cenko}, S.~B.,
  {Clubb}, K.~I., {Cooper}, M.~C., {Diamond-Stanic}, A.~M., {Hiner}, K.~D.,
  {H{\"o}nig}, S.~F., {Joner}, M.~D., {Kandrashoff}, M.~T., {Laney}, C.~D.,
  {Lazarova}, M.~S., {Nierenberg}, A.~M., {Park}, D., {Silverman}, J.~M.,
  {Son}, D., {Sonnenfeld}, A., {Tollerud}, E.~J., {Walsh}, J.~L., {Walters},
  R., {da Silva}, R.~L., {Fumagalli}, M., {Gregg}, M.~D., {Harris}, C.~E.,
  {Hsiao}, E.~Y., {Lee}, J., {Lopez}, L., {Rex}, J., {Suzuki}, N., {Trump},
  J.~R., {Tytler}, D., {Worseck}, G., \& {Yesuf}, H.~M. 2011, \apjl, 743, L4

\bibitem[{{Bell} {et~al.}(2004){Bell}, {Wolf}, {Meisenheimer}, {Rix}, {Borch},
  {Dye}, {Kleinheinrich}, {Wisotzki}, \& {McIntosh}}]{bell2004}
{Bell}, E.~F., {Wolf}, C., {Meisenheimer}, K., {Rix}, H.-W., {Borch}, A.,
  {Dye}, S., {Kleinheinrich}, M., {Wisotzki}, L., \& {McIntosh}, D.~H. 2004,
  \apj, 608, 752

\bibitem[{{Bentz} {et~al.}(2009){Bentz}, {Walsh}, {Barth}, {Baliber},
  {Bennert}, {Canalizo}, {Filippenko}, {Ganeshalingam}, {Gates}, {Greene},
  {Hidas}, {Hiner}, {Lee}, {Li}, {Malkan}, {Minezaki}, {Sakata}, {Serduke},
  {Silverman}, {Steele}, {Stern}, {Street}, {Thornton}, {Treu}, {Wang}, {Woo},
  \& {Yoshii}}]{bentz2009}
{Bentz}, M.~C., {Walsh}, J.~L., {Barth}, A.~J., {Baliber}, N., {Bennert},
  V.~N., {Canalizo}, G., {Filippenko}, A.~V., {Ganeshalingam}, M., {Gates},
  E.~L., {Greene}, J.~E., {Hidas}, M.~G., {Hiner}, K.~D., {Lee}, N., {Li}, W.,
  {Malkan}, M.~A., {Minezaki}, T., {Sakata}, Y., {Serduke}, F.~J.~D.,
  {Silverman}, J.~M., {Steele}, T.~N., {Stern}, D., {Street}, R.~A.,
  {Thornton}, C.~E., {Treu}, T., {Wang}, X., {Woo}, J.-H., \& {Yoshii}, Y.
  2009, \apj, 705, 199

\bibitem[{{Bertin} \& {Arnouts}(1996)}]{Bertin1996}
{Bertin}, E. \& {Arnouts}, S. 1996, \aaps, 117, 393

\bibitem[{{Best} {et~al.}(2005){Best}, {Kauffmann}, {Heckman}, {Brinchmann},
  {Charlot}, {Ivezi{\'c}}, \& {White}}]{best2005}
{Best}, P.~N., {Kauffmann}, G., {Heckman}, T.~M., {Brinchmann}, J., {Charlot},
  S., {Ivezi{\'c}}, {\v Z}., \& {White}, S.~D.~M. 2005, \mnras, 362, 25

\bibitem[{{Bluck} {et~al.}(2011){Bluck}, {Conselice}, {Almaini}, {Laird},
  {Nandra}, \& {Gr{\"u}tzbauch}}]{bluck2011}
{Bluck}, A.~F.~L., {Conselice}, C.~J., {Almaini}, O., {Laird}, E.~S., {Nandra},
  K., \& {Gr{\"u}tzbauch}, R. 2011, \mnras, 410, 1174

\bibitem[{{Brinchmann} {et~al.}(2004){Brinchmann}, {Charlot}, {White},
  {Tremonti}, {Kauffmann}, {Heckman}, \& {Brinkmann}}]{brinchmann2004}
{Brinchmann}, J., {Charlot}, S., {White}, S.~D.~M., {Tremonti}, C.,
  {Kauffmann}, G., {Heckman}, T., \& {Brinkmann}, J. 2004, \mnras, 351, 1151

\bibitem[{{Brusa} {et~al.}(2009){Brusa}, {Fiore}, {Santini}, {Grazian},
  {Comastri}, {Zamorani}, {Hasinger}, {Merloni}, {Civano}, {Fontana}, \&
  {Mainieri}}]{brusa2009}
{Brusa}, M., {Fiore}, F., {Santini}, P., {Grazian}, A., {Comastri}, A.,
  {Zamorani}, G., {Hasinger}, G., {Merloni}, A., {Civano}, F., {Fontana}, A.,
  \& {Mainieri}, V. 2009, \aap, 507, 1277

\bibitem[{{Bruzual} \& {Charlot}(2003)}]{bc2003}
{Bruzual}, G. \& {Charlot}, S. 2003, \mnras, 344, 1000

\bibitem[{{Bundy} {et~al.}(2008){Bundy}, {Georgakakis}, {Nandra}, {Ellis},
  {Conselice}, {Laird}, {Coil}, {Cooper}, {Faber}, {Newman}, {Pierce},
  {Primack}, \& {Yan}}]{bundy2008}
{Bundy}, K., {Georgakakis}, A., {Nandra}, K., {Ellis}, R.~S., {Conselice},
  C.~J., {Laird}, E., {Coil}, A., {Cooper}, M.~C., {Faber}, S.~M., {Newman},
  J.~A., {Pierce}, C.~M., {Primack}, J.~R., \& {Yan}, R. 2008, \apj, 681, 931

\bibitem[{{Calzetti} {et~al.}(2000){Calzetti}, {Armus}, {Bohlin}, {Kinney},
  {Koornneef}, \& {Storchi-Bergmann}}]{calzetti2000}
{Calzetti}, D., {Armus}, L., {Bohlin}, R.~C., {Kinney}, A.~L., {Koornneef}, J.,
  \& {Storchi-Bergmann}, T. 2000, \apj, 533, 682

\bibitem[{{Cardamone} {et~al.}(2010){Cardamone}, {Urry}, {Schawinski},
  {Treister}, {Brammer}, \& {Gawiser}}]{cardamone2010}
{Cardamone}, C.~N., {Urry}, C.~M., {Schawinski}, K., {Treister}, E., {Brammer},
  G., \& {Gawiser}, E. 2010, \apjl, 721, L38

\bibitem[{{Cardelli} {et~al.}(1989){Cardelli}, {Clayton}, \&
  {Mathis}}]{cardelli1989}
{Cardelli}, J.~A., {Clayton}, G.~C., \& {Mathis}, J.~S. 1989, \apj, 345, 245

\bibitem[{{Cassata} {et~al.}(2008){Cassata}, {Cimatti}, {Kurk}, {Rodighiero},
  {Pozzetti}, {Bolzonella}, {Daddi}, {Mignoli}, {Berta}, {Dickinson},
  {Franceschini}, {Halliday}, {Renzini}, {Rosati}, \& {Zamorani}}]{cassata2008}
{Cassata}, P., {Cimatti}, A., {Kurk}, J., {Rodighiero}, G., {Pozzetti}, L.,
  {Bolzonella}, M., {Daddi}, E., {Mignoli}, M., {Berta}, S., {Dickinson}, M.,
  {Franceschini}, A., {Halliday}, C., {Renzini}, A., {Rosati}, P., \&
  {Zamorani}, G. 2008, \aap, 483, L39

\bibitem[{{Chabrier}(2003)}]{chabrier2003}
{Chabrier}, G. 2003, \pasp, 115, 763

\bibitem[{{Coil} {et~al.}(2009){Coil}, {Georgakakis}, {Newman}, {Cooper},
  {Croton}, {Davis}, {Koo}, {Laird}, {Nandra}, {Weiner}, {Willmer}, \&
  {Yan}}]{coil2009}
{Coil}, A.~L., {Georgakakis}, A., {Newman}, J.~A., {Cooper}, M.~C., {Croton},
  D., {Davis}, M., {Koo}, D.~C., {Laird}, E.~S., {Nandra}, K., {Weiner}, B.~J.,
  {Willmer}, C.~N.~A., \& {Yan}, R. 2009, \apj, 701, 1484

\bibitem[{{Croton} {et~al.}(2006){Croton}, {Springel}, {White}, {De Lucia},
  {Frenk}, {Gao}, {Jenkins}, {Kauffmann}, {Navarro}, \& {Yoshida}}]{croton2006}
{Croton}, D.~J., {Springel}, V., {White}, S.~D.~M., {De Lucia}, G., {Frenk},
  C.~S., {Gao}, L., {Jenkins}, A., {Kauffmann}, G., {Navarro}, J.~F., \&
  {Yoshida}, N. 2006, \mnras, 365, 11

\bibitem[{{Daddi} {et~al.}(2007){Daddi}, {Dickinson}, {Morrison}, {Chary},
  {Cimatti}, {Elbaz}, {Frayer}, {Renzini}, {Pope}, {Alexander}, {Bauer},
  {Giavalisco}, {Huynh}, {Kurk}, \& {Mignoli}}]{daddi2007}
{Daddi}, E., {Dickinson}, M., {Morrison}, G., {Chary}, R., {Cimatti}, A.,
  {Elbaz}, D., {Frayer}, D., {Renzini}, A., {Pope}, A., {Alexander}, D.~M.,
  {Bauer}, F.~E., {Giavalisco}, M., {Huynh}, M., {Kurk}, J., \& {Mignoli}, M.
  2007, \apj, 670, 156

\bibitem[{{Donley} {et~al.}(2012){Donley}, {Koekemoer}, {Brusa}, {Capak},
  {Cardamone}, {Civano}, {Ilbert}, {Impey}, {Kartaltepe}, {Miyaji}, {Salvato},
  {Sanders}, {Trump}, \& {Zamorani}}]{donley2012}
{Donley}, J.~L., {Koekemoer}, A.~M., {Brusa}, M., {Capak}, P., {Cardamone},
  C.~N., {Civano}, F., {Ilbert}, O., {Impey}, C.~D., {Kartaltepe}, J.~S.,
  {Miyaji}, T., {Salvato}, M., {Sanders}, D.~B., {Trump}, J.~R., \& {Zamorani},
  G. 2012, ArXiv e-prints

\bibitem[{{Elbaz} {et~al.}(2007){Elbaz}, {Daddi}, {Le Borgne}, {Dickinson},
  {Alexander}, {Chary}, {Starck}, {Brandt}, {Kitzbichler}, {MacDonald},
  {Nonino}, {Popesso}, {Stern}, \& {Vanzella}}]{elbaz2007}
{Elbaz}, D., {Daddi}, E., {Le Borgne}, D., {Dickinson}, M., {Alexander}, D.~M.,
  {Chary}, R.-R., {Starck}, J.-L., {Brandt}, W.~N., {Kitzbichler}, M.,
  {MacDonald}, E., {Nonino}, M., {Popesso}, P., {Stern}, D., \& {Vanzella}, E.
  2007, \aap, 468, 33

\bibitem[{{Elvis} {et~al.}(1994){Elvis}, {Wilkes}, {McDowell}, {Green},
  {Bechtold}, {Willner}, {Oey}, {Polomski}, \& {Cutri}}]{elvis1994}
{Elvis}, M., {Wilkes}, B.~J., {McDowell}, J.~C., {Green}, R.~F., {Bechtold},
  J., {Willner}, S.~P., {Oey}, M.~S., {Polomski}, E., \& {Cutri}, R. 1994,
  \apjs, 95, 1

\bibitem[{{Eminian} {et~al.}(2008){Eminian}, {Kauffmann}, {Charlot}, {Wild},
  {Bruzual}, {Rettura}, \& {Loveday}}]{eminian2008}
{Eminian}, C., {Kauffmann}, G., {Charlot}, S., {Wild}, V., {Bruzual}, G.,
  {Rettura}, A., \& {Loveday}, J. 2008, \mnras, 384, 930

\bibitem[{{Erb} {et~al.}(2006{\natexlab{a}}){Erb}, {Shapley}, {Pettini},
  {Steidel}, {Reddy}, \& {Adelberger}}]{erb2006a}
{Erb}, D.~K., {Shapley}, A.~E., {Pettini}, M., {Steidel}, C.~C., {Reddy},
  N.~A., \& {Adelberger}, K.~L. 2006{\natexlab{a}}, \apj, 644, 813

\bibitem[{{Erb} {et~al.}(2006{\natexlab{b}}){Erb}, {Steidel}, {Shapley},
  {Pettini}, {Reddy}, \& {Adelberger}}]{erb2006b}
{Erb}, D.~K., {Steidel}, C.~C., {Shapley}, A.~E., {Pettini}, M., {Reddy},
  N.~A., \& {Adelberger}, K.~L. 2006{\natexlab{b}}, \apj, 646, 107

\bibitem[{{Fadda} {et~al.}(2004){Fadda}, {Jannuzi}, {Ford}, \&
  {Storrie-Lombardi}}]{fadda2004}
{Fadda}, D., {Jannuzi}, B.~T., {Ford}, A., \& {Storrie-Lombardi}, L.~J. 2004,
  \aj, 128, 1

\bibitem[{{Fazio} {et~al.}(2004){Fazio}, {Hora}, {Allen}, {Ashby}, {Barmby},
  {Deutsch}, {Huang}, {Kleiner}, {Marengo}, {Megeath}, {Melnick}, {Pahre},
  {Patten}, {Polizotti}, {Smith}, {Taylor}, {Wang}, {Willner}, {Hoffmann},
  {Pipher}, {Forrest}, {McMurty}, {McCreight}, {McKelvey}, {McMurray}, {Koch},
  {Moseley}, {Arendt}, {Mentzell}, {Marx}, {Losch}, {Mayman}, {Eichhorn},
  {Krebs}, {Jhabvala}, {Gezari}, {Fixsen}, {Flores}, {Shakoorzadeh}, {Jungo},
  {Hakun}, {Workman}, {Karpati}, {Kichak}, {Whitley}, {Mann}, {Tollestrup},
  {Eisenhardt}, {Stern}, {Gorjian}, {Bhattacharya}, {Carey}, {Nelson},
  {Glaccum}, {Lacy}, {Lowrance}, {Laine}, {Reach}, {Stauffer}, {Surace},
  {Wilson}, {Wright}, {Hoffman}, {Domingo}, \& {Cohen}}]{fazio2004}
{Fazio}, G.~G., {Hora}, J.~L., {Allen}, L.~E., {Ashby}, M.~L.~N., {Barmby}, P.,
  {Deutsch}, L.~K., {Huang}, J.-S., {Kleiner}, S., {Marengo}, M., {Megeath},
  S.~T., {Melnick}, G.~J., {Pahre}, M.~A., {Patten}, B.~M., {Polizotti}, J.,
  {Smith}, H.~A., {Taylor}, R.~S., {Wang}, Z., {Willner}, S.~P., {Hoffmann},
  W.~F., {Pipher}, J.~L., {Forrest}, W.~J., {McMurty}, C.~W., {McCreight},
  C.~R., {McKelvey}, M.~E., {McMurray}, R.~E., {Koch}, D.~G., {Moseley}, S.~H.,
  {Arendt}, R.~G., {Mentzell}, J.~E., {Marx}, C.~T., {Losch}, P., {Mayman}, P.,
  {Eichhorn}, W., {Krebs}, D., {Jhabvala}, M., {Gezari}, D.~Y., {Fixsen},
  D.~J., {Flores}, J., {Shakoorzadeh}, K., {Jungo}, R., {Hakun}, C., {Workman},
  L., {Karpati}, G., {Kichak}, R., {Whitley}, R., {Mann}, S., {Tollestrup},
  E.~V., {Eisenhardt}, P., {Stern}, D., {Gorjian}, V., {Bhattacharya}, B.,
  {Carey}, S., {Nelson}, B.~O., {Glaccum}, W.~J., {Lacy}, M., {Lowrance},
  P.~J., {Laine}, S., {Reach}, W.~T., {Stauffer}, J.~A., {Surace}, J.~A.,
  {Wilson}, G., {Wright}, E.~L., {Hoffman}, A., {Domingo}, G., \& {Cohen}, M.
  2004, \apjs, 154, 10

\bibitem[{{Ferrarese} \& {Merritt}(2000)}]{ferrarese2000}
{Ferrarese}, L. \& {Merritt}, D. 2000, \apjl, 539, L9

\bibitem[{{Gebhardt} {et~al.}(2000){Gebhardt}, {Bender}, {Bower}, {Dressler},
  {Faber}, {Filippenko}, {Green}, {Grillmair}, {Ho}, {Kormendy}, {Lauer},
  {Magorrian}, {Pinkney}, {Richstone}, \& {Tremaine}}]{gebhardt2000}
{Gebhardt}, K., {Bender}, R., {Bower}, G., {Dressler}, A., {Faber}, S.~M.,
  {Filippenko}, A.~V., {Green}, R., {Grillmair}, C., {Ho}, L.~C., {Kormendy},
  J., {Lauer}, T.~R., {Magorrian}, J., {Pinkney}, J., {Richstone}, D., \&
  {Tremaine}, S. 2000, \apjl, 539, L13

\bibitem[{{Gordon} \& {Clayton}(1998)}]{gordon1998}
{Gordon}, K.~D. \& {Clayton}, G.~C. 1998, \apj, 500, 816

\bibitem[{{Grogin} {et~al.}(2011){Grogin}, {Kocevski}, {Faber}, {Ferguson},
  {Koekemoer}, {Riess}, {Acquaviva}, {Alexander}, {Almaini}, {Ashby}, {Barden},
  {Bell}, {Bournaud}, {Brown}, {Caputi}, {Casertano}, {Cassata}, {Castellano},
  {Challis}, {Chary}, {Cheung}, {Cirasuolo}, {Conselice}, {Roshan Cooray},
  {Croton}, {Daddi}, {Dahlen}, {Dav{\'e}}, {de Mello}, {Dekel}, {Dickinson},
  {Dolch}, {Donley}, {Dunlop}, {Dutton}, {Elbaz}, {Fazio}, {Filippenko},
  {Finkelstein}, {Fontana}, {Gardner}, {Garnavich}, {Gawiser}, {Giavalisco},
  {Grazian}, {Guo}, {Hathi}, {H{\"a}ussler}, {Hopkins}, {Huang}, {Huang},
  {Jha}, {Kartaltepe}, {Kirshner}, {Koo}, {Lai}, {Lee}, {Li}, {Lotz}, {Lucas},
  {Madau}, {McCarthy}, {McGrath}, {McIntosh}, {McLure}, {Mobasher},
  {Moustakas}, {Mozena}, {Nandra}, {Newman}, {Niemi}, {Noeske}, {Papovich},
  {Pentericci}, {Pope}, {Primack}, {Rajan}, {Ravindranath}, {Reddy}, {Renzini},
  {Rix}, {Robaina}, {Rodney}, {Rosario}, {Rosati}, {Salimbeni}, {Scarlata},
  {Siana}, {Simard}, {Smidt}, {Somerville}, {Spinrad}, {Straughn}, {Strolger},
  {Telford}, {Teplitz}, {Trump}, {van der Wel}, {Villforth}, {Wechsler},
  {Weiner}, {Wiklind}, {Wild}, {Wilson}, {Wuyts}, {Yan}, \& {Yun}}]{grogin2011}
{Grogin}, N.~A., {Kocevski}, D.~D., {Faber}, S.~M., {Ferguson}, H.~C.,
  {Koekemoer}, A.~M., {Riess}, A.~G., {Acquaviva}, V., {Alexander}, D.~M.,
  {Almaini}, O., {Ashby}, M.~L.~N., {Barden}, M., {Bell}, E.~F., {Bournaud},
  F., {Brown}, T.~M., {Caputi}, K.~I., {Casertano}, S., {Cassata}, P.,
  {Castellano}, M., {Challis}, P., {Chary}, R.-R., {Cheung}, E., {Cirasuolo},
  M., {Conselice}, C.~J., {Roshan Cooray}, A., {Croton}, D.~J., {Daddi}, E.,
  {Dahlen}, T., {Dav{\'e}}, R., {de Mello}, D.~F., {Dekel}, A., {Dickinson},
  M., {Dolch}, T., {Donley}, J.~L., {Dunlop}, J.~S., {Dutton}, A.~A., {Elbaz},
  D., {Fazio}, G.~G., {Filippenko}, A.~V., {Finkelstein}, S.~L., {Fontana}, A.,
  {Gardner}, J.~P., {Garnavich}, P.~M., {Gawiser}, E., {Giavalisco}, M.,
  {Grazian}, A., {Guo}, Y., {Hathi}, N.~P., {H{\"a}ussler}, B., {Hopkins},
  P.~F., {Huang}, J.-S., {Huang}, K.-H., {Jha}, S.~W., {Kartaltepe}, J.~S.,
  {Kirshner}, R.~P., {Koo}, D.~C., {Lai}, K., {Lee}, K.-S., {Li}, W., {Lotz},
  J.~M., {Lucas}, R.~A., {Madau}, P., {McCarthy}, P.~J., {McGrath}, E.~J.,
  {McIntosh}, D.~H., {McLure}, R.~J., {Mobasher}, B., {Moustakas}, L.~A.,
  {Mozena}, M., {Nandra}, K., {Newman}, J.~A., {Niemi}, S.-M., {Noeske}, K.~G.,
  {Papovich}, C.~J., {Pentericci}, L., {Pope}, A., {Primack}, J.~R., {Rajan},
  A., {Ravindranath}, S., {Reddy}, N.~A., {Renzini}, A., {Rix}, H.-W.,
  {Robaina}, A.~R., {Rodney}, S.~A., {Rosario}, D.~J., {Rosati}, P.,
  {Salimbeni}, S., {Scarlata}, C., {Siana}, B., {Simard}, L., {Smidt}, J.,
  {Somerville}, R.~S., {Spinrad}, H., {Straughn}, A.~N., {Strolger}, L.-G.,
  {Telford}, O., {Teplitz}, H.~I., {Trump}, J.~R., {van der Wel}, A.,
  {Villforth}, C., {Wechsler}, R.~H., {Weiner}, B.~J., {Wiklind}, T., {Wild},
  V., {Wilson}, G., {Wuyts}, S., {Yan}, H.-J., \& {Yun}, M.~S. 2011, \apjs,
  197, 35

\bibitem[{{G{\"u}ltekin} {et~al.}(2009){G{\"u}ltekin}, {Richstone}, {Gebhardt},
  {Lauer}, {Tremaine}, {Aller}, {Bender}, {Dressler}, {Faber}, {Filippenko},
  {Green}, {Ho}, {Kormendy}, {Magorrian}, {Pinkney}, \&
  {Siopis}}]{gultekin2009}
{G{\"u}ltekin}, K., {Richstone}, D.~O., {Gebhardt}, K., {Lauer}, T.~R.,
  {Tremaine}, S., {Aller}, M.~C., {Bender}, R., {Dressler}, A., {Faber}, S.~M.,
  {Filippenko}, A.~V., {Green}, R., {Ho}, L.~C., {Kormendy}, J., {Magorrian},
  J., {Pinkney}, J., \& {Siopis}, C. 2009, \apj, 698, 198

\bibitem[{{Hainline} {et~al.}(2011{\natexlab{a}}){Hainline}, {Shapley},
  {Greene}, \& {Steidel}}]{hainline2011}
{Hainline}, K.~N., {Shapley}, A.~E., {Greene}, J.~E., \& {Steidel}, C.~C.
  2011{\natexlab{a}}, \apj, 733, 31

\bibitem[{{Hainline} {et~al.}(2011{\natexlab{b}}){Hainline}, {Blain}, {Smail},
  {Alexander}, {Armus}, {Chapman}, \& {Ivison}}]{hainlinel2011}
{Hainline}, L.~J., {Blain}, A.~W., {Smail}, I., {Alexander}, D.~M., {Armus},
  L., {Chapman}, S.~C., \& {Ivison}, R.~J. 2011{\natexlab{b}}, \apj, 740, 96

\bibitem[{{H{\"a}ring} \& {Rix}(2004)}]{haring2004}
{H{\"a}ring}, N. \& {Rix}, H. 2004, \apjl, 604, L89

\bibitem[{{Hickox} {et~al.}(2007){Hickox}, {Jones}, {Forman}, {Murray},
  {Brodwin}, {Brown}, {Eisenhardt}, {Stern}, {Kochanek}, {Eisenstein}, {Cool},
  {Jannuzi}, {Dey}, {Brand}, {Gorjian}, \& {Caldwell}}]{hickox2007}
{Hickox}, R.~C., {Jones}, C., {Forman}, W.~R., {Murray}, S.~S., {Brodwin}, M.,
  {Brown}, M.~J.~I., {Eisenhardt}, P.~R., {Stern}, D., {Kochanek}, C.~S.,
  {Eisenstein}, D., {Cool}, R.~J., {Jannuzi}, B.~T., {Dey}, A., {Brand}, K.,
  {Gorjian}, V., \& {Caldwell}, N. 2007, \apj, 671, 1365

\bibitem[{{Hickox} {et~al.}(2009){Hickox}, {Jones}, {Forman}, {Murray},
  {Kochanek}, {Eisenstein}, {Jannuzi}, {Dey}, {Brown}, {Stern}, {Eisenhardt},
  {Gorjian}, {Brodwin}, {Narayan}, {Cool}, {Kenter}, {Caldwell}, \&
  {Anderson}}]{hickox2009}
{Hickox}, R.~C., {Jones}, C., {Forman}, W.~R., {Murray}, S.~S., {Kochanek},
  C.~S., {Eisenstein}, D., {Jannuzi}, B.~T., {Dey}, A., {Brown}, M.~J.~I.,
  {Stern}, D., {Eisenhardt}, P.~R., {Gorjian}, V., {Brodwin}, M., {Narayan},
  R., {Cool}, R.~J., {Kenter}, A., {Caldwell}, N., \& {Anderson}, M.~E. 2009,
  \apj, 696, 891

\bibitem[{{Hopkins} {et~al.}(2006){Hopkins}, {Hernquist}, {Cox}, {Di Matteo},
  {Robertson}, \& {Springel}}]{hopkins2006}
{Hopkins}, P.~F., {Hernquist}, L., {Cox}, T.~J., {Di Matteo}, T., {Robertson},
  B., \& {Springel}, V. 2006, \apjs, 163, 1

\bibitem[{{Hopkins} {et~al.}(2008){Hopkins}, {Hernquist}, {Cox}, \& {Kere{\v
  s}}}]{hopkins2008}
{Hopkins}, P.~F., {Hernquist}, L., {Cox}, T.~J., \& {Kere{\v s}}, D. 2008,
  \apjs, 175, 356

\bibitem[{{Hopkins} {et~al.}(2007){Hopkins}, {Richards}, \&
  {Hernquist}}]{hopkins2007}
{Hopkins}, P.~F., {Richards}, G.~T., \& {Hernquist}, L. 2007, \apj, 654, 731

\bibitem[{{Kauffmann} {et~al.}(2003){Kauffmann}, {Heckman}, {Tremonti},
  {Brinchmann}, {Charlot}, {White}, {Ridgway}, {Brinkmann}, {Fukugita}, {Hall},
  {Ivezi{\' c}}, {Richards}, \& {Schneider}}]{kauffmann2003b}
{Kauffmann}, G., {Heckman}, T.~M., {Tremonti}, C., {Brinchmann}, J., {Charlot},
  S., {White}, S.~D.~M., {Ridgway}, S.~E., {Brinkmann}, J., {Fukugita}, M.,
  {Hall}, P.~B., {Ivezi{\' c}}, {\v Z}., {Richards}, G.~T., \& {Schneider},
  D.~P. 2003, \mnras, 346, 1055

\bibitem[{{Khalatyan} {et~al.}(2008){Khalatyan}, {Cattaneo}, {Schramm},
  {Gottl{\"o}ber}, {Steinmetz}, \& {Wisotzki}}]{khalatyan2008}
{Khalatyan}, A., {Cattaneo}, A., {Schramm}, M., {Gottl{\"o}ber}, S.,
  {Steinmetz}, M., \& {Wisotzki}, L. 2008, \mnras, 387, 13

\bibitem[{{Kocevski} {et~al.}(2012){Kocevski}, {Faber}, {Mozena}, {Koekemoer},
  {Nandra}, {Rangel}, {Laird}, {Brusa}, {Wuyts}, {Trump}, {Koo}, {Somerville},
  {Bell}, {Lotz}, {Alexander}, {Bournaud}, {Conselice}, {Dahlen}, {Dekel},
  {Donley}, {Dunlop}, {Finoguenov}, {Georgakakis}, {Giavalisco}, {Guo},
  {Grogin}, {Hathi}, {Juneau}, {Kartaltepe}, {Lucas}, {McGrath}, {McIntosh},
  {Mobasher}, {Robaina}, {Rosario}, {Straughn}, {van der Wel}, \&
  {Villforth}}]{kocevski2012}
{Kocevski}, D.~D., {Faber}, S.~M., {Mozena}, M., {Koekemoer}, A.~M., {Nandra},
  K., {Rangel}, C., {Laird}, E.~S., {Brusa}, M., {Wuyts}, S., {Trump}, J.~R.,
  {Koo}, D.~C., {Somerville}, R.~S., {Bell}, E.~F., {Lotz}, J.~M., {Alexander},
  D.~M., {Bournaud}, F., {Conselice}, C.~J., {Dahlen}, T., {Dekel}, A.,
  {Donley}, J.~L., {Dunlop}, J.~S., {Finoguenov}, A., {Georgakakis}, A.,
  {Giavalisco}, M., {Guo}, Y., {Grogin}, N.~A., {Hathi}, N.~P., {Juneau}, S.,
  {Kartaltepe}, J.~S., {Lucas}, R.~A., {McGrath}, E.~J., {McIntosh}, D.~H.,
  {Mobasher}, B., {Robaina}, A.~R., {Rosario}, D., {Straughn}, A.~N., {van der
  Wel}, A., \& {Villforth}, C. 2012, \apj, 744, 148

\bibitem[{{Koekemoer} {et~al.}(2011){Koekemoer}, {Faber}, {Ferguson}, {Grogin},
  {Kocevski}, {Koo}, {Lai}, {Lotz}, {Lucas}, {McGrath}, {Ogaz}, {Rajan},
  {Riess}, {Rodney}, {Strolger}, {Casertano}, {Castellano}, {Dahlen},
  {Dickinson}, {Dolch}, {Fontana}, {Giavalisco}, {Grazian}, {Guo}, {Hathi},
  {Huang}, {van der Wel}, {Yan}, {Acquaviva}, {Alexander}, {Almaini}, {Ashby},
  {Barden}, {Bell}, {Bournaud}, {Brown}, {Caputi}, {Cassata}, {Challis},
  {Chary}, {Cheung}, {Cirasuolo}, {Conselice}, {Roshan Cooray}, {Croton},
  {Daddi}, {Dav{\'e}}, {de Mello}, {de Ravel}, {Dekel}, {Donley}, {Dunlop},
  {Dutton}, {Elbaz}, {Fazio}, {Filippenko}, {Finkelstein}, {Frazer}, {Gardner},
  {Garnavich}, {Gawiser}, {Gruetzbauch}, {Hartley}, {H{\"a}ussler},
  {Herrington}, {Hopkins}, {Huang}, {Jha}, {Johnson}, {Kartaltepe},
  {Khostovan}, {Kirshner}, {Lani}, {Lee}, {Li}, {Madau}, {McCarthy},
  {McIntosh}, {McLure}, {McPartland}, {Mobasher}, {Moreira}, {Mortlock},
  {Moustakas}, {Mozena}, {Nandra}, {Newman}, {Nielsen}, {Niemi}, {Noeske},
  {Papovich}, {Pentericci}, {Pope}, {Primack}, {Ravindranath}, {Reddy},
  {Renzini}, {Rix}, {Robaina}, {Rosario}, {Rosati}, {Salimbeni}, {Scarlata},
  {Siana}, {Simard}, {Smidt}, {Snyder}, {Somerville}, {Spinrad}, {Straughn},
  {Telford}, {Teplitz}, {Trump}, {Vargas}, {Villforth}, {Wagner}, {Wandro},
  {Wechsler}, {Weiner}, {Wiklind}, {Wild}, {Wilson}, {Wuyts}, \&
  {Yun}}]{koekemoer2011}
{Koekemoer}, A.~M., {Faber}, S.~M., {Ferguson}, H.~C., {Grogin}, N.~A.,
  {Kocevski}, D.~D., {Koo}, D.~C., {Lai}, K., {Lotz}, J.~M., {Lucas}, R.~A.,
  {McGrath}, E.~J., {Ogaz}, S., {Rajan}, A., {Riess}, A.~G., {Rodney}, S.~A.,
  {Strolger}, L., {Casertano}, S., {Castellano}, M., {Dahlen}, T., {Dickinson},
  M., {Dolch}, T., {Fontana}, A., {Giavalisco}, M., {Grazian}, A., {Guo}, Y.,
  {Hathi}, N.~P., {Huang}, K.-H., {van der Wel}, A., {Yan}, H.-J., {Acquaviva},
  V., {Alexander}, D.~M., {Almaini}, O., {Ashby}, M.~L.~N., {Barden}, M.,
  {Bell}, E.~F., {Bournaud}, F., {Brown}, T.~M., {Caputi}, K.~I., {Cassata},
  P., {Challis}, P.~J., {Chary}, R.-R., {Cheung}, E., {Cirasuolo}, M.,
  {Conselice}, C.~J., {Roshan Cooray}, A., {Croton}, D.~J., {Daddi}, E.,
  {Dav{\'e}}, R., {de Mello}, D.~F., {de Ravel}, L., {Dekel}, A., {Donley},
  J.~L., {Dunlop}, J.~S., {Dutton}, A.~A., {Elbaz}, D., {Fazio}, G.~G.,
  {Filippenko}, A.~V., {Finkelstein}, S.~L., {Frazer}, C., {Gardner}, J.~P.,
  {Garnavich}, P.~M., {Gawiser}, E., {Gruetzbauch}, R., {Hartley}, W.~G.,
  {H{\"a}ussler}, B., {Herrington}, J., {Hopkins}, P.~F., {Huang}, J.-S.,
  {Jha}, S.~W., {Johnson}, A., {Kartaltepe}, J.~S., {Khostovan}, A.~A.,
  {Kirshner}, R.~P., {Lani}, C., {Lee}, K.-S., {Li}, W., {Madau}, P.,
  {McCarthy}, P.~J., {McIntosh}, D.~H., {McLure}, R.~J., {McPartland}, C.,
  {Mobasher}, B., {Moreira}, H., {Mortlock}, A., {Moustakas}, L.~A., {Mozena},
  M., {Nandra}, K., {Newman}, J.~A., {Nielsen}, J.~L., {Niemi}, S., {Noeske},
  K.~G., {Papovich}, C.~J., {Pentericci}, L., {Pope}, A., {Primack}, J.~R.,
  {Ravindranath}, S., {Reddy}, N.~A., {Renzini}, A., {Rix}, H.-W., {Robaina},
  A.~R., {Rosario}, D.~J., {Rosati}, P., {Salimbeni}, S., {Scarlata}, C.,
  {Siana}, B., {Simard}, L., {Smidt}, J., {Snyder}, D., {Somerville}, R.~S.,
  {Spinrad}, H., {Straughn}, A.~N., {Telford}, O., {Teplitz}, H.~I., {Trump},
  J.~R., {Vargas}, C., {Villforth}, C., {Wagner}, C.~R., {Wandro}, P.,
  {Wechsler}, R.~H., {Weiner}, B.~J., {Wiklind}, T., {Wild}, V., {Wilson}, G.,
  {Wuyts}, S., \& {Yun}, M.~S. 2011, \apjs, 197, 36

\bibitem[{{Kriek} {et~al.}(2008){Kriek}, {van der Wel}, {van Dokkum}, {Franx},
  \& {Illingworth}}]{kriek2008}
{Kriek}, M., {van der Wel}, A., {van Dokkum}, P.~G., {Franx}, M., \&
  {Illingworth}, G.~D. 2008, \apj, 682, 896

\bibitem[{{Kriek} {et~al.}(2007){Kriek}, {van Dokkum}, {Franx}, {Illingworth},
  {Coppi}, {F{\"o}rster Schreiber}, {Gawiser}, {Labb{\'e}}, {Lira},
  {Marchesini}, {Quadri}, {Rudnick}, {Taylor}, {Urry}, \& {van der
  Werf}}]{kriek2007}
{Kriek}, M., {van Dokkum}, P.~G., {Franx}, M., {Illingworth}, G.~D., {Coppi},
  P., {F{\"o}rster Schreiber}, N.~M., {Gawiser}, E., {Labb{\'e}}, I., {Lira},
  P., {Marchesini}, D., {Quadri}, R., {Rudnick}, G., {Taylor}, E.~N., {Urry},
  C.~M., \& {van der Werf}, P.~P. 2007, \apj, 669, 776

\bibitem[{{Labb{\'e}} {et~al.}(2007){Labb{\'e}}, {Franx}, {Rudnick},
  {Schreiber}, {van Dokkum}, {Moorwood}, {Rix}, {R{\"o}ttgering}, {Trujillo},
  \& {van der Werf}}]{labbe2007}
{Labb{\'e}}, I., {Franx}, M., {Rudnick}, G., {Schreiber}, N.~M.~F., {van
  Dokkum}, P.~G., {Moorwood}, A., {Rix}, H.-W., {R{\"o}ttgering}, H.,
  {Trujillo}, I., \& {van der Werf}, P. 2007, \apj, 665, 944

\bibitem[{{Lacy} {et~al.}(2004){Lacy}, {Storrie-Lombardi}, {Sajina},
  {Appleton}, {Armus}, {Chapman}, {Choi}, {Fadda}, {Fang}, {Frayer},
  {Heinrichsen}, {Helou}, {Im}, {Marleau}, {Masci}, {Shupe}, {Soifer},
  {Surace}, {Teplitz}, {Wilson}, \& {Yan}}]{lacy2004}
{Lacy}, M., {Storrie-Lombardi}, L.~J., {Sajina}, A., {Appleton}, P.~N.,
  {Armus}, L., {Chapman}, S.~C., {Choi}, P.~I., {Fadda}, D., {Fang}, F.,
  {Frayer}, D.~T., {Heinrichsen}, I., {Helou}, G., {Im}, M., {Marleau}, F.~R.,
  {Masci}, F., {Shupe}, D.~L., {Soifer}, B.~T., {Surace}, J., {Teplitz}, H.~I.,
  {Wilson}, G., \& {Yan}, L. 2004, \apjs, 154, 166

\bibitem[{{Lacy} {et~al.}(2005){Lacy}, {Wilson}, {Masci}, {Storrie-Lombardi},
  {Appleton}, {Armus}, {Chapman}, {Choi}, {Fadda}, {Fang}, {Frayer},
  {Heinrichsen}, {Helou}, {Im}, {Laine}, {Marleau}, {Shupe}, {Soifer},
  {Squires}, {Surace}, {Teplitz}, \& {Yan}}]{lacy2005}
{Lacy}, M., {Wilson}, G., {Masci}, F., {Storrie-Lombardi}, L.~J., {Appleton},
  P.~N., {Armus}, L., {Chapman}, S.~C., {Choi}, P.~I., {Fadda}, D., {Fang}, F.,
  {Frayer}, D.~T., {Heinrichsen}, I., {Helou}, G., {Im}, M., {Laine}, S.,
  {Marleau}, F.~R., {Shupe}, D.~L., {Soifer}, B.~T., {Squires}, G.~K.,
  {Surace}, J., {Teplitz}, H.~I., \& {Yan}, L. 2005, \apjs, 161, 41

\bibitem[{{Lauer} {et~al.}(2007){Lauer}, {Faber}, {Richstone}, {Gebhardt},
  {Tremaine}, {Postman}, {Dressler}, {Aller}, {Filippenko}, {Green}, {Ho},
  {Kormendy}, {Magorrian}, \& {Pinkney}}]{lauer2007}
{Lauer}, T.~R., {Faber}, S.~M., {Richstone}, D., {Gebhardt}, K., {Tremaine},
  S., {Postman}, M., {Dressler}, A., {Aller}, M.~C., {Filippenko}, A.~V.,
  {Green}, R., {Ho}, L.~C., {Kormendy}, J., {Magorrian}, J., \& {Pinkney}, J.
  2007, \apj, 662, 808

\bibitem[{{Law} {et~al.}(2007){Law}, {Steidel}, {Erb}, {Pettini}, {Reddy},
  {Shapley}, {Adelberger}, \& {Simenc}}]{law2007}
{Law}, D.~R., {Steidel}, C.~C., {Erb}, D.~K., {Pettini}, M., {Reddy}, N.~A.,
  {Shapley}, A.~E., {Adelberger}, K.~L., \& {Simenc}, D.~J. 2007, \apj, 656, 1

\bibitem[{{Madau}(1995)}]{madau1995}
{Madau}, P. 1995, \apj, 441, 18

\bibitem[{{Mainieri} {et~al.}(2011){Mainieri}, {Bongiorno}, {Merloni}, {Aller},
  {Carollo}, {Iwasawa}, {Koekemoer}, {Mignoli}, {Silverman}, {Bolzonella},
  {Brusa}, {Comastri}, {Gilli}, {Halliday}, {Ilbert}, {Lusso}, {Salvato},
  {Vignali}, {Zamorani}, {Contini}, {Kneib}, {Le F{\`e}vre}, {Lilly},
  {Renzini}, {Scodeggio}, {Balestra}, {Bardelli}, {Caputi}, {Coppa},
  {Cucciati}, {de la Torre}, {de Ravel}, {Franzetti}, {Garilli}, {Iovino},
  {Kampczyk}, {Knobel}, {Kova{\v c}}, {Lamareille}, {Le Borgne}, {Le Brun},
  {Maier}, {Nair}, {Pello}, {Peng}, {Perez Montero}, {Pozzetti},
  {Ricciardelli}, {Tanaka}, {Tasca}, {Tresse}, {Vergani}, {Zucca}, {Aussel},
  {Capak}, {Cappelluti}, {Elvis}, {Fiore}, {Hasinger}, {Impey}, {Le Floc'h},
  {Scoville}, {Taniguchi}, \& {Trump}}]{mainieri2011}
{Mainieri}, V., {Bongiorno}, A., {Merloni}, A., {Aller}, M., {Carollo}, M.,
  {Iwasawa}, K., {Koekemoer}, A.~M., {Mignoli}, M., {Silverman}, J.~D.,
  {Bolzonella}, M., {Brusa}, M., {Comastri}, A., {Gilli}, R., {Halliday}, C.,
  {Ilbert}, O., {Lusso}, E., {Salvato}, M., {Vignali}, C., {Zamorani}, G.,
  {Contini}, T., {Kneib}, J.-P., {Le F{\`e}vre}, O., {Lilly}, S., {Renzini},
  A., {Scodeggio}, M., {Balestra}, I., {Bardelli}, S., {Caputi}, K., {Coppa},
  G., {Cucciati}, O., {de la Torre}, S., {de Ravel}, L., {Franzetti}, P.,
  {Garilli}, B., {Iovino}, A., {Kampczyk}, P., {Knobel}, C., {Kova{\v c}}, K.,
  {Lamareille}, F., {Le Borgne}, J.-F., {Le Brun}, V., {Maier}, C., {Nair}, P.,
  {Pello}, R., {Peng}, Y., {Perez Montero}, E., {Pozzetti}, L., {Ricciardelli},
  E., {Tanaka}, M., {Tasca}, L., {Tresse}, L., {Vergani}, D., {Zucca}, E.,
  {Aussel}, H., {Capak}, P., {Cappelluti}, N., {Elvis}, M., {Fiore}, F.,
  {Hasinger}, G., {Impey}, C., {Le Floc'h}, E., {Scoville}, N., {Taniguchi},
  Y., \& {Trump}, J. 2011, \aap, 535, A80

\bibitem[{{Ma{\'{\i}}z Apell{\'a}niz}(2006)}]{maiz2006}
{Ma{\'{\i}}z Apell{\'a}niz}, J. 2006, \aj, 131, 1184

\bibitem[{{Maraston}(2005)}]{maraston2005}
{Maraston}, C. 2005, \mnras, 362, 799

\bibitem[{{Maraston} {et~al.}(2006){Maraston}, {Daddi}, {Renzini}, {Cimatti},
  {Dickinson}, {Papovich}, {Pasquali}, \& {Pirzkal}}]{maraston2006}
{Maraston}, C., {Daddi}, E., {Renzini}, A., {Cimatti}, A., {Dickinson}, M.,
  {Papovich}, C., {Pasquali}, A., \& {Pirzkal}, N. 2006, \apj, 652, 85

\bibitem[{{Maraston} {et~al.}(2010){Maraston}, {Pforr}, {Renzini}, {Daddi},
  {Dickinson}, {Cimatti}, \& {Tonini}}]{maraston2010}
{Maraston}, C., {Pforr}, J., {Renzini}, A., {Daddi}, E., {Dickinson}, M.,
  {Cimatti}, A., \& {Tonini}, C. 2010, \mnras, 407, 830

\bibitem[{{Martini} {et~al.}(2004){Martini}, {Persson}, {Murphy}, {Birk},
  {Shectman}, {Gunnels}, \& {Koch}}]{martini2004}
{Martini}, P., {Persson}, S.~E., {Murphy}, D.~C., {Birk}, C., {Shectman},
  S.~A., {Gunnels}, S.~M., \& {Koch}, E. 2004, in Society of Photo-Optical
  Instrumentation Engineers (SPIE) Conference Series, Vol. 5492, Society of
  Photo-Optical Instrumentation Engineers (SPIE) Conference Series, ed.
  {A.~F.~M.~Moorwood \& M.~Iye}, 1653--1660

\bibitem[{{Merloni} {et~al.}(2010){Merloni}, {Bongiorno}, {Bolzonella},
  {Brusa}, {Civano}, {Comastri}, {Elvis}, {Fiore}, {Gilli}, {Hao}, {Jahnke},
  {Koekemoer}, {Lusso}, {Mainieri}, {Mignoli}, {Miyaji}, {Renzini}, {Salvato},
  {Silverman}, {Trump}, {Vignali}, {Zamorani}, {Capak}, {Lilly}, {Sanders},
  {Taniguchi}, {Bardelli}, {Carollo}, {Caputi}, {Contini}, {Coppa}, {Cucciati},
  {de la Torre}, {de Ravel}, {Franzetti}, {Garilli}, {Hasinger}, {Impey},
  {Iovino}, {Iwasawa}, {Kampczyk}, {Kneib}, {Knobel}, {Kova{\v c}},
  {Lamareille}, {Le Borgne}, {Le Brun}, {Le F{\`e}vre}, {Maier}, {Pello},
  {Peng}, {Perez Montero}, {Ricciardelli}, {Scodeggio}, {Tanaka}, {Tasca},
  {Tresse}, {Vergani}, \& {Zucca}}]{merloni2010}
{Merloni}, A., {Bongiorno}, A., {Bolzonella}, M., {Brusa}, M., {Civano}, F.,
  {Comastri}, A., {Elvis}, M., {Fiore}, F., {Gilli}, R., {Hao}, H., {Jahnke},
  K., {Koekemoer}, A.~M., {Lusso}, E., {Mainieri}, V., {Mignoli}, M., {Miyaji},
  T., {Renzini}, A., {Salvato}, M., {Silverman}, J., {Trump}, J., {Vignali},
  C., {Zamorani}, G., {Capak}, P., {Lilly}, S.~J., {Sanders}, D., {Taniguchi},
  Y., {Bardelli}, S., {Carollo}, C.~M., {Caputi}, K., {Contini}, T., {Coppa},
  G., {Cucciati}, O., {de la Torre}, S., {de Ravel}, L., {Franzetti}, P.,
  {Garilli}, B., {Hasinger}, G., {Impey}, C., {Iovino}, A., {Iwasawa}, K.,
  {Kampczyk}, P., {Kneib}, J.-P., {Knobel}, C., {Kova{\v c}}, K., {Lamareille},
  F., {Le Borgne}, J.-F., {Le Brun}, V., {Le F{\`e}vre}, O., {Maier}, C.,
  {Pello}, R., {Peng}, Y., {Perez Montero}, E., {Ricciardelli}, E.,
  {Scodeggio}, M., {Tanaka}, M., {Tasca}, L.~A.~M., {Tresse}, L., {Vergani},
  D., \& {Zucca}, E. 2010, \apj, 708, 137

\bibitem[{{Nandra} {et~al.}(2007){Nandra}, {Georgakakis}, {Willmer}, {Cooper},
  {Croton}, {Davis}, {Faber}, {Koo}, {Laird}, \& {Newman}}]{nandra2007}
{Nandra}, K., {Georgakakis}, A., {Willmer}, C.~N.~A., {Cooper}, M.~C.,
  {Croton}, D.~J., {Davis}, M., {Faber}, S.~M., {Koo}, D.~C., {Laird}, E.~S.,
  \& {Newman}, J.~A. 2007, \apjl, 660, L11

\bibitem[{{Neugebauer} {et~al.}(1979){Neugebauer}, {Oke}, {Becklin}, \&
  {Matthews}}]{neugebauer1979}
{Neugebauer}, G., {Oke}, J.~B., {Becklin}, E.~E., \& {Matthews}, K. 1979, \apj,
  230, 79

\bibitem[{{Noeske} {et~al.}(2007){Noeske}, {Weiner}, {Faber}, {Papovich},
  {Koo}, {Somerville}, {Bundy}, {Conselice}, {Newman}, {Schiminovich}, {Le
  Floc'h}, {Coil}, {Rieke}, {Lotz}, {Primack}, {Barmby}, {Cooper}, {Davis},
  {Ellis}, {Fazio}, {Guhathakurta}, {Huang}, {Kassin}, {Martin}, {Phillips},
  {Rich}, {Small}, {Willmer}, \& {Wilson}}]{noeske2007}
{Noeske}, K.~G., {Weiner}, B.~J., {Faber}, S.~M., {Papovich}, C., {Koo}, D.~C.,
  {Somerville}, R.~S., {Bundy}, K., {Conselice}, C.~J., {Newman}, J.~A.,
  {Schiminovich}, D., {Le Floc'h}, E., {Coil}, A.~L., {Rieke}, G.~H., {Lotz},
  J.~M., {Primack}, J.~R., {Barmby}, P., {Cooper}, M.~C., {Davis}, M., {Ellis},
  R.~S., {Fazio}, G.~G., {Guhathakurta}, P., {Huang}, J., {Kassin}, S.~A.,
  {Martin}, D.~C., {Phillips}, A.~C., {Rich}, R.~M., {Small}, T.~A., {Willmer},
  C.~N.~A., \& {Wilson}, G. 2007, \apjl, 660, L43

\bibitem[{{Pannella} {et~al.}(2009){Pannella}, {Carilli}, {Daddi}, {McCracken},
  {Owen}, {Renzini}, {Strazzullo}, {Civano}, {Koekemoer}, {Schinnerer},
  {Scoville}, {Smol{\v c}i{\'c}}, {Taniguchi}, {Aussel}, {Kneib}, {Ilbert},
  {Mellier}, {Salvato}, {Thompson}, \& {Willott}}]{pannella2009}
{Pannella}, M., {Carilli}, C.~L., {Daddi}, E., {McCracken}, H.~J., {Owen},
  F.~N., {Renzini}, A., {Strazzullo}, V., {Civano}, F., {Koekemoer}, A.~M.,
  {Schinnerer}, E., {Scoville}, N., {Smol{\v c}i{\'c}}, V., {Taniguchi}, Y.,
  {Aussel}, H., {Kneib}, J.~P., {Ilbert}, O., {Mellier}, Y., {Salvato}, M.,
  {Thompson}, D., \& {Willott}, C.~J. 2009, \apjl, 698, L116

\bibitem[{{Papovich} {et~al.}(2001){Papovich}, {Dickinson}, \&
  {Ferguson}}]{papovich2001}
{Papovich}, C., {Dickinson}, M., \& {Ferguson}, H.~C. 2001, \apj, 559, 620

\bibitem[{{Papovich} {et~al.}(2011){Papovich}, {Finkelstein}, {Ferguson},
  {Lotz}, \& {Giavalisco}}]{papovich2011}
{Papovich}, C., {Finkelstein}, S.~L., {Ferguson}, H.~C., {Lotz}, J.~M., \&
  {Giavalisco}, M. 2011, \mnras, 412, 1123

\bibitem[{{Park} {et~al.}(2010){Park}, {Barmby}, {Willner}, {Ashby}, {Fazio},
  {Georgakakis}, {Ivison}, {Konidaris}, {Miyazaki}, {Nandra}, \&
  {Rosario}}]{park2010}
{Park}, S.~Q., {Barmby}, P., {Willner}, S.~P., {Ashby}, M.~L.~N., {Fazio},
  G.~G., {Georgakakis}, A., {Ivison}, R.~J., {Konidaris}, N.~P., {Miyazaki},
  S., {Nandra}, K., \& {Rosario}, D.~J. 2010, \apj, 717, 1181

\bibitem[{{Persson} {et~al.}(1998){Persson}, {Murphy}, {Krzeminski}, {Roth}, \&
  {Rieke}}]{persson1998}
{Persson}, S.~E., {Murphy}, D.~C., {Krzeminski}, W., {Roth}, M., \& {Rieke},
  M.~J. 1998, \aj, 116, 2475

\bibitem[{{Polletta} {et~al.}(2007){Polletta}, {Tajer}, {Maraschi},
  {Trinchieri}, {Lonsdale}, {Chiappetti}, {Andreon}, {Pierre}, {Le F{\`e}vre},
  {Zamorani}, {Maccagni}, {Garcet}, {Surdej}, {Franceschini}, {Alloin},
  {Shupe}, {Surace}, {Fang}, {Rowan-Robinson}, {Smith}, \&
  {Tresse}}]{polletta2007}
{Polletta}, M., {Tajer}, M., {Maraschi}, L., {Trinchieri}, G., {Lonsdale},
  C.~J., {Chiappetti}, L., {Andreon}, S., {Pierre}, M., {Le F{\`e}vre}, O.,
  {Zamorani}, G., {Maccagni}, D., {Garcet}, O., {Surdej}, J., {Franceschini},
  A., {Alloin}, D., {Shupe}, D.~L., {Surace}, J.~A., {Fang}, F.,
  {Rowan-Robinson}, M., {Smith}, H.~E., \& {Tresse}, L. 2007, \apj, 663, 81

\bibitem[{{Reddy} {et~al.}(2010){Reddy}, {Erb}, {Pettini}, {Steidel}, \&
  {Shapley}}]{reddy2010}
{Reddy}, N.~A., {Erb}, D.~K., {Pettini}, M., {Steidel}, C.~C., \& {Shapley},
  A.~E. 2010, \apj, 712, 1070

\bibitem[{{Reddy} {et~al.}(2012){Reddy}, {Pettini}, {Steidel}, {Shapley},
  {Erb}, \& {Law}}]{reddy2012}
{Reddy}, N.~A., {Pettini}, M., {Steidel}, C.~C., {Shapley}, A.~E., {Erb},
  D.~K., \& {Law}, D.~R. 2012, ArXiv e-prints

\bibitem[{{Reddy} \& {Steidel}(2004)}]{reddy2004}
{Reddy}, N.~A. \& {Steidel}, C.~C. 2004, \apjl, 603, L13

\bibitem[{{Reddy} {et~al.}(2006{\natexlab{a}}){Reddy}, {Steidel}, {Erb},
  {Shapley}, \& {Pettini}}]{reddy2006b}
{Reddy}, N.~A., {Steidel}, C.~C., {Erb}, D.~K., {Shapley}, A.~E., \& {Pettini},
  M. 2006{\natexlab{a}}, \apj, 653, 1004

\bibitem[{{Reddy} {et~al.}(2006{\natexlab{b}}){Reddy}, {Steidel}, {Fadda},
  {Yan}, {Pettini}, {Shapley}, {Erb}, \& {Adelberger}}]{reddy2006a}
{Reddy}, N.~A., {Steidel}, C.~C., {Fadda}, D., {Yan}, L., {Pettini}, M.,
  {Shapley}, A.~E., {Erb}, D.~K., \& {Adelberger}, K.~L. 2006{\natexlab{b}},
  \apj, 644, 792

\bibitem[{{Reddy} {et~al.}(2008){Reddy}, {Steidel}, {Pettini}, {Adelberger},
  {Shapley}, {Erb}, \& {Dickinson}}]{reddy2008}
{Reddy}, N.~A., {Steidel}, C.~C., {Pettini}, M., {Adelberger}, K.~L.,
  {Shapley}, A.~E., {Erb}, D.~K., \& {Dickinson}, M. 2008, \apjs, 175, 48

\bibitem[{{Rees} {et~al.}(1969){Rees}, {Silk}, {Werner}, \&
  {Wickramasinghe}}]{rees1969}
{Rees}, M.~J., {Silk}, J.~I., {Werner}, M.~W., \& {Wickramasinghe}, N.~C. 1969,
  \nat, 223, 788

\bibitem[{{Richards} {et~al.}(2006){Richards}, {Lacy}, {Storrie-Lombardi},
  {Hall}, {Gallagher}, {Hines}, {Fan}, {Papovich}, {Vanden Berk}, {Trammell},
  {Schneider}, {Vestergaard}, {York}, {Jester}, {Anderson}, {Budav{\'a}ri}, \&
  {Szalay}}]{richards2006}
{Richards}, G.~T., {Lacy}, M., {Storrie-Lombardi}, L.~J., {Hall}, P.~B.,
  {Gallagher}, S.~C., {Hines}, D.~C., {Fan}, X., {Papovich}, C., {Vanden Berk},
  D.~E., {Trammell}, G.~B., {Schneider}, D.~P., {Vestergaard}, M., {York},
  D.~G., {Jester}, S., {Anderson}, S.~F., {Budav{\'a}ri}, T., \& {Szalay},
  A.~S. 2006, \apjs, 166, 470

\bibitem[{{Salim} {et~al.}(2007){Salim}, {Rich}, {Charlot}, {Brinchmann},
  {Johnson}, {Schiminovich}, {Seibert}, {Mallery}, {Heckman}, {Forster},
  {Friedman}, {Martin}, {Morrissey}, {Neff}, {Small}, {Wyder}, {Bianchi},
  {Donas}, {Lee}, {Madore}, {Milliard}, {Szalay}, {Welsh}, \& {Yi}}]{salim2007}
{Salim}, S., {Rich}, R.~M., {Charlot}, S., {Brinchmann}, J., {Johnson}, B.~D.,
  {Schiminovich}, D., {Seibert}, M., {Mallery}, R., {Heckman}, T.~M.,
  {Forster}, K., {Friedman}, P.~G., {Martin}, D.~C., {Morrissey}, P., {Neff},
  S.~G., {Small}, T., {Wyder}, T.~K., {Bianchi}, L., {Donas}, J., {Lee}, Y.-W.,
  {Madore}, B.~F., {Milliard}, B., {Szalay}, A.~S., {Welsh}, B.~Y., \& {Yi},
  S.~K. 2007, \apjs, 173, 267

\bibitem[{{Santini} {et~al.}(2012){Santini}, {Rosario}, {Shao}, {Lutz},
  {Maiolino}, {Alexander}, {Altieri}, {Andreani}, {Aussel}, {Bauer}, {Berta},
  {Bongiovanni}, {Brandt}, {Brusa}, {Cepa}, {Cimatti}, {Daddi}, {Elbaz},
  {Fontana}, {Forster Schreiber}, {Genzel}, {Grazian}, {Le Floc'h}, {Magnelli},
  {Mainieri}, {Nordon}, {Perez Garcia}, {Poglitsch}, {Popesso}, {Pozzi},
  {Riguccini}, {Rodighiero}, {Salvato}, {Sanchez-Portal}, {Sturm}, {Tacconi},
  {Valtchanov}, \& {Wuyts}}]{santini2012}
{Santini}, P., {Rosario}, D., {Shao}, L., {Lutz}, D., {Maiolino}, R.,
  {Alexander}, D.~M., {Altieri}, B., {Andreani}, P., {Aussel}, H., {Bauer},
  F.~E., {Berta}, S., {Bongiovanni}, A., {Brandt}, W.~N., {Brusa}, M., {Cepa},
  J., {Cimatti}, A., {Daddi}, E., {Elbaz}, D., {Fontana}, A., {Forster
  Schreiber}, N.~M., {Genzel}, R., {Grazian}, A., {Le Floc'h}, E., {Magnelli},
  B., {Mainieri}, V., {Nordon}, R., {Perez Garcia}, A.~M., {Poglitsch}, A.,
  {Popesso}, P., {Pozzi}, F., {Riguccini}, L., {Rodighiero}, G., {Salvato}, M.,
  {Sanchez-Portal}, M., {Sturm}, E., {Tacconi}, L.~J., {Valtchanov}, I., \&
  {Wuyts}, S. 2012, ArXiv e-prints

\bibitem[{{Sawicki} \& {Yee}(1998)}]{sawicki1998}
{Sawicki}, M. \& {Yee}, H.~K.~C. 1998, \aj, 115, 1329

\bibitem[{{Schaerer} \& {de Barros}(2011)}]{schaerer2011}
{Schaerer}, D. \& {de Barros}, S. 2011, ArXiv e-prints

\bibitem[{{Schawinski} {et~al.}(2007){Schawinski}, {Thomas}, {Sarzi},
  {Maraston}, {Kaviraj}, {Joo}, {Yi}, \& {Silk}}]{schawinski2007}
{Schawinski}, K., {Thomas}, D., {Sarzi}, M., {Maraston}, C., {Kaviraj}, S.,
  {Joo}, S.-J., {Yi}, S.~K., \& {Silk}, J. 2007, \mnras, 382, 1415

\bibitem[{{Schawinski} {et~al.}(2010){Schawinski}, {Urry}, {Virani}, {Coppi},
  {Bamford}, {Treister}, {Lintott}, {Sarzi}, {Keel}, {Kaviraj}, {Cardamone},
  {Masters}, {Ross}, {Andreescu}, {Murray}, {Nichol}, {Raddick}, {Slosar},
  {Szalay}, {Thomas}, \& {Vandenberg}}]{schawinski2010}
{Schawinski}, K., {Urry}, C.~M., {Virani}, S., {Coppi}, P., {Bamford}, S.~P.,
  {Treister}, E., {Lintott}, C.~J., {Sarzi}, M., {Keel}, W.~C., {Kaviraj}, S.,
  {Cardamone}, C.~N., {Masters}, K.~L., {Ross}, N.~P., {Andreescu}, D.,
  {Murray}, P., {Nichol}, R.~C., {Raddick}, M.~J., {Slosar}, A., {Szalay},
  A.~S., {Thomas}, D., \& {Vandenberg}, J. 2010, \apj, 711, 284

\bibitem[{{Shapley}(2011)}]{shapley2011}
{Shapley}, A.~E. 2011, \araa, 49, 525

\bibitem[{{Shapley} {et~al.}(2004){Shapley}, {Erb}, {Pettini}, {Steidel}, \&
  {Adelberger}}]{shapley2004}
{Shapley}, A.~E., {Erb}, D.~K., {Pettini}, M., {Steidel}, C.~C., \&
  {Adelberger}, K.~L. 2004, \apj, 612, 108

\bibitem[{{Shapley} {et~al.}(2001){Shapley}, {Steidel}, {Adelberger},
  {Dickinson}, {Giavalisco}, \& {Pettini}}]{shapley2001}
{Shapley}, A.~E., {Steidel}, C.~C., {Adelberger}, K.~L., {Dickinson}, M.,
  {Giavalisco}, M., \& {Pettini}, M. 2001, \apj, 562, 95

\bibitem[{{Shapley} {et~al.}(2005){Shapley}, {Steidel}, {Erb}, {Reddy},
  {Adelberger}, {Pettini}, {Barmby}, \& {Huang}}]{shapley2005a}
{Shapley}, A.~E., {Steidel}, C.~C., {Erb}, D.~K., {Reddy}, N.~A., {Adelberger},
  K.~L., {Pettini}, M., {Barmby}, P., \& {Huang}, J. 2005, \apj, 626, 698

\bibitem[{{Shapley} {et~al.}(2003){Shapley}, {Steidel}, {Pettini}, \&
  {Adelberger}}]{shapley2003}
{Shapley}, A.~E., {Steidel}, C.~C., {Pettini}, M., \& {Adelberger}, K.~L. 2003,
  \apj, 588, 65

\bibitem[{{Shuder}(1981)}]{shuder1981}
{Shuder}, J.~M. 1981, \apj, 244, 12

\bibitem[{{Silverman} {et~al.}(2008){Silverman}, {Green}, {Barkhouse}, {Kim},
  {Kim}, {Wilkes}, {Cameron}, {Hasinger}, {Jannuzi}, {Smith}, {Smith}, \&
  {Tananbaum}}]{silverman2008}
{Silverman}, J.~D., {Green}, P.~J., {Barkhouse}, W.~A., {Kim}, D., {Kim}, M.,
  {Wilkes}, B.~J., {Cameron}, R.~A., {Hasinger}, G., {Jannuzi}, B.~T., {Smith},
  M.~G., {Smith}, P.~S., \& {Tananbaum}, H. 2008, \apj, 679, 118

\bibitem[{{Silverman} {et~al.}(2009){Silverman}, {Lamareille}, {Maier},
  {Lilly}, {Mainieri}, {Brusa}, {Cappelluti}, {Hasinger}, {Zamorani},
  {Scodeggio}, {Bolzonella}, {Contini}, {Carollo}, {Jahnke}, {Kneib}, {Le
  F{\`e}vre}, {Merloni}, {Bardelli}, {Bongiorno}, {Brunner}, {Caputi},
  {Civano}, {Comastri}, {Coppa}, {Cucciati}, {de la Torre}, {de Ravel},
  {Elvis}, {Finoguenov}, {Fiore}, {Franzetti}, {Garilli}, {Gilli}, {Iovino},
  {Kampczyk}, {Knobel}, {Kova{\v c}}, {Le Borgne}, {Le Brun}, {Mignoli},
  {Pello}, {Peng}, {Perez Montero}, {Ricciardelli}, {Tanaka}, {Tasca},
  {Tresse}, {Vergani}, {Vignali}, {Zucca}, {Bottini}, {Cappi}, {Cassata},
  {Fumana}, {Griffiths}, {Kartaltepe}, {Koekemoer}, {Marinoni}, {McCracken},
  {Memeo}, {Meneux}, {Oesch}, {Porciani}, \& {Salvato}}]{silverman2009a}
{Silverman}, J.~D., {Lamareille}, F., {Maier}, C., {Lilly}, S.~J., {Mainieri},
  V., {Brusa}, M., {Cappelluti}, N., {Hasinger}, G., {Zamorani}, G.,
  {Scodeggio}, M., {Bolzonella}, M., {Contini}, T., {Carollo}, C.~M., {Jahnke},
  K., {Kneib}, J.-P., {Le F{\`e}vre}, O., {Merloni}, A., {Bardelli}, S.,
  {Bongiorno}, A., {Brunner}, H., {Caputi}, K., {Civano}, F., {Comastri}, A.,
  {Coppa}, G., {Cucciati}, O., {de la Torre}, S., {de Ravel}, L., {Elvis}, M.,
  {Finoguenov}, A., {Fiore}, F., {Franzetti}, P., {Garilli}, B., {Gilli}, R.,
  {Iovino}, A., {Kampczyk}, P., {Knobel}, C., {Kova{\v c}}, K., {Le Borgne},
  J.-F., {Le Brun}, V., {Mignoli}, M., {Pello}, R., {Peng}, Y., {Perez
  Montero}, E., {Ricciardelli}, E., {Tanaka}, M., {Tasca}, L., {Tresse}, L.,
  {Vergani}, D., {Vignali}, C., {Zucca}, E., {Bottini}, D., {Cappi}, A.,
  {Cassata}, P., {Fumana}, M., {Griffiths}, R., {Kartaltepe}, J., {Koekemoer},
  A., {Marinoni}, C., {McCracken}, H.~J., {Memeo}, P., {Meneux}, B., {Oesch},
  P., {Porciani}, C., \& {Salvato}, M. 2009, \apj, 696, 396

\bibitem[{{Somerville} {et~al.}(2008){Somerville}, {Hopkins}, {Cox},
  {Robertson}, \& {Hernquist}}]{somerville2008}
{Somerville}, R.~S., {Hopkins}, P.~F., {Cox}, T.~J., {Robertson}, B.~E., \&
  {Hernquist}, L. 2008, \mnras, 391, 481

\bibitem[{{Steidel} {et~al.}(2003){Steidel}, {Adelberger}, {Shapley},
  {Pettini}, {Dickinson}, \& {Giavalisco}}]{steidel2003}
{Steidel}, C.~C., {Adelberger}, K.~L., {Shapley}, A.~E., {Pettini}, M.,
  {Dickinson}, M., \& {Giavalisco}, M. 2003, \apj, 592, 728

\bibitem[{{Steidel} {et~al.}(2002){Steidel}, {Hunt}, {Shapley}, {Adelberger},
  {Pettini}, {Dickinson}, \& {Giavalisco}}]{steidel2002}
{Steidel}, C.~C., {Hunt}, M.~P., {Shapley}, A.~E., {Adelberger}, K.~L.,
  {Pettini}, M., {Dickinson}, M., \& {Giavalisco}, M. 2002, \apj, 576, 653

\bibitem[{{Steidel} {et~al.}(2004){Steidel}, {Shapley}, {Pettini},
  {Adelberger}, {Erb}, {Reddy}, \& {Hunt}}]{steidel2004}
{Steidel}, C.~C., {Shapley}, A.~E., {Pettini}, M., {Adelberger}, K.~L., {Erb},
  D.~K., {Reddy}, N.~A., \& {Hunt}, M.~P. 2004, \apj, 604, 534

\bibitem[{{Stern} {et~al.}(2005){Stern}, {Eisenhardt}, {Gorjian}, {Kochanek},
  {Caldwell}, {Eisenstein}, {Brodwin}, {Brown}, {Cool}, {Dey}, {Green},
  {Jannuzi}, {Murray}, {Pahre}, \& {Willner}}]{stern2005}
{Stern}, D., {Eisenhardt}, P., {Gorjian}, V., {Kochanek}, C.~S., {Caldwell},
  N., {Eisenstein}, D., {Brodwin}, M., {Brown}, M.~J.~I., {Cool}, R., {Dey},
  A., {Green}, P., {Jannuzi}, B.~T., {Murray}, S.~S., {Pahre}, M.~A., \&
  {Willner}, S.~P. 2005, \apj, 631, 163

\bibitem[{{Targett} {et~al.}(2012){Targett}, {Dunlop}, \&
  {McLure}}]{targett2012}
{Targett}, T.~A., {Dunlop}, J.~S., \& {McLure}, R.~J. 2012, \mnras, 420, 3621

\bibitem[{{Tremonti} {et~al.}(2007){Tremonti}, {Moustakas}, \&
  {Diamond-Stanic}}]{tremonti2007}
{Tremonti}, C.~A., {Moustakas}, J., \& {Diamond-Stanic}, A.~M. 2007, \apjl,
  663, L77

\bibitem[{{Vestergaard} \& {Osmer}(2009)}]{vo2009}
{Vestergaard}, M. \& {Osmer}, P.~S. 2009, \apj, 699, 800

\bibitem[{{Williams} {et~al.}(2009){Williams}, {Quadri}, {Franx}, {van Dokkum},
  \& {Labb{\'e}}}]{williams2009}
{Williams}, R.~J., {Quadri}, R.~F., {Franx}, M., {van Dokkum}, P., \&
  {Labb{\'e}}, I. 2009, \apj, 691, 1879

\bibitem[{{Xue} {et~al.}(2010){Xue}, {Brandt}, {Luo}, {Rafferty}, {Alexander},
  {Bauer}, {Lehmer}, {Schneider}, \& {Silverman}}]{xue2010}
{Xue}, Y.~Q., {Brandt}, W.~N., {Luo}, B., {Rafferty}, D.~A., {Alexander},
  D.~M., {Bauer}, F.~E., {Lehmer}, B.~D., {Schneider}, D.~P., \& {Silverman},
  J.~D. 2010, \apj, 720, 368

\bibitem[{{Yee}(1980)}]{yee1980}
{Yee}, H.~K.~C. 1980, \apj, 241, 894

\bibitem[{{Zakamska} {et~al.}(2006){Zakamska}, {Strauss}, {Krolik}, {Ridgway},
  {Schmidt}, {Smith}, {Heckman}, {Schneider}, {Hao}, \&
  {Brinkmann}}]{zakamska2006}
{Zakamska}, N.~L., {Strauss}, M.~A., {Krolik}, J.~H., {Ridgway}, S.~E.,
  {Schmidt}, G.~D., {Smith}, P.~S., {Heckman}, T.~M., {Schneider}, D.~P.,
  {Hao}, L., \& {Brinkmann}, J. 2006, \aj, 132, 1496

\end{thebibliography}


\begin{deluxetable}{lcccccc}
\tabletypesize{\tiny}
\tablecaption{UV-Selected AGN Photometry\label{tab:photometry}}
\tablewidth{0pt}
\setlength{\tabcolsep}{0.04in} 
\tablehead{
\colhead{OBJECT} & \colhead{RA (J2000)} & \colhead{DEC (J2000)} & \colhead{$z_{Ly\alpha}$} & \colhead{$\cal{R}$\tablenotemark{a}} & 
\colhead{$U-G$\tablenotemark{a}} & \colhead{$G-\cal{R}$\tablenotemark{a}} 
}
\startdata

Q0000-C7 & 00:03:28.85 & -26:03:53.3 & 3.431 & 24.37 $\pm$ 0.19 & - & 0.07 $\pm$ 0.09 \\ 
Q0000-C14 & 00:03:30.39 & -26:01:20.7 & 3.057 & 24.47 $\pm$ 0.17 & - & 0.86 $\pm$ 0.18 \\ 
CDFB-D3 & 00:53:43.02 & \phm{-}12:22:02.5 & 2.777 & 24.82 $\pm$ 0.20 & - & 0.47 $\pm$ 0.14 \\ 
Q0100-BX172 & 01:03:08.46 & \phm{-}13:16:41.7 & 2.312 & 23.50 $\pm$ 0.09 & 0.81 $\pm$ 0.27 & 0.26 $\pm$ 0.07 \\ 
Q0142-BX195 & 01:45:17.68 & -09:44:54.2 & 2.382 & 23.56 $\pm$ 0.08 & 0.98 $\pm$ 0.23 & 0.53 $\pm$ 0.09 \\ 
Q0142-BX256 & 01:45:15.74 & -09:42:12.5 & 2.321 & 23.91 $\pm$ 0.08 & 0.88 $\pm$ 0.16 & 0.32 $\pm$ 0.07 \\ 
Q0201-OC12 & 02:03:56.16 & \phm{-}11:36:30.1 & 2.357 & 25.01 $\pm$ 0.34 & - & 0.56 $\pm$ 0.13 \\ 
Q0256-MD37 & 02:59:02.21 & \phm{-}00:12:03.4 & 2.803 & 24.36 $\pm$ 0.18 & - & 0.63 $\pm$ 0.14 \\ 
Q0933-MD38 & 09:33:48.60 & \phm{-}28:44:32.3 & 2.763 & 22.61 $\pm$ 0.06 & - & 0.04 $\pm$ 0.09 \\ 
Q1217-BX46 & 12:19:19.94 & \phm{-}49:40:22.7 & 1.980 & 23.85 $\pm$ 0.17 & 0.76 $\pm$ 0.29 & 0.44 $\pm$ 0.14 \\ 
HDF-MMD12 & 12:37:19.80 & \phm{-}62:09:56.0 & 2.648 & 24.36 $\pm$ 0.17 & - & 1.02 $\pm$ 0.20 \\ 
HDF-BMZ1156 & 12:37:04.34 & \phm{-}62:14:46.3 & 2.211 & 24.62 $\pm$ 0.20 & - & 0.37 $\pm$ 0.08 \\ 
HDF-BMZ1384 & 12:37:23.15 & \phm{-}62:15:38.0 & 2.243 & 23.98 $\pm$ 0.14 & 0.45 $\pm$ 0.06 & 0.49 $\pm$ 0.08 \\ 
HDF-BX160 & 12:37:20.07 & \phm{-}62:12:22.7 & 2.461 & 24.02 $\pm$ 0.17 & 0.92 $\pm$ 0.15 & 0.74 $\pm$ 0.12 \\ 
Westphal-MM47 & 14:17:57.39 & \phm{-}52:31:04.5 & 3.027 & 24.30 $\pm$ 0.13 & - & 1.26 $\pm$ 0.10 \\ 
Q1422-C73 & 14:24:46.41 & \phm{-}22:55:45.5 & 3.382 & 24.88 $\pm$ 0.14 & - & 1.05 $\pm$ 0.17 \\ 
Q1422-MD109 & 14:24:42.58 & \phm{-}22:54:46.6 & 2.229 & 23.69 $\pm$ 0.07 & 1.61 $\pm$ 0.14 & 0.56 $\pm$ 0.08 \\ 
Q1623-BX454 & 16:25:51.42 & \phm{-}26:43:46.3 & 2.422 & 23.89 $\pm$ 0.13 & 0.50 $\pm$ 0.14 & 0.22 $\pm$ 0.06 \\ 
Q1623-BX663 & 16:26:04.58 & \phm{-}26:48:00.2 & 2.435 & 24.14 $\pm$ 0.15 & 1.02 $\pm$ 0.17 & 0.24 $\pm$ 0.07 \\ 
Q1623-BX747 & 16:26:13.46 & \phm{-}26:45:53.2 & 2.441 & 22.55 $\pm$ 0.14 & 0.42 $\pm$ 0.04 & 0.12 $\pm$ 0.03 \\ 
Q1700-MD157 & 17:00:52.19 & \phm{-}64:15:29.3 & 2.295 & 24.35 $\pm$ 0.13 & 1.65 $\pm$ 0.28 & 0.35 $\pm$ 0.07 \\ 
Q1700-MD174 & 17:00:54.54 & \phm{-}64:16:24.8 & 2.347 & 24.56 $\pm$ 0.17 & 1.50 $\pm$ 0.28 & 0.32 $\pm$ 0.08 \\ 
Q2233-D3 & 22:36:16.12 & \phm{-}13:55:19.2 & 2.795 & 24.08 $\pm$ 0.18 & - & 1.08 $\pm$ 0.16 \\ 
Q2233-MD21 & 22:36:35.83 & \phm{-}13:55:42.0 & 2.549 & 25.72 $\pm$ 0.32 & - & 0.85 $\pm$ 0.21 \\ 
DSF2237A-D11 & 22:40:02.99 & \phm{-}11:52:13.9 & 2.959 & 25.09 $\pm$ 0.26 & - & 0.41 $\pm$ 0.13 \\ 
DSF2237B-MD53 & 22:39:28.67 & \phm{-}11:52:09.5 & 2.292 & 24.05 $\pm$ 0.13 & - & 0.35 $\pm$ 0.10 \\ 
Q2343-BX333 & 23:46:21.51 & \phm{-}12:47:03.2 & 2.397 & 24.12 $\pm$ 0.14 & 0.65 $\pm$ 0.13 & 0.30 $\pm$ 0.08 \\ 
Q2346-BX445 & 23:48:13.20 & \phm{-}00:25:15.8 & 2.330 & 23.81 $\pm$ 0.13 & 1.00 $\pm$ 0.36 & 0.39 $\pm$ 0.09 \\ 

 \\
\hline \hline
 \\
OBJECT & $\cal{R}$$-J$\tablenotemark{b} & $\cal{R}$$-K$\tablenotemark{b} & [3.6 $\mu$m]\tablenotemark{a} 
& [4.5 $\mu$m]\tablenotemark{a} & [5.8 $\mu$m]\tablenotemark{a} & [8.0 $\mu$m]\tablenotemark{a} \\
 \\
\hline \hline
 \\
 
 Q0000-C7 & 2.17 $\pm$ 0.43 & 4.50 $\pm$ 0.20 & - & - & - & - \\ 
Q0000-C14 & 4.43 $\pm$ 0.14 & - & - & - & - \\ 
CDFB-D3 & 4.97 $\pm$ 0.34 & - & - & - & - \\ 
Q0100-BX172 & - & 3.36 $\pm$ 0.22 & 21.43 $\pm$ 0.10 & 21.24 $\pm$ 0.10 & 21.00 $\pm$ 0.23 & 20.64 $\pm$ 0.10 \\ 
Q0142-BX195 & - & 3.48 $\pm$ 0.18 & 21.22 $\pm$ 0.10 & 21.01 $\pm$ 0.10 & 20.75 $\pm$ 0.24 & 20.26 $\pm$ 0.10 \\ 
Q0142-BX256 & - & 4.02 $\pm$ 0.13 & 21.45 $\pm$ 0.10 & 21.21 $\pm$ 0.18 & 20.15 $\pm$ 0.34 & - \\ 
Q0201-OC12 & 2.61 $\pm$ 0.40 & 5.51 $\pm$ 0.20 & - & - & - & - \\ 
Q0256-MD37 & - & 3.68 $\pm$ 0.21 & - & - & - & - \\ 
Q0933-MD38 & - & 2.37 $\pm$ 0.20 & - & - & - & - \\ 
Q1217-BX46 & 2.36 $\pm$ 0.14 & 3.56 $\pm$ 0.53 & 21.65 $\pm$ 0.10 & - & - & - \\ 
HDF-MMD12 & 1.85 $\pm$ 0.27 & 4.07 $\pm$ 0.21 & 21.40 $\pm$ 0.07 & 20.95 $\pm$ 0.07 & 20.24 $\pm$ 0.07 & 19.55 $\pm$ 0.07 \\ 
HDF-BMZ1156 & 2.58 $\pm$ 0.20 & 4.29 $\pm$ 0.17 & 22.01 $\pm$ 0.07 & 21.85 $\pm$ 0.07 & 20.96 $\pm$ 0.07 & 19.75 $\pm$ 0.07 \\ 
HDF-BMZ1384 & 2.09 $\pm$ 0.16 & 4.11 $\pm$ 0.14 & 21.97 $\pm$ 0.07 & 21.60 $\pm$ 0.07 & 21.06 $\pm$ 0.07 & 20.49 $\pm$ 0.07 \\ 
HDF-BX160 & 2.09 $\pm$ 0.18 & 3.97 $\pm$ 0.20 & 21.39 $\pm$ 0.07 & 21.22 $\pm$ 0.07 & 20.90 $\pm$ 0.07 & 20.85 $\pm$ 0.07 \\ 
Westphal-MM47 & - & - & 21.83 $\pm$ 0.10 & 21.84 $\pm$ 0.10 & - & - \\ 
Q1422-C73 & - & 3.82 $\pm$ 0.21 & - & - & - & - \\ 
Q1422-MD109 & - & 4.48 $\pm$ 0.10 & - & - & - & - \\ 
Q1623-BX454 & - & - & 22.87 $\pm$ 0.15 & 22.64 $\pm$ 0.15 & - & - \\ 
Q1623-BX663 & 1.63 $\pm$ 0.26 & 4.22 $\pm$ 0.14 & 20.66 $\pm$ 0.10 & 19.86 $\pm$ 0.10 & 19.03 $\pm$ 0.10 & 18.13 $\pm$ 0.10 \\ 
Q1623-BX747 & 1.51 $\pm$ 0.11 & 2.19 $\pm$ 0.23 & 22.33 $\pm$ 0.16 & - & - & - \\ 
Q1700-MD157 & 2.24 $\pm$ 0.10 & 4.12 $\pm$ 0.19 & - & 21.61 $\pm$ 0.10 & - & 21.25 $\pm$ 0.11 \\ 
Q1700-MD174 & - & 4.66 $\pm$ 0.15 & - & 20.41 $\pm$ 0.10 & - & 18.26 $\pm$ 0.10 \\ 
Q2233-D3 & - & 4.53 $\pm$ 0.19 & - & - & - & - \\ 
Q2233-MD21 & - & 4.37 $\pm$ 0.31 & 22.89 $\pm$ 0.11 & - & - & - \\ 
DSF2237A-D11 & - & 3.00 $\pm$ 0.50 & - & - & - & - \\ 
DSF2237B-MD53 & - & 4.68 $\pm$ 0.14 & - & - & - & - \\ 
Q2343-BX333 & 1.93 $\pm$ 0.14 & 4.21 $\pm$ 0.16 & - & 21.01 $\pm$ 0.10 & - & 20.47 $\pm$ 0.10 \\ 
Q2346-BX445 & - & 4.34 $\pm$ 0.13 & - & - & - & - \\ 
\enddata
\tablenotetext{a}{$U$, $G$, $\cal{R}$, and IRAC magnitudes are on the AB system.}
\tablenotetext{b}{$\cal{R}$$-J$ and $\cal{R}$$-K$ colors are AB$-$Vega.}

\end{deluxetable}
\clearpage

\begin{deluxetable}{lccccc}
\tabletypesize{\scriptsize}
\tablecaption{Best Fit Parameters, AGNs with IRAC Coverage\label{tab:iracparam}}
\tablewidth{0pt}
\tablehead{
\colhead{OBJECT} & \colhead{log[Age]} & \colhead{log(M$_{*}$)} & \colhead{$E(B-V)_{SF}$} & \colhead{SFR} & \colhead{$E(B-V)_{AGN}$} \\
 & \colhead{Myr} & \colhead{M$_{\sun}$} & & \colhead{M$_{\sun}$ yr$^{-1}$} &  
}
\startdata
 
Q0100-BX172 & 3.21 $\pm$ 0.18 & 10.77 $\pm$ 0.09 & 0.140 $\pm$ 0.030 & \phantom{0}37 $\pm$ 9\phantom{00} & 2.600 $\pm$ 2.433  \\ 
Q0142-BX195 & 2.76 $\pm$ 0.31 & 10.76 $\pm$ 0.15 & 0.260 $\pm$ 0.053 & 101 $\pm$ 62\phantom{0} & 4.600 $\pm$ 2.066  \\ 
Q0142-BX256 &  3.44 $\pm$ 0.08 & 10.96 $\pm$ 0.06 & 0.180 $\pm$ 0.021 & \phantom{0}33 $\pm$ 6\phantom{00} & 5.600 $\pm$ 1.010  \\ 
HDF-MMD12 &  2.31 $\pm$ 0.11 & 10.60 $\pm$ 0.12 & 0.400 $\pm$ 0.033 & 196 $\pm$ 49\phantom{0} & 2.000 $\pm$ 0.614 \\ 
HDF-BMZ1156 &  3.34 $\pm$ 0.10 & 10.64 $\pm$ 0.05 & 0.240 $\pm$ 0.024 & \phantom{0}20 $\pm$ 3\phantom{00} & 6.200 $\pm$ 0.646 \\ 
HDF-BMZ1384 &  3.44 $\pm$ 0.00 & 10.69 $\pm$ 0.07 & 0.140 $\pm$ 0.018 & \phantom{0}18 $\pm$ 3\phantom{00} & 2.800 $\pm$ 1.254 \\ 
HDF-BX160 &  3.11 $\pm$ 0.21 & 10.91 $\pm$ 0.08 & 0.260 $\pm$ 0.029 & \phantom{0}64 $\pm$ 15\phantom{0} & 6.400 $\pm$ 1.231 \\ 
Q1623-BX663 &  2.76 $\pm$ 0.81 & 10.26 $\pm$ 0.85 & 0.180 $\pm$ 0.020 & \phantom{0}32 $\pm$ 4\phantom{00} & 1.000 $\pm$ 0.078 \\ 
Q1700-MD157 &  3.16 $\pm$ 0.17 & 10.74 $\pm$ 0.09 & 0.260 $\pm$ 0.025 & \phantom{0}38 $\pm$ 8\phantom{00} & 7.400 $\pm$ 0.616 \\ 
Q1700-MD174 &  3.44 $\pm$ 0.00 & 11.15 $\pm$ 0.17 & 0.300 $\pm$ 0.042 & \phantom{0}51 $\pm$ 16\phantom{0} & 3.600 $\pm$ 1.361 \\ 
Q2343-BX333 & 3.41 $\pm$ 0.00 & 10.96 $\pm$ 0.09 & 0.200 $\pm$ 0.019 & \phantom{0}35 $\pm$ 7\phantom{00} & 1.000 $\pm$ 2.461 \\ 

 \\
\hline \hline
 \\ 
OBJECT & log($N_{AGN}$)\tablenotemark{a} & $f_{AGN, 1 \mu m}$\tablenotemark{b} & log($\mathrm{L}_{\mathrm{bol,AGN}}$) & $\chi^{2}$,$\chi^{2}_{red}$\tablenotemark{c}  \\
 &  &  & $\mathrm{L}_{\mathrm{\sun}}$ & \\
 \\
\hline \hline
 \\
 
Q0100-BX172 &  \phm{1}-1.00 $\pm$ 0.55 & \phm{$<$}0.02 & 10.82 $\pm$ 0.57 & 0.26, 0.09 & \\ 
Q0142-BX195 & \phm{1}-1.00 $\pm$ 0.27 & $<$0.01 & 11.26 $\pm$ 0.35 & 1.03, 0.34 & \\ 
Q0142-BX256 &  \phm{-1}0.18 $\pm$ 0.00 & \phm{$<$}0.01 & 11.95 $\pm$ 0.09 & 6.30, 3.15 & \\ 
HDF-MMD12 & \phm{1}-1.00 $\pm$ 0.00 & \phm{$<$}0.16 & 11.55 $\pm$ 0.08 & 4.63, 1.54 & \\ 
HDF-BMZ1156 & \phm{-1}0.15 $\pm$ 0.07 & \phm{$<$}0.01 & 11.70 $\pm$ 0.08 & 19.06, 6.35 & \\ 
HDF-BMZ1384 & \phm{1}-0.52 $\pm$ 0.11 & \phm{$<$}0.04 & 10.98 $\pm$ 0.15 & 45.06, 11.26 & \\ 
HDF-BX160 & \phm{1}-1.00 $\pm$ 0.00 & $<$0.01 & 11.06 $\pm$ 0.11 & 15.39, 3.85 & \\ 
Q1623-BX663 & \phm{-1}0.18 $\pm$ 0.02 & \phm{$<$}0.79 & 11.94 $\pm$ 0.04 & 4.91, 1.23 & \\ 
Q1700-MD157 & \phm{1}-1.00 $\pm$ 0.20 & $<$0.01 & 10.84 $\pm$ 0.18 & 22.97, 11.48 & \\ 
Q1700-MD174 & \phm{-1}0.15 $\pm$ 0.02 & \phm{$<$}0.09 & 12.11 $\pm$ 0.14 & 13.33, 13.33 & \\ 
Q2343-BX333 & \phm{1}-1.00 $\pm$ 0.00 & \phm{$<$}0.11 & 10.80 $\pm$ 0.10 & 14.11, 7.06 & \\
\enddata

\tablenotetext{a}{$N_{AGN} = L_{bol} / (\mathrm{SFR} \times 10^{10})$}
\tablenotetext{b}{The fraction of AGN emission at rest-frame 1 $\mu$m.}
\tablenotetext{c}{Raw and reduced $\chi^{2}$ values for each best-fit model.}

\end{deluxetable}

\begin{deluxetable}{lccccc}
\tabletypesize{\scriptsize}
\tablecaption{Best Fit Parameters, AGNs with IRAC Coverage, SPS-only, through $K$-band\label{tab:iracparamspsonly}}
\tablewidth{0pt}
\tablehead{
\colhead{OBJECT} & \colhead{log[Age]} & \colhead{log(M$_{*}$)} & \colhead{$E(B-V)_{SF}$}
& \colhead{SFR} & \colhead{$\chi^{2}$,$\chi^{2}_{red}$\tablenotemark{a}} \\
 & \colhead{Myr} & \colhead{M$_{\sun}$} & & \colhead{M$_{\sun}$ yr$^{-1}$} & 
}
\startdata

Q0100-BX172 & 3.21 $\pm$ 0.23 & 10.77 $\pm$ 0.20 & 0.140 $\pm$ 0.032 & \phm{0}37 $\pm$ 12\phm{0} & 0.19, 0.19 \\ 
Q0142-BX195 & 2.66 $\pm$ 0.32 & 10.68 $\pm$ 0.17 & 0.265 $\pm$ 0.056 & 106 $\pm$ 50\phm{0} & 0.77, 0.77 \\ 
Q0142-BX256 & 3.44 $\pm$ 0.00 & 11.09 $\pm$ 0.11 & 0.210 $\pm$ 0.019 & \phm{0}45 $\pm$ 9\phm{00} & 1.81, 1.81 \\ 
HDF-MMD12 & 1.86 $\pm$ 0.72 & 10.56 $\pm$ 0.27 & 0.485 $\pm$ 0.112 & 505 $\pm$ 757 & 0.16, 0.16 \\ 
HDF-BMZ1156 & 3.44 $\pm$ 0.00 & 10.95 $\pm$ 0.26 & 0.265 $\pm$ 0.024 & \phm{0}32 $\pm$ 10\phm{0} & 8.44, 8.44 \\ 
HDF-BMZ1384 & 3.44 $\pm$ 0.00 & 10.91 $\pm$ 0.14 & 0.175 $\pm$ 0.016 & \phm{0}29 $\pm$ 6\phm{00} & 34.12, 17.06 \\ 
HDF-BX160 & 3.34 $\pm$ 0.13 & 11.17 $\pm$ 0.26 & 0.255 $\pm$ 0.028 & \phm{0}68 $\pm$ 24\phm{0} & 12.73, 6.37 \\ 
Q1623-BX663 & 3.41 $\pm$ 0.00 & 11.06 $\pm$ 0.66 & 0.220 $\pm$ 0.019 & \phm{0}44 $\pm$ 11\phm{0} & 9.83, 4.91 \\ 
Q1700-MD157 & 3.44 $\pm$ 0.00 & 11.10 $\pm$ 0.25 & 0.265 $\pm$ 0.021 & \phm{0}45 $\pm$ 12\phm{0} & 16.13, 8.07 \\ 
Q1700-MD174 & 3.44 $\pm$ 0.00 & 11.14 $\pm$ 0.12 & 0.295 $\pm$ 0.022 & \phm{0}51 $\pm$ 14\phm{0} & 13.31, 13.31 \\ 
Q2343-BX333 & 3.41 $\pm$ 0.00 & 11.02 $\pm$ 0.11 & 0.210 $\pm$ 0.018 & \phm{0}40 $\pm$ 9\phm{00} & 14.91, 7.45 

\enddata

\tablenotetext{a}{Raw and reduced $\chi^{2}$ values for each best-fit model.}
\end{deluxetable}

\begin{deluxetable}{lllll}
\tabletypesize{\scriptsize}
\tablecaption{AGN Correction Factors\label{tab:corrections}}
\tablewidth{0pt}
\tablehead{
\colhead{OBJECT} & \colhead{Mass\tablenotemark{a}} & \colhead{SFR\tablenotemark{a}} & \colhead{Age\tablenotemark{a}} & \colhead{$E(B-V)_{SF}$\tablenotemark{a}}
}
\startdata
$\mathrm{Q0100-BX172}$ & 1.0 $\pm$ 0.5 & 1.0 $\pm$ 0.4 & 1.0 $\pm$ 0.6 & \phm{0}1.0 $\pm$ 0.3 \\ 
$\mathrm{Q0142-BX195}$ & 1.2 $\pm$ 0.6 & 1.1 $\pm$ 0.7 & 1.3 $\pm$ 1.0 & \phm{0}1.0 $\pm$ 0.3 \\ 
$\mathrm{Q0142-BX256}$ & 0.7 $\pm$ 0.2 & 0.8 $\pm$ 0.3 & 0.8 $\pm$ 0.2 & \phm{0}0.9 $\pm$ 0.1 \\ 
$\mathrm{HDF-MMD12}$ & 0.9 $\pm$ 0.7 & 0.2 $\pm$ 1.1 & 2.1 $\pm$ 1.4 & \phm{0}0.8 $\pm$ 0.2 \\ 
$\mathrm{HDF-BMZ1156}$ & 0.5 $\pm$ 0.3 & 0.6 $\pm$ 0.3 & 0.8 $\pm$ 0.2 & \phm{0}0.9 $\pm$ 0.1 \\ 
$\mathrm{HDF-BMZ1384}$ & 0.6 $\pm$ 0.3 & 0.6 $\pm$ 0.4 & 1.0 $\pm$ 0.0 & \phm{0}0.8 $\pm$ 0.2 \\ 
$\mathrm{HDF-BX160}$ & 0.6 $\pm$ 0.4 & 0.8 $\pm$ 0.4 & 0.8 $\pm$ 0.5 & \phm{0}1.0 $\pm$ 0.2 \\ 
$\mathrm{Q1623-BX663}$ & 0.3 $\pm$ 1.0 & 0.8 $\pm$ 0.3 & 0.4 $\pm$ 0.9 & \phm{0}0.8 $\pm$ 0.1 \\ 
$\mathrm{Q1700-MD157}$ & 0.5 $\pm$ 0.3 & 0.8 $\pm$ 0.3 & 0.6 $\pm$ 0.4 & \phm{0}1.0 $\pm$ 0.1 \\ 
$\mathrm{Q1700-MD174}$ & 0.9 $\pm$ 0.5 & 0.9 $\pm$ 0.4 & 1.0 $\pm$ 0.0 & \phm{0}1.0 $\pm$ 0.2 \\ 
$\mathrm{Q2343-BX333}$ & 1.0 $\pm$ 0.3 & 1.0 $\pm$ 0.3 & 1.0 $\pm$ 0.0 & \phm{0}1.0 $\pm$ 0.1 \\
\hline 
\textbf{average\tablenotemark{b}} &  0.7 $\pm$ 0.3 & 0.8 $\pm$ 0.2 & 1.0 $\pm$ 0.4 & 0.92 $\pm$ 0.07
\enddata
\tablenotetext{a}{Correction values represent the derived parameters from the AGN+SF modeling fit divided by the SF modeling fit through the $K$ band data.}
\tablenotetext{b}{Uncertainties for the average correction factors represent the standard deviation of the distributions.}

\end{deluxetable}

\begin{deluxetable}{lcccccc}
\tabletypesize{\scriptsize}
\tablecaption{Best Fit Parameters, AGNs without IRAC Coverage\label{tab:noiracparam}}
\tablewidth{0pt}
\tablehead{
\colhead{OBJECT} & \colhead{log[Age]} & \colhead{log[Age]$_{corr}$} & \colhead{log(M$_{*}$)} &  \colhead{log(M$_{*})_{corr}$} 
 \\
& \colhead{Myr} & \colhead{Myr}  &  &   
}
\startdata

Q0000-C7 & 3.26 $\pm$ 0.00 & 3.25 $\pm$ 0.19 & 10.35 $\pm$ 0.14 & 10.21 $\pm$ 0.29 &   \\ 
Q0000-C14 & 3.30 $\pm$ 0.00 & 3.30 $\pm$ 0.19 & 11.43 $\pm$ 0.12 & 11.30 $\pm$ 0.27 &   \\ 
CDFB-D3 & 3.36 $\pm$ 0.00 & 3.36 $\pm$ 0.19 & 11.15 $\pm$ 0.21 & 11.02 $\pm$ 0.36 &   \\ 
Q0201-OC12 & 3.41 $\pm$ 0.00 & 3.41 $\pm$ 0.19 & 11.53 $\pm$ 0.19 & 11.39 $\pm$ 0.34 &  \\ 
Q0256-MD37 & 3.06 $\pm$ 0.29 & 3.05 $\pm$ 0.35 & 10.78 $\pm$ 0.21 & 10.65 $\pm$ 0.36 &  \\ 
Q0933-MD38 & 2.96 $\pm$ 0.26 & 2.95 $\pm$ 0.33 & 10.52 $\pm$ 0.25 & 10.39 $\pm$ 0.41 &   \\ 
Q1217-BX46 & 2.86 $\pm$ 0.14 & 2.85 $\pm$ 0.24 & 10.49 $\pm$ 0.08 & 10.35 $\pm$ 0.25 &   \\ 
WEST-MM47 & 2.06 $\pm$ 0.35 & 2.05 $\pm$ 0.41 & 10.41 $\pm$ 0.13 & 10.27 $\pm$ 0.28 &   \\ 
Q1422-C73 & 3.26 $\pm$ 0.00 & 3.25 $\pm$ 0.19 & 10.82 $\pm$ 0.15 & 10.69 $\pm$ 0.30 &   \\ 
Q1422-MD109 & 3.01 $\pm$ 0.27 & 3.00 $\pm$ 0.34 & 11.29 $\pm$ 0.14 & 11.16 $\pm$ 0.29 &   \\ 
Q1623-BX454 & 2.76 $\pm$ 0.31 & 2.75 $\pm$ 0.36 & 10.03 $\pm$ 0.19 & 9.90 $\pm$ 0.33 &   \\ 
Q1623-BX747 & 2.81 $\pm$ 0.12 & 2.80 $\pm$ 0.23 & 10.32 $\pm$ 0.09 & 10.19 $\pm$ 0.25 &   \\ 
Q2233-D3 & 2.71 $\pm$ 0.63 & 2.70 $\pm$ 0.66 & 11.36 $\pm$ 0.25 & 11.22 $\pm$ 0.41 &   \\ 
Q2233-MD21 & 2.66 $\pm$ 0.71 & 2.65 $\pm$ 0.74 & 10.20 $\pm$ 0.19 & 10.07 $\pm$ 0.34 &   \\ 
DSF2237A-D11 & 3.32 $\pm$ 0.16 & 3.32 $\pm$ 0.25 & 10.20 $\pm$ 0.20 & 10.06 $\pm$ 0.35 &  \\ 
DSF2237B-MD53 & 3.44 $\pm$ 0.00 & 3.44 $\pm$ 0.19 & 11.37 $\pm$ 0.10 & 11.24 $\pm$ 0.26 &   \\ 
Q2346-BX445 & 3.44 $\pm$ 0.00 & 3.44 $\pm$ 0.19 & 11.33 $\pm$ 0.10 & 11.19 $\pm$ 0.26 &   \\ 

 \\
\hline \hline
 \\ 
OBJECT & $E(B-V)_{SF}$ & $E(B-V)_{SF,corr}$ & SFR & SFR$_{corr}$ & $\chi^{2}$,$\chi^{2}_{red}$$^{a,b}$ \\
 & \colhead{M$_{\sun}$} & \colhead{M$_{\sun}$} & M$_{\sun}$ yr$^{-1}$ & M$_{\sun}$ yr$^{-1}$ \\
 \\
\hline \hline
 \\ 

Q0000-C7 & 0.015 $\pm$ 0.017 & 0.014 $\pm$ 1.133 & \phantom{0}12 $\pm$ 4\phantom{00} & \phantom{0}10 $\pm$ 5\phantom{00} & 151.65, 151.65 \\ 
Q0000-C14 & 0.315 $\pm$ 0.024 & 0.289 $\pm$ 0.076 & 135 $\pm$ 37\phantom{0} & 104 $\pm$ 55\phantom{0} & 2.56, \phantom{00}-\phantom{0} \\ 
CDFB-D3 & 0.295 $\pm$ 0.044 & 0.271 $\pm$ 0.149 &  \phantom{0}62 $\pm$ 30\phantom{0} & \phantom{0}48 $\pm$ 36\phantom{0} & 8.20, \phantom{00}-\phantom{0} \\ 
Q0201-OC12 & 0.460 $\pm$ 0.032 & 0.422 $\pm$ 0.070 & 129 $\pm$ 57\phantom{0} & \phantom{0}99 $\pm$ 68\phantom{0} & 7.44, 7.44 \\ 
Q0256-MD37 & 0.225 $\pm$ 0.052 & 0.206 $\pm$ 0.231 & \phantom{0}53 $\pm$ 31\phantom{0} & \phantom{0}41 $\pm$ 35\phantom{0} & 0.00, \phantom{00}-\phantom{0} \\ 
Q0933-MD38 & 0.000 $\pm$ 0.000 & 0.000 $\pm$ 0.000 & \phantom{0}37 $\pm$ 2\phantom{00} & \phantom{0}28 $\pm$ 11\phantom{0} & 0.94, \phantom{00}-\phantom{0} \\ 
Q1217-BX46 & 0.255 $\pm$ 0.026 & 0.234 $\pm$ 0.102 & \phantom{0}43 $\pm$ 10\phantom{0} & \phantom{0}33 $\pm$ 16\phantom{0} & 3.88, 1.94 \\ 
WEST-MM47 & 0.355 $\pm$ 0.026 & 0.326 $\pm$ 0.073 & 225 $\pm$ 34\phantom{0} & 174 $\pm$ 75\phantom{0} & 6.84, 6.84 \\ 
Q1422-C73 & 0.185 $\pm$ 0.031 & 0.170 $\pm$ 0.168 &  \phantom{0}37 $\pm$ 13\phantom{0} & \phantom{0}28 $\pm$ 17\phantom{0} & 2.84, \phantom{00}-\phantom{0} \\ 
Q1422-MD109 & 0.380 $\pm$ 0.037 & 0.348 $\pm$ 0.097 & 194 $\pm$ 70\phantom{0} & 150 $\pm$ 91\phantom{0} & 10.67, 10.67 \\ 
Q1623-BX454 & 0.090 $\pm$ 0.045 & 0.083 $\pm$ 0.500 & \phantom{0}19 $\pm$ 5\phantom{00} & \phantom{0}15 $\pm$ 7\phantom{00} & 3.68, 1.84 \\ 
Q1623-BX747 & 0.030 $\pm$ 0.010 & 0.028 $\pm$ 0.333 & \phantom{0}33 $\pm$ 5\phantom{00} & \phantom{0}25 $\pm$ 11\phantom{0} & 43.23, 14.41 \\ 
Q2233-D3 & 0.440 $\pm$ 0.096 & 0.403 $\pm$ 0.218 & 446 $\pm$ 303 & 344 $\pm$ 331 & 0.00, \phantom{00}-\phantom{0} \\ 
Q2233-MD21 & 0.380 $\pm$ 0.046 & 0.348 $\pm$ 0.121 & \phantom{0}35 $\pm$ 7\phantom{00} & \phantom{0}27 $\pm$ 13\phantom{0} & 1.47, 1.47 \\ 
DSF2237A-D11 & 0.065 $\pm$ 0.050 & 0.060 $\pm$ 0.769 & \phantom{00}8 $\pm$ 6\phantom{00} & \phantom{00}6 $\pm$ 6\phantom{00} & 0.00, \phantom{00}-\phantom{0} \\ 
DSF2237B-MD53 & 0.310 $\pm$ 0.022 & 0.284 $\pm$ 0.071 & \phantom{0}85 $\pm$ 21\phantom{0} & \phantom{0}66 $\pm$ 33\phantom{0} & 8.21, \phantom{00}-\phantom{0} \\ 
Q2346-BX445 & 0.265 $\pm$ 0.021 & 0.243 $\pm$ 0.079 & \phantom{0}77 $\pm$ 19\phantom{0} & \phantom{0}60 $\pm$ 30\phantom{0} & 2.26, 2.26 \\ 

\enddata
\tablenotetext{a}{Raw and reduced $\chi^{2}$ values for each best-fit model.}
\tablenotetext{b}{Objects with only three photometric data points have reduced $\chi^{2}$ values given by a dash.}
\end{deluxetable}

\begin{deluxetable}{lcc}
\tabletypesize{\scriptsize}
\tablecaption{UV Emission Features for Stellar Mass Subsamples\label{tab:mass_separated}}
\tablewidth{0pt}
\tablehead{
  & \colhead{$\mathrm{log}(M_{*}/M_{\sun}) > 10.7$} & \colhead{$\mathrm{log}(M_{*}/M_{\sun}) < 10.7$}
}
\startdata

$W_{\mathrm{Ly\alpha}}\tablenotemark{a}$     & \phm{0.}56$\pm$15\phm{.0}       & \phm{.}58$\pm$18\phm{.0}      \\
$W_{\mathrm{NV,1240}}\tablenotemark{a}$      & \phm{0}4.6$\pm$1.4\phm{0}      & 3.7$\pm$1.6\phm{0}   \\
$W_{\mathrm{CIV,1549}}\tablenotemark{a}$     & 17.6$\pm$3.3\phm{0}     & 7.4$\pm$3.3\phm{0}    \\
$W_{\mathrm{HeII,1640}}\tablenotemark{a}$    & 10.2$\pm$2.4\phm{0}      & 5.0$\pm$1.1\phm{0}    

\enddata
\tablenotetext{a}{Rest-frame EW in \AA, measured from the composite spectra. Uncertainties are calculated as described in
\S \ref{sec:uvcomposite}.}

\end{deluxetable}

\clearpage


\clearpage

\end{document}